\documentclass[useAMS,usenatbib]{article}
\usepackage{times}
\usepackage{fullpage}
\usepackage{graphicx}
\usepackage{latexsym}
\usepackage{gensymb}
\usepackage{hyperref}

\usepackage{amssymb,amsmath}
\usepackage{subfig}
\usepackage{psfig}
\usepackage{caption}
\usepackage{setspace}

\begin{document}
\def\mnras{MNRAS}
\def\apj{ApJ}
\def\aj{AJ}
\def\aap{A\&A}
\def\apjl{ApJL}
\def\apjs{ApJS}
\def\araa{ARA\&A}
\def\pasj{PASJ}
\def\na{NewA}
\def\pasp{PASP}
\def\apss{Ap\&SS}
\def\nat{Nature}
\def\amp{\&}

\begin{center}
\huge{\sc the modelling of feedback in star formation simulations}\\
\vspace{0.5in}
\large{James E. Dale}\\
\large{\url{dale.james.e@gmail.com}}\\
\large{University Observatory/Excellence Cluster `Universe'}\\
\large{Scheinerstrasse 1, 81679 M\"{u}nchen, Germany}
\end{center}
\sloppy
\section*{Abstract}
I review the current state of numerical simulations of stellar feedback in the context of star formation at scales ranging from the formation of individual stars to models of galaxy formation including cosmic reionisation. I survey the wealth of algorithms developed recently to solve the radiative transfer problem and to simulate stellar winds, supernovae and protostellar jets. I discuss the results of these simulations with regard to star formation in molecular clouds, the interaction of different feedback mechanisms with each other and with magnetic fields, and in the wider context of galactic-- and cosmological--scale simulations.
\section*{Keywords}
Star formation; feedback; numerical simulations
\section{Introduction and scope of this review}
\indent The formation of stars is arguably the most important process in astrophysics, impacting virtually every theoretical and observational subfield. Despite its prominence and decades of intensive study, there is still much about star formation that is not understood. One reason for this slow progress is the fact that the conversion of gas to stars is a non--linear process. There are a variety of reasons for this, such as the non--linearity of the self--gravitational forces which lead to the collapse of individual stars, but the issue of most interest here is that the rate of star formation is influenced by stars themselves via feedback.\\
\indent There are many other phenomena that affect the rate and morphology of star formation, such as mergers between galaxies (e.g. \cite{2010AJ....140...75W}, \cite{2013MNRAS.430.1901H}), collisions between molecular clouds (e.g. \cite{2009ApJ...696L.115F}, \cite{2009ApJ...700..358T}) or the passage of spiral shocks through galactic discs (e.g. \cite{2013ApJ...779...45M}, \cite{2013MNRAS.430.1790B}). However, while a given star--forming region may or may not have experienced these particular perturbations, feedback from the stars themselves is, by definition, \emph{always} present, and it is this broad range of processes that are the focus of this review\\
\indent Stellar feedback has been invoked, with varying degrees of success, to solve a wide range of issues and problems in astrophysics, including the slow and inefficient star formation observed in molecular clouds and on galactic and cosmological scales, the triggering of star formation, the formation of disc galaxies and the suppression of excess dwarf galaxy formation in cosmological simulations. A glance at almost any image from \emph{HST}, \emph{Spitzer}, \emph{Herschel}, \emph{WISE} and many ground--based images reveals that the structure of the ISM is riddled with bubbles, shells, pillars and outflows, none of which can be explained without invoking feedback. Since one of the main purposes of astrophysical simulations is to help explain what is observed in the Universe, it is clear that feedback is a critical component of such simulations, and of any general model of star formation.\\
\indent More specifically, this is a review of \emph{numerical simulations of feedback}. A self--gravitating pure--hydrodynamics problem would already by of sufficient complexity to require the use of high--dimensional computer simulations. The inclusion of additional physical processes, particularly the transfer of radiation, only makes the problem more complex and the necessity of using simulations all the greater. However, progress in this field is very rapid and  the existence of a varied set of mutually--interacting feedback processes has resulted in a bewildering number of recent studies. A timely summary will therefore be of benefit to specialists and non--specialists alike. In this review, the fundamental physical processes will be only briefly rehearsed; a detailed overview can be found in \cite{2014prpl.conf..243K}. The spotlight will instead fall on the algorithms that have been written to model them, and of simulations which have been performed including them, and what we have learned from these simulations. Numerical studies of feedback have a long and rich history (e.g. \cite{1979A&A....71...59T}, \cite{1985A&A...145...70T}, \cite{1989A&A...216..207Y}, \cite{1996ApJ...469..171G}). No single review could encompass all this work, and this article will concentrate on articles published for the most part in the last ten years, and on two-- or three--dimensional simulations.\\
\indent Although feedback in the galactic context will be discussed, this review deals with \emph{stellar} feedback and, despite its evident importance, AGN feedback will not be covered. Interested readers are referred to \cite{2012ARA&A..50..455F}. Similarly, since we are here concerned with the connection between stellar feedback and star formation, the focus will be on feedback from low-- and intermediate--mass proto-- and pre--main stars, and on O--stars, whose entire life--cycle is comparable to the lifetimes of GMCs. Planetary nebulae and feedback from accretion onto compact objects will not be examined, and readers are instead directed to \cite{2002ARA&A..40..439B}. For the most part, this review also does not cover feedback from Population III stars, an area of research which has grown substantially in recent years, and which is eloquently reviewed by \cite{2015ComAC...2....3G}.\\
\indent Magnetic fields are not commonly regarded as a type of feedback and will not receive dedicated attention here. However, the presence of a magnetic field will likely alter the response of a fluid to some or all of the feedback mechanisms under consideration and several authors have performed simulations including both feedback and magnetic fields. This work will be discussed, but the algorithms used to model the magnetic field will not be described.\\
\indent The structure of this article is as follows. Section 2 gives a brief introduction to the major feedback mechanisms, namely photoionisation, stellar winds, supernovae, accretion heating, radiation pressure and protostellar jets. Section 3 briefly introduces the major classes of astrophysical hydrodynamics codes -- particle--based schemes such as SPH, grid--based schemes such as AMR, and the new generation of moving--mech codes. Section 4 surveys the algorithms used for modelling radiation transport, winds, supernova and jets. Section 5 discusses the science which has been done with the codes described in Section 4 with reference to particular astrophysical problems, including the fragmentation and destruction of molecular clouds and the formation and evolution of spiral and dwarf galaxies. Section 6 contains a short summary and outlook for the future.\\ 
\section{Brief introduction to stellar feedback physics}
\indent Stellar feedback involves the insertion of matter, momentum and energy from stars into the surrounding fluid, from which the stars may also still be accreting gas. In terms of material, momentum and energy emitted \emph{per star}, massive OB--type stars far outweigh their lower--mass brethren in importance. In clouds where there are no O--stars (either because the cloud mass is too small to support massive star formation, or because there has not been time for O--stars to form), feedback from low-- and intermediate--mass stars in the form of jets and outflows and the conversion of gravitational potential energy to heat are dominant processes. On larger lengthscales and longer timescales, encompassing the formation and evolution of galaxies and cosmic star formation, it is again the O--stars that dominate the \emph{stellar} contribution to feedback, but the other smaller--scale processes may still have influence, since they help determine the environments in which the O--stars are born.\\
\subsection{Photoionisation}
\indent The physics of photoionisation were first elucidated in detail by \cite{1939ApJ....89..526S} and an excellent modern introduction can be found in \cite{2006agna.book.....O}. If a source of $Q_{\rm H}$ ionising photons per second ignites in a cloud with initial number density $n_{0}$ atoms cm$^{-3}$, the number density of ions $n_{\rm i}$ will also initially be equal to $n_{0}$. If the cloud is pure hydrogen and overall electrically neutral, the electron number density $n_{\rm e}$ must equal the ion number density, so we have $n_{\rm e}=n_{\rm i}=n_{0}$.\\
\indent As the ionisation front moves outwards, more photons are consumed by recombinations behind it. The recombination rate per unit volume is given by $\alpha n_{\rm e}n_{\rm i}=\alpha n_{0}^{2}$. Recombinations directly to the ground state produce photons which are able to ionise another atom elsewhere in the nebula. However, since a recombination of this kind quickly produces another ion, their overall effect can be neglected to first order, which is known as the `on--the--spot' (OTS) approximation. Only recombinations to states other than the ground state consume stellar photons, and this rate is given by $\alpha_{\rm B} n_{0}^{2}$, $\alpha_{\rm B}$ being the recombination rate to all states above the ground state.\\
\indent Eventually, the total recombination rate behind the ionisation front equals the rate at which the source produces photons, so no more neutral gas can be ionised. This state is known as the Str\"omgren sphere, described by the Str\"omgren radius $R_{\rm s}$, satisfying\\
\begin{eqnarray}
R_{\rm s}=\left(\frac{3Q_{\rm H}}{4\pi \alpha_{\rm B}n_{0}^{2}}\right)^{\frac{1}{3}}.
\label{eqn:str}
\end{eqnarray}
The ionised gas inside the Str\"omgren sphere has a temperature determined by the equilibrium of heating and cooling processes. The main heating process is the absorption of stellar photons. In a nebula with solar metallicity, the main coolants are optical line emission from metals, with some contribution from free--free emission. These process equilibrate at 8 000 -- 10 000K, although this temperature may be higher in low-metallicity environments where line cooling is less efficient. Since the temperatures of neutral clouds lie in the range 10--100 K, the Str\"omgren sphere will be vastly overpressured and will expand supersonically. This process was first studied by \cite{1978ppim.book.....S}, who derived the well--known relation for the radial evolution of the shock driven by the expanding HII region
\begin{eqnarray}
R(t)=R_{\rm s}\left(1+\frac{7}{4}\frac{c_{\rm II}t}{R_{\rm s}}\right)^{\frac{4}{7}}.
\end{eqnarray}
\indent Later, \cite{1990ApJ...349..126F} considered the case of HII regions expanding in clouds with radial density gradients $\rho(r)\propto r^{-n}$. They showed, in particular, that if $n>3/2$, the ionisation front will inevitably overtake the shock front and ionise the whole cloud, regardless of its mass. Early 2D numerical work by \cite{1996ApJ...469..171G} found that expansion of ionisation fronts in density gradients, and in uniform clouds, was accompanied by the formation of fingers of dense neutral material reaching into the HII region, reminiscent of observed pillars/elephants' trunks. They interpreted these structures in terms of the generic ionisation front instability analysed by \cite{1979ApJ...233..280G}.\\
\subsection{Main--sequence line--driven winds}
\indent The powerful fluxes of energetic photons emitted by OB stars are able to accelerate line--driven winds in their atmospheres. Material leaves the surface of the star at a velocity which asymptotically approaches the wind terminal velocity $v_{\infty}$, of order $10^{3}$ km s$^{-1}$ for a main--sequence O--type star (\cite{1999isw..book.....L}). The wind mass fluxes $\dot{M}$ for such stars are typically $\sim10^{-5}$M$_{\odot}$ yr$^{-1}$ but can approach $\sim10^{-4}$M$_{\odot}$ yr$^{-1}$.\\
\indent The effect of the wind depends sensitively on the thermodynamics of the shocked gas inside the wind bubble. The extremal assumptions are (i) that all the mechanical energy is retained by the bubble and the expansion is adiabatic, or pressure driven, in which case $R(t)=(L/\rho_{0})^{\frac{1}{5}}t^{\frac{3}{5}}$, and (ii) that cooling is maximally efficient, whence $R(t)=[2\dot{M}v_{\infty}/(3\rho_{0})]^{\frac{1}{4}}t^{\frac{1}{2}}$ Note in both cases the very weak dependence on the both the stellar properties and the density of the background medium. The reality is much more complex and lies somewhere between these extremes. More sophisticated models were first computed in 1D by \cite{1975ApJ...200L.107C,1977ApJ...218..377W} who examined the influence of thermal conduction between the hot shocked wind and the cool swept--up ISM and the corresponding evaporation of ISM material into the wind bubble, which rapidly comes to dominate its mass. \cite{1992ApJ...388...93K} and \cite{1992ApJ...388..103K} discuss in exhaustive detail the realistic case of partially radiative bubbles.\\
\indent Recently, \cite{2014MNRAS.442.2701R} have attempted to evaluate the importance of these processes \emph{observationally} by totting up the energy inserted by winds and that lost by various physical processes (radiative cooling, mechanical work, thermal conduction, gas--grain interactions) in four star--forming regions -- 30 Doradus, Carina, NGC 3603 and M17. They concluded that radiative cooling and mechanical work are unimportant, except in the case of M17 where work done on the cold gas can account for 38$\%$ of the injected wind energy. Adding thermal conduction and gas--dust cooling can account for the remainder of the input energy, but only if rather extreme assumptions are made. They instead infer that large fractions of the wind energy is lost via bulk leakage of hot wind gas, or small--scale mixing with the cold gas.\\
\subsection{Stellar evolution}
\indent The lifetimes of the most massive stars are comparable to or shorter than the inferred lifetimes of GMCs, so at least some of these stars are likely to enter the final stages of their lives while embedded in their natal clouds. The effects of exotic evolutionary phases such as LBV and WR stars have generally not been considered in GMC--scale simulations owing to the expense of simulating the long timescales involved.\\
\indent The WR stage profoundly alters the properties of the wind of a massive star. The mass loss rate increases dramatically to $\sim10^{-4}-10^{-3}$M$_{\odot}$ yr$^{-1}$ as the star sheds its outer layers, and the terminal velocity correspondingly declines to $\sim100$km s$^{-1}$. A thorough introduction to these and other kinds of stellar wind can be found in \cite{1999isw..book.....L}\\
\indent Understanding the effects of photoionisation and winds clearly relies on knowing how the ionising photon luminosity and wind mass loss rate and terminal velocity vary as a function of stellar age and mass. There are several observational studies of this issue, such as \cite{2006MNRAS.367..763S} or \cite{2010A&A...524A..98W}.\\
\indent Photoionisation and winds have traditionally been the most popular feedback mechanisms, perhaps since their effects are readily observable as $\sim10$pc bubble structures in atomic emission lines, radio continuum and infrared dust emission. Several authors have considered which of the two should be more important (e.g.\cite{2001PASP..113..677C,2002ApJ...566..302M}), generally concluding that expanding HIIRs are more damaging.\\
\subsection{Supernovae}
\indent When massive ($>$ 8M$_{\odot}$) stars exhaust their core hydrogen, a chain of events ensues which eventually results in a supernova explosion. The timescale on which hydrogen exhaustion occurs depends on the stellar mass. For a 10M$_{\odot}$ star, it is $\approx$30 Myr, but for the most massive stars, it may be as short as $\approx$ 5Myr, comparable to or shorter than the lifetimes of GMCs. The supernova explosion results in the ejection of $\sim$10 M$_{\odot}$ of metal--enriched material at speeds of $\approx3\times10^{3}$km s$^{-1}$, carrying approximately 10$^{51}$erg of total energy. In the classic problem of the point deposition of energy in a uniform medium, the supernova remnant passes though a brief `free--expansion' phase until the mass swept up becomes comparable to the ejecta mass, before entering the adiabatic Sedov--Taylor phase, during which the blast wave radius evolves with time as
\begin{eqnarray}
R(t)=\beta\left(\frac{E}{\rho_{0}}\right)^{\frac{1}{5}}t^{\frac{2}{5}}
\end{eqnarray}
where $\beta$ is an order--unity constant. The Sedov--Taylor phase ends when the cooling timescale becomes shorter than the expansion timescale and the supernovae remnant enters the radiative phase. In reality, the evolution of a supernova blastwave is likely to be much more complex, since it in reality encounters the interior of a wind bubble/HII region, rather than a smooth ambient medium. A comprehensive review of the expansion of astrophysical shock-- and blast--  waves, including the cases of nonuniform background density fields, is given by \cite{1988RvMP...60....1O}.\\
\indent Much of our understanding of the evolution of massive stars is predicated on the assumption of single stars evolving in isolation, owing to the extreme difficulty of assuming otherwise. However, a large fraction (45--75$\%$) of massive stars are members of binaries (usually with other massive stars), most of which are sufficiently close that the evolution of the two members is expected to be influenced by mass exchange or outright mergers (see \cite{2012ARA&A..50..107L} for a recent review of massive single and binary star evolution). How binary evolution would affect the production of ionising photons, the emission of winds, and the timing and energy of supernovae is a field with a lot of work still ahead of it.\\
\subsection{Accretion heating}
\indent In the early stages of formation, all protostars are accreting gas from their host GMCs via an envelope and a disc. As the gas falls into the potential well of the protostar and is eventually accreted by it, gravitational potential energy is converted to heat, which the protostar radiates away. Although not readily absorbed by the gas, this radiation is absorbed by dust. If the gas density is large enough, thermal coupling of the gas and dust may then transfer heat to the gas, effectively coupling the radiation field to the gas. \cite{2009ApJ...703..131O} point out that accretion luminosity dominates the energy budget of GMCs unless and until O--stars are born.\\
\subsection{Radiation pressure}
\indent All photons emitted by all stellar objects transfer momentum, as well as thermal energy, to the surrounding gas and dust. The accretion heating process described above also exerts radiation pressure forces on dust grains, although these are likely to be small and less important than the thermal pressures generated by dust and gas heating. However, radiation pressure from much more luminous massive stars is likely to be dynamically important in very dense clouds (e.g. \cite{2009ApJ...703.1352K,2010ApJ...709..191M,2010ApJ...710L.142F}).\\
\indent Although a conceptually simple idea, radiation pressure is difficult to compute in practice. The momentum \emph{carried} by the photons of a source of luminosity $L$ is $L/c$, giving (for a point source) a \emph{radiative momentum flux} at radius $r$ of $L/(4\pi r^{2}c)$. However, to compute the radiation pressure, one needs to know how much of this momentum is actually absorbed by the gas. Clearly, if the gas is optically thin, the absorbed momentum can be zero, even if the radiative momentum flux is very large. \cite{2009ApJ...703.1352K} introduce a parameter $f_{\rm trap}$ which encapsulates this uncertainty, so that
\begin{eqnarray}
P_{\rm rad}(r)=f_{\rm trap}\frac{L}{4\pi r^{2}c}.
\end{eqnarray}
If $f_{\rm trap}=1$, all the emitted photons are absorbed once before escaping. However, if the shell is moderately optically thick, photons are likely to interact several times, depositing more momentum, before escaping the shell and $f_{\rm trap}$ exceeds unity. Clearly, computing $f_{\rm trap}$ self--consistently is a very difficult radiative transfer problem.\\ 
\subsection{Jets}
\indent As well as thermal feedback, accretion also drives emission of stellar jets -- collimated high--velocity outflows emerging bidirectionally along the stellar rotation axis (for a recent view, see \cite{2014prpl.conf..451F}). The origin of jets is likely a magneto--hydrodynamic interaction between the star and its accretion disc, producing a magnetic field configuration which acts like a particle accelerator or collimator. The details of this process are still much debated and I will not  address them here. From the point of view of view of feedback, it is useful to know that initial jet velocities fall in the range 100--1 000 kms$^{-1}$, while material swept up by the jet bowshocks has typical velocities in the range 1--30 kms$^{-1}$. The mass--loss rate via jets is typically $\sim$10$^{-8}$M$_{\odot}$yr$^{-1}$. The shocks produced by jets are highly radiative, so to a good approximation they may be considered as sources of momentum only.\\
\section{Brief introduction to astrophysical fluid dynamics codes}
\indent Star formation takes place in the interstellar medium (ISM), the thin and usually hot gas which occupies much of the volume of most galaxies. The mean free paths of ions, atoms and molecules in the ISM tend to be small compared with the sizes of the structures which they belong to. It is therefore reasonable to approximate the ISM as a smoothly--varying fluid. However, in order to model the behaviour of a fluid on a computer, it is necessary to discretise it in some way into individual \emph{fluid elements}. How the discretisation is done inevitably has some effect on how physical processes are treated. A very brief summary of the three main types of astrophysical hydrodynamics codes and their advantages and disadvantages is therefore necessary before specific feedback algorithms are discussed. Since this review covers simulations of star formation, particular attention will be paid for each type of code to the way in which the formation of individual stars is modelled.\\
\subsection{Grid codes}
\indent Grid codes break fluids up into \emph{volume elements} which fill the space inside a set of boundaries delimiting the computational domain. The simulation is evolved by calculating the forces on the fluid in each volume element and moving material to or from that element's adjoining neighbours. Matter which crosses one of the boundaries is either destroyed (open or outflow boundaries), sent back into the domain as though bouncing off a wall (reflective boundaries), or reinserted at the opposite side of the domain (periodic boundaries). Material may also be created at the boundaries and allowed to flow into the domain (inflow boundaries). Some codes make do with a single fixed grid, but most allow a given grid cell to be subdivided or refined if higher resolution is required at that location, or incorporated into larger grid cells if the resolution at that place is higher than needed. Codes such as {\sc flash} \cite{2000ApJS..131..273F} and {\sc ramses} which are able to do this on the fly are known as Adaptive Mesh Refinement codes.\\
\indent The hydrodynamic equations themselves are solved by a wide variety of methods. Finite difference methods discretise the differential equations connecting quantities in adjacent cells. Finite volume methods integrate quantities over the volumes of the grid cells to compute fluxes between them, often by solving the Riemann problem. Exhaustive reviews of these various methods can be found in many textbooks, e.g. \cite{2007nmai.conf.....B}.\\
\indent The advantages of grid codes include being able to use virtually any criterion to decide when and where to refine or derefine the grid, and that the mass contained in individual grid cells may become very small or very large, so that very large dynamic ranges in the masses of objects are possible. Disadvantages are that some form of boundary condition \emph{must} be specified, which limits the \emph{volume} that can be studied, all grid cells must contain non--zero quantities of gas, so that some computational power may be wasted on simulating regions where very little is happening, and that fluid advection is almost inevitably slightly more efficient along the principal grid axes, leading to artefacts (often known as `carbuncles'). Tracing fluid flows in grid codes can be done by advecting passive scalars, but this is somewhat cumbersome and cannot be used to trace completely arbitrary flows. Modelling gravity in grid codes is non--trivial and is usually done by solving Poisson's equation using multigrid methods, although there are also implementations of tree algorithms similar to those used in particle--based codes. In addition, grid codes are not Galilean invariant and moving objects through the grid at speeds in excess of the local sound speed can lead to problems such as unphysical diffusion.\\
\indent If it occurs that the density of any particular grid cell becomes so high that its integration time becomes prohibitively short, some of the mass in the cell may be converted into a Lagrangian sink particle which is allowed to move between grid cells and to accrete further material. \cite{2010ApJ...713..269F} describe in detail their implementation of sink particles in {\sc flash}. As with SPH codes, the first criterion is that the density of the grid cell in question must exceed a given threshold. Six tests are then applied for all cells within a sink accretion radius of the dense cell: (i) the cells must all be on the highest allowed AMR refinement level; (ii) the flow at that location must be converging; (iii) the densest cell must lie at a local potential minimum; (iv) the gas within the accretion radius must be Jeans unstable; (v) the gas within the accretion radius must be bound; (vi) the volume must not overlap the accretion volume of a pre--existing sink. Once a sink is created, it may later accrete gas above the threshold density in the grid cell in which it finds itself, provided that gas is bound to it.\\
\subsection{Particle codes}
\indent Particle codes discretise fluids into \emph{mass elements} with a total mass equal to that of the whole fluid. The fluid is evolved by calculating the forces on each particle and moving the particles relative to one another. Since the particles can in principle go anywhere, it is not obligatory to have boundaries of any kind in a particle simulation, although reflective and periodic boundaries are commonly used. Open boundaries at which particles are destroyed, or inflow boundaries where new particles are inserted are also possible but uncommon.\\
\indent By far the most popular particle method is Smoothed Particle Hydrodynamics (SPH, e.g. \cite{2010ARA&A..48..391S}). SPH codes represent fluids as particles whose masses are smoothed or smeared out over a volume called the \emph{smoothing kernel}. The kernel is spherical, but the smoothing is done such that most of the mass is concentrated near the particle centre. The radius of the smoothing kernel is continuously recomputed so that it contains the centres of, on average (or, in some codes, exactly) a given number (usually $\approx50$) of `neighbour' particles. This is achieved in many codes (e.g. {\sc seren}, \cite{2011A&A...529A..27H}) by iteratively solving for each particle $i$ the expression $\rho_{i}h_{\i}^{3}=$constant (i.e. the fixed particle mass), as described in\cite{2010ARA&A..48..391S}. Fluid quantities such as density are computed as averages over a particle and all of its neighbours. The requirement that the number of neighbours be fixed automatically results in the smoothing kernels, which set the resolution of an SPH code, being smallest where the gas density is highest.\\
\indent Gravitational forces could in principle be computed directly between particles, but this scales extremely poorly as the number of particles increases so is rarely if ever used. Many codes reduce the expense of computing the gravitational forces by grouping particles into a tree structure, and computing gravitational forces using tree nodes instead of individual particles, provided that the tree node subtends a sufficiently small angle at the location where the forces are to be computed. Alternatively some codes use the particle--mesh method where the particle densities are converted to densities on a mesh and gravitational forces are computed by solving Poisson's equation.\\
\indent Not needing boundaries, the relative ease of computing self--gravitational forces, and the possibility of having parts of the computational domain genuinely empty are the main advantages of particle codes. Since each particle has a unique and preserved identity, it is also trivial to trace \emph{fluid flows} in particle simulations, which is very helpful when, for example, studying triggered star formation or pollution by supernovae. Particle codes generally do not treat shocks as well as grid codes, and they are restrictive in the sense that only the mass density can be used to control the resolution and other quantities cannot be refined on as needed. Traditionally, particle code also have problems dealing with contact discontinuities (\cite{2007MNRAS.380..963A}), although there are now several solutions to this issue available (e.g.\cite{2010MNRAS.405.1513R,2013ApJ...768...44S}).\\
\indent Since they are Lagrangian codes, implementing sink particles in SPH schemes is somewhat more straightforward than in grid codes, since the sinks can be treated like gas particles except that they do not feel or exert pressure forces. An early implementation was described by \cite{1995MNRAS.277..362B}. A density threshold is defined and any gas particles exceeding this threshold, along with their neighbours, are considered for sink formation. Four criteria must be met: (i) the ratio $\alpha$ of thermal to gravitational potential energy of the group must be $<0.5$ (ii) the sum of $\alpha$ and $\beta$, the ratio of rotational energy to gravitational potential energy must be $<1$ (iii) the total energy of the group must be negative (iv) the divergence of the acceleration must be negative. If these tests are passed, a sink is created with the total mass and momentum of the seed gas particles.\\
\indent Accretion onto the sink is achieved by assigning it an accretion radius and testing particles which pass within it. Particles which are bound to the sink with a specific angular momentum less than that required to form a circular orbit at the accretion radius are accreted. A much more sophisticated SPH sink particle algorithm was recently presented by \cite{2013MNRAS.430.3261H}. The creation criteria are again that the gas particle being considered for promotion must have a density exceeding a threshold, the putative sink particle would not overlap any pre--existing sinks, it must sit at a local potential minimum, and the candidate particle's density must be such that it is smaller than the Hill sphere defined by itself and any pre--existing sink.\\
\indent Once created, their sinks do not immediately accrete gas particles entering the accretion radius (which they term the `interaction zone'). Instead they are added to an interaction list (from which they may be struck off if they exit the interaction zone) and are gradually accreted over a physically--motivated timescale, while still being permitted to interact with other SPH particles. The smooth accretion and the continued interaction with gas particles outside the sink interaction zone, particularly in respect of angular momentum transfer, results in more physically--motivated and robust sink behaviour.\\
\indent In practice, most of these conditions are usually dropped in large--scale simulations where sink masses are too big for them to considered as single stars. In these cases, sink particles are often created simply from particles whose densities exceed a threshold.\\
\subsection{Moving--mesh codes}
\begin{figure*}
     \centering
   \subfloat[SPH]{\includegraphics[width=0.30\textwidth]{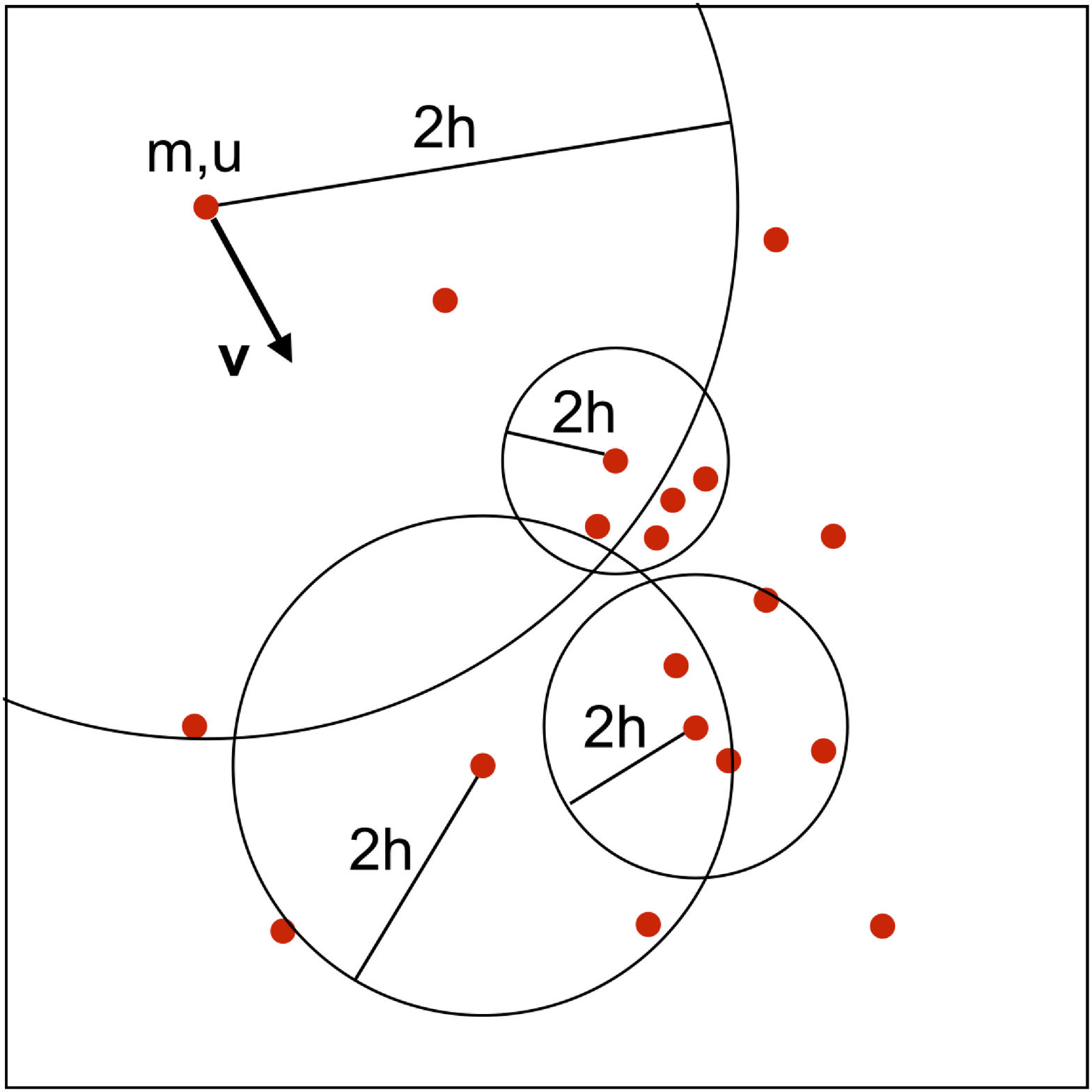}}
   \hspace{0.1in}
      \subfloat[AMR]{\includegraphics[width=0.30\textwidth]{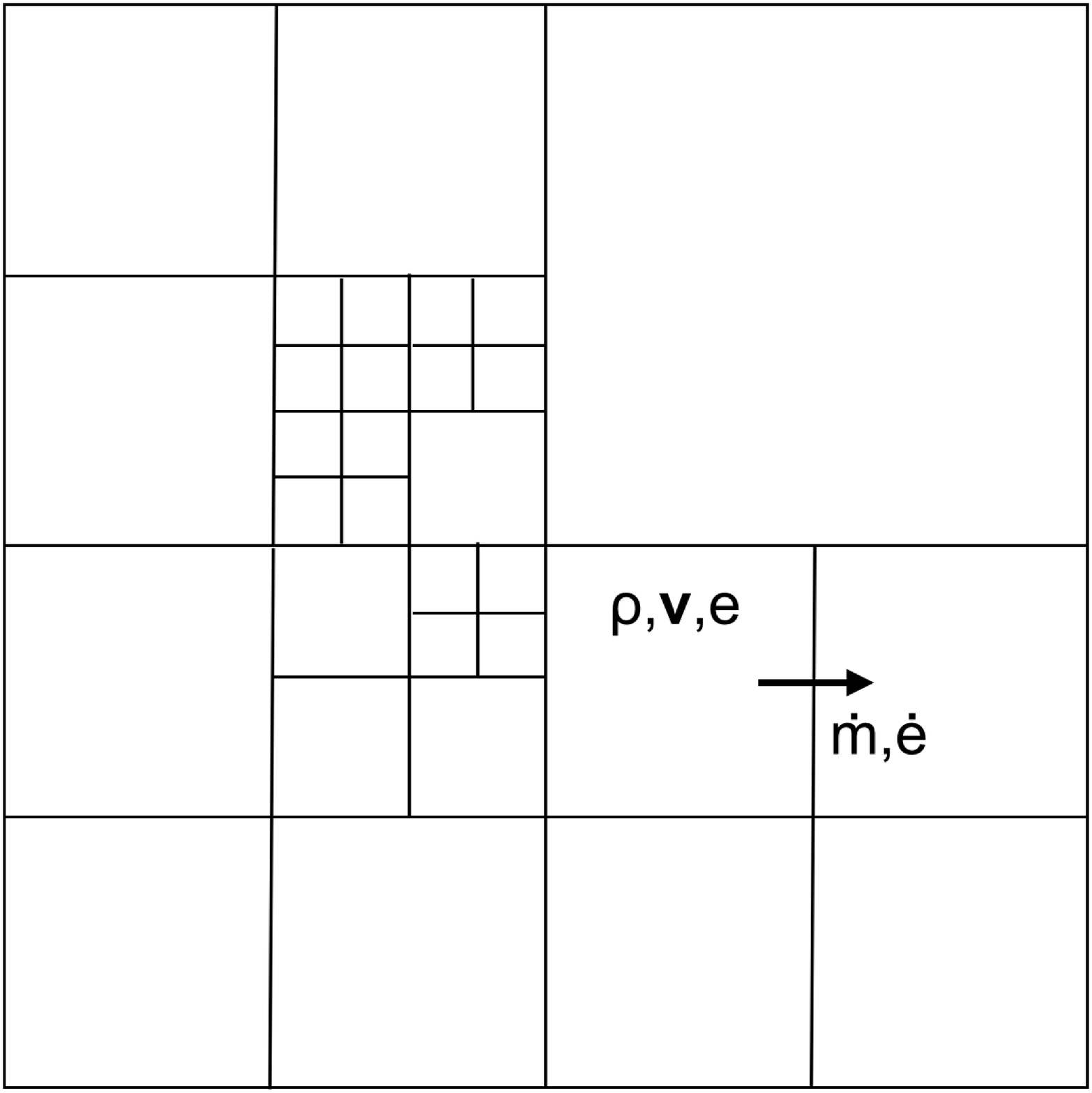}}
         \hspace{0.1in}
      \subfloat[Moving mesh]{\includegraphics[width=0.31\textwidth]{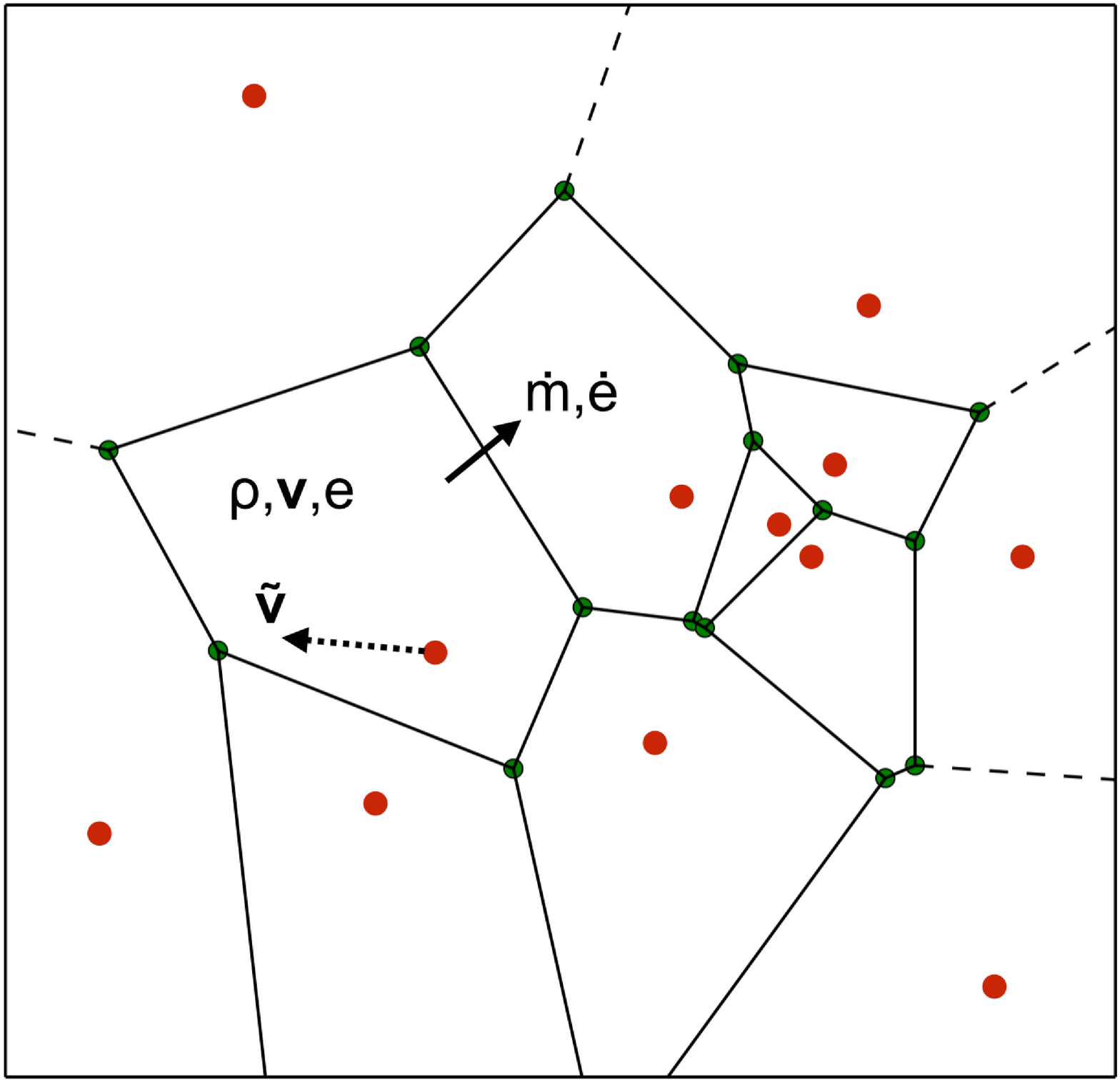}}
\caption{Simple illustration of the ideas behind the major types of astrophysical hydrodynamics code. The left panel shows how an SPH code represents a fluid, with red dots being individual particles carrying a mass $m$ and specific internal energy $u$. Each particle has a velocity $v$ and feels forces from other particles. Denser regions of gas are represented by higher concentrations of particles. The size of a particle is defined by the radius $2h$ which encloses a constant number of neighbours (here, only 4 for clarity). The middle panel shows an AMR representation of a fluid with only three levels of refinement. The gas in each cell has a density $\rho$, velocity $v$ and internal energy $e$. Cells exchange mass, momentum and energy with their neighbours. The right panel depicts a moving mesh code. The fluid mass is again represented by particles, shown as red dots, and the fluid volume is partitioned around these by a Voronoi tessellation. Cells exchange matter, momentum and energy, but the particles move with the fluid flow.}
\label{fig:3types}
\end{figure*}

\indent The most recent additions to the stable of astrophysical fluid codes are the moving--mesh codes (e.g. \cite{2011MNRAS.414..129G,2015MNRAS.450...53H}). The most widely used to date is {\sc arepo} (\cite{2010MNRAS.401..791S}) which solves the hydrodynamical equations using a finite--volume Godunov method on a 3D Voronoi mesh dynamically created around a population of Lagrangian particles which follow the fluid flow. The code is then in some sense a hybrid between traditional grid-- and particle--based codes, and shares most of the advantages and few of the drawbacks of both alternatives. Moving--mesh codes are Galilean--invariant like SPH codes but capture shocks and contact discontinuities as well as grid codes. They also share the ability to refine or derefine their resolution based on arbitrary conditions by splitting or merging their Lagrangian tracer particles. Sink particles can be implemented in ways analogous to those employed by grid--based codes, in which sinks remove mass from grid cells. The only disadvantages of moving--mesh codes are their relative complexity and novelty. Figure \ref{fig:3types} illustrates schematically the differences between these three types of code.\\
\section{Feedback algorithms}
This section briefly surveys some of the algorithms used to model stellar feedback mechanisms. The focus is on the algorithms themselves and the assumptions that underlie them. The results gained from using them will be discussed in a later section.\\
\subsection{Radiative transfer algorithms}
\indent During and after their formation, stars -- even low--mass objects -- are strong sources of radiation which deposit energy and momentum into the surrounding gas. The interaction of radiation and matter is an immensely difficult problem to solve and a great deal of effort has been expended on it. The summary given here is of necessity brief -- a more detailed and wide--ranging review can be found in \cite{2011ASL.....4..228T}. The two main processes of interest here are the radiation emitted mostly in the infrared by protostars (deriving from the loss of gravitational potential energy by accreting material and by the contracting protostar itself, and sometimes due to the burning of deuterium before hydrogen burning gets underway), and emission of ultraviolet photoionising photons by massive stars as they quickly settle onto the main sequence.\\
\indent In common with self--gravitational forces, radiative transfer (RT) in principle allows every part of the computational domain to communicate with every other part. However, the latter issue is much worse, since the communication between any two fluid elements depends on all the intervening material through which any radiation must pass, which of course does not apply to gravitational forces. This problem is in general too demanding to be solved explicitly but numerous physical and numerical approximations have been made to render it tractable.\\
\subsubsection{Ray--tracing methods}
\begin{figure*}
     \centering
   \subfloat[]{\includegraphics[width=0.30\textwidth]{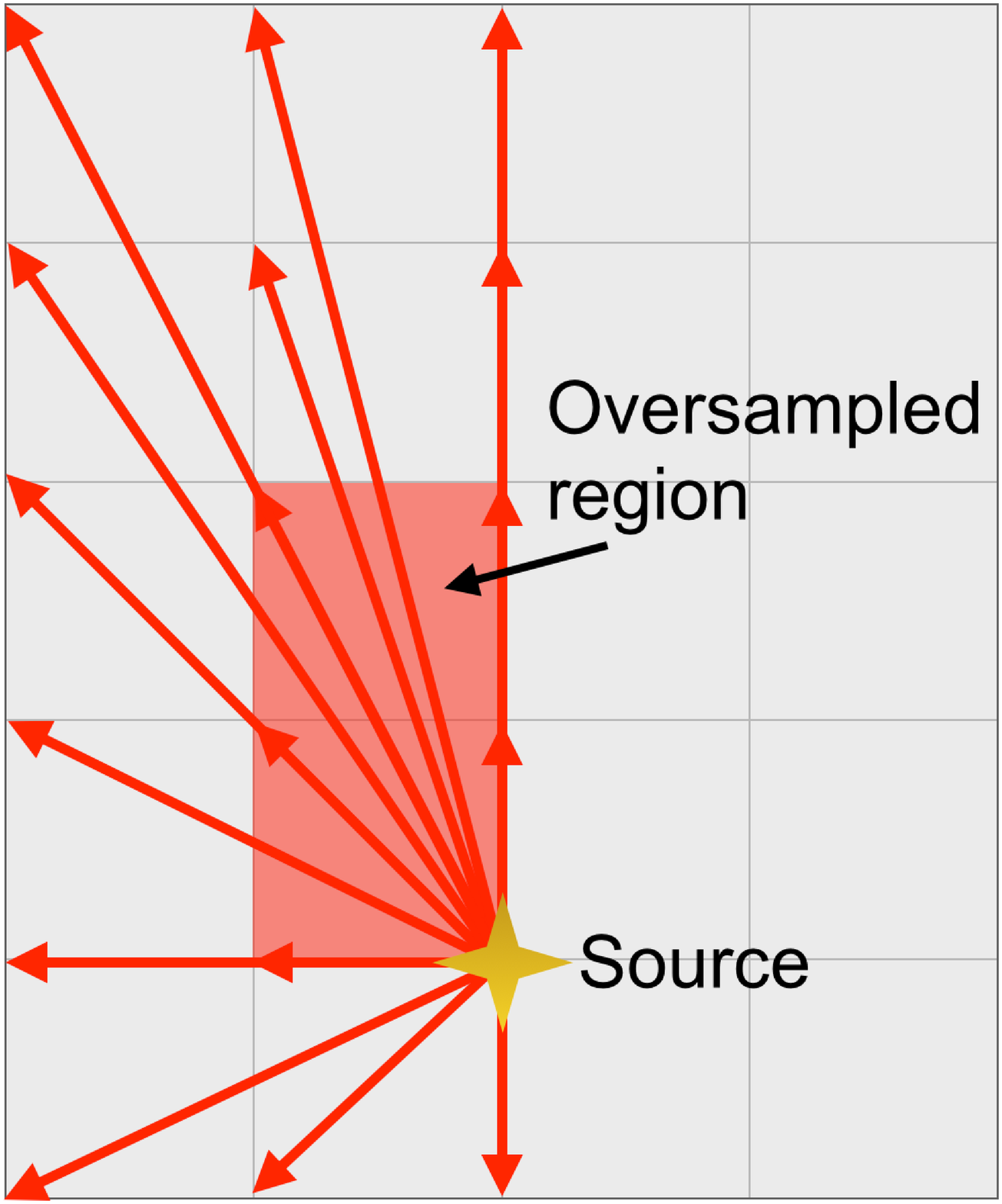}}
   \hspace{0.1in}
      \subfloat[]{\includegraphics[width=0.30\textwidth]{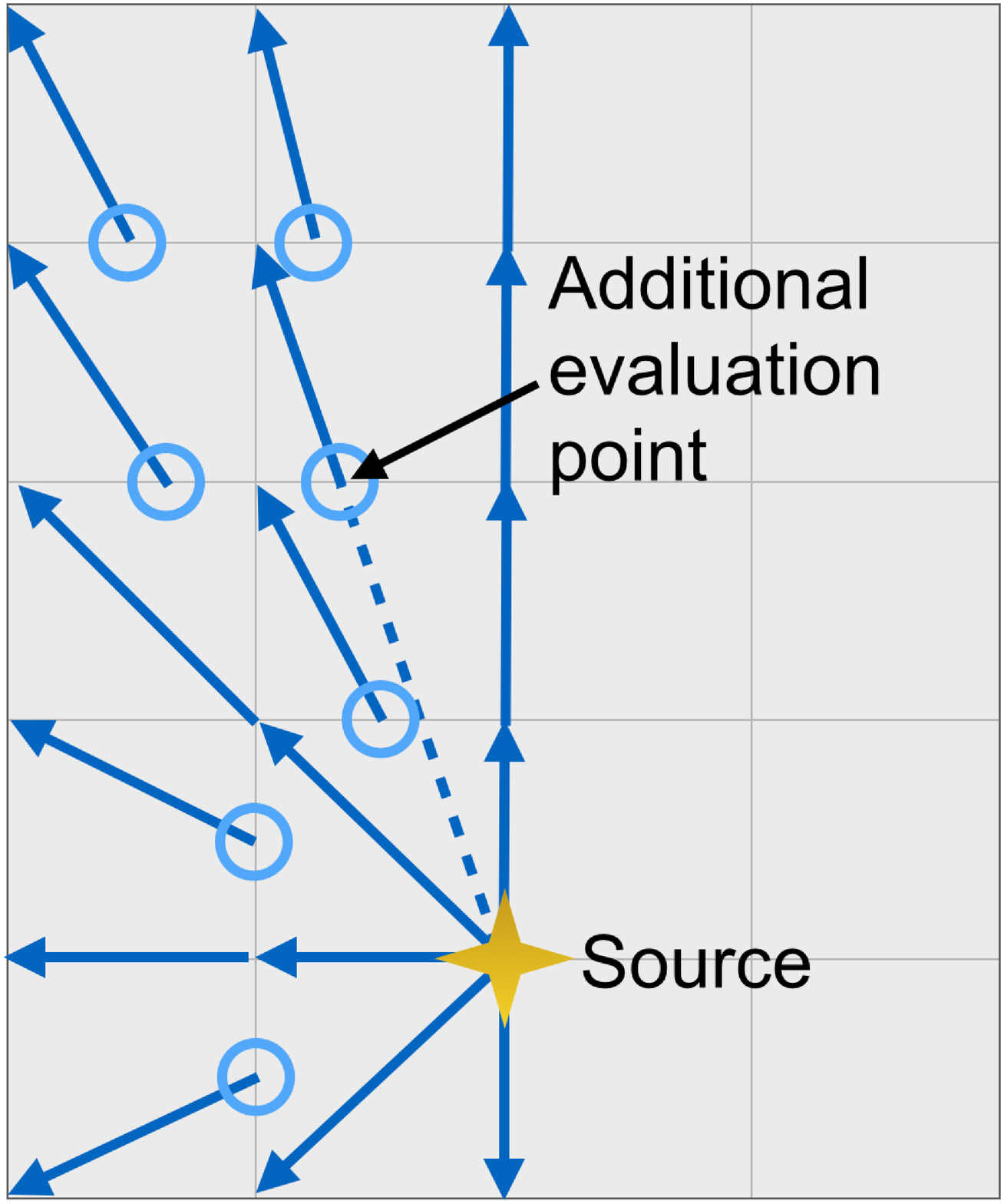}}
\caption{Illustration of the difference between long and short characteristics. Long characteristics (left panel) are drawn from the source to every grid vertex, so that regions close to the source are crossed by many rays. Short characteristics (right panel) are drawn across single cells, connecting the point where a vector from the source enters the cell (dotted line) to the nearest grid vertex. The radiation field at the entry points of cells (blue circles) must be interpolated from values at neighbouring vertices, making this method more diffusive.}
\label{fig:chars}
\end{figure*}

\indent The most intuitive approach to RT computations is ray--tracing, that is, drawing lines from radiation sources to target fluid elements and solving the radiation transfer equation along them. We shall limit ourselves to the consideration of unpolarised radiation. The time--dependent RT equation takes the form
\begin{eqnarray}
\frac{1}{c}\frac{\partial I_{\nu}}{\partial t}+{\bf n}.\nabla I_{\nu}=\epsilon_{\nu}-\kappa_{\nu}\rho I_{\nu},
\label{eqn:rte}
\end{eqnarray}
with $I_{\nu}$ the specific intensity at frequency $\nu$, ${\bf n}$ a unit vector pointing in the direction of radiation propagation, $\epsilon_{\nu}$ the emissivity of the medium and $\kappa_{\nu}$ the specific absorption coefficient. This is an equation in seven variables and solving it is a formidable problem. This is particularly true if large frequency ranges are of interest, for which the frequency dependence of the emissivity and opacity is likely to be significant and the problem has to be solved for many different values of $\nu$. It is common to avoid this issue by computing effective average emissivities and opacities, commonly referred to as the `grey' approximation.\\
\indent It is also common to simplify Equation \ref{eqn:rte} by various approximations. If the time--dependence can be neglected, equivalent to finding a radiative equilibrium, the time--independent RT equation results:
\begin{eqnarray}
{\bf n}.\nabla I_{\nu}=\epsilon_{\nu}-\kappa_{\nu}\rho I_{\nu}.
\label{eqn:rte_1}
\end{eqnarray}
Furthermore, it is often the case that the radiation field is dominated by a small number of very bright sources (e.g. stars), and that the emissivity of the gas may be neglected, yielding
\begin{eqnarray}
{\bf n}.\nabla I_{\nu}=-\kappa_{\nu}\rho I_{\nu}.
\label{eqn:rte_2}
\end{eqnarray}
\indent Several authors have made use of ray--tracing algorithms to attack the problem of photoionisation, differences being mainly in how they choose to cast their rays. Under the OTS approximation, any number of independent rays may be drawn emanating from an ionising source and the thermodynamic state of the gas can be found by locating the ionisation front along each ray using a generalisation of Equation \ref{eqn:str}. If the radius of the ionisation front is a function of direction $R_{\rm IF}(\theta,\phi)$, one can write
\begin{eqnarray}
\frac{Q_{\rm H}}{4\pi}=\int_{r'=0}^{r'=R_{\rm IF}(\theta,\phi)}n(r,\theta,\phi)^{2}\alpha_{\rm B}r'^{2}{\rm d}r'.
\label{eqn:rif_theta}
\end{eqnarray}
\indent \cite{2000MNRAS.315..713K} and \cite{2007MNRAS.382.1759D} using SPH codes defined rays connecting the ionising source to all active gas particles (a similar method was used by \cite{2007ApJ...665...85J}, except they first constructed a spherical grid with 10$^{5}$ rays, each divided into 500 radial segments). Particles' neighbours are tested to find the one closest (in an angular sense) to the ray leading back source. This process is repeated until the source is reached, generating a list of particles along the ray, whose densities are then used to calculate the integral in Equation \ref{eqn:rif_theta}. This can be used in a time--independent way to locate the ionisation front assuming ionisation equilibrium and heat the gas behind the ionisation front. However, \cite{2007MNRAS.382.1759D} use it to compute the photon flux at each particle to determine whether it is sufficient to keep an ionised particle in that state, or to (partially or completely) ionise a neutral particle, during the current timestep.\\
\indent This algorithm was modified in \cite{2011MNRAS.414..321D} and \cite{2012MNRAS.424..377D} to allow for multiple sources ionising the same HII regions. In the former paper, this was achieved by identifying all particles illuminated by more than one source and dividing their recombination rates by the number of sources illuminating them. The solution for the radiation field is iterated until the number of ionised particles converges. In the latter paper, a more sophisticated approach was adopted where the total photon flux at each particle is evaluated and the fraction of the recombination rate that each source is expected to pay for at a given particle is set by the flux striking it from that source as a fraction of the total. These methods give similar results in practice.\\
\indent In the Eulerian {\sc heracles} code by \cite{2012A&A...538A..31T}, solve a differential form of Equation \ref{eqn:rif_theta}, taking the photon flux $F$ as the variable of interest, writing
\begin{eqnarray}
\frac{1}{r^{2}}\frac{{\rm d}}{{\rm d}r}(r^{2}F)=Q_{\rm H}-n_{\rm H_{0}}\sigma F,
\end{eqnarray}
where $\sigma$ is the absorption cross section to ionising photons and $n_{\rm H_{0}}$ is the density of neutral hydrogen atoms. This equation is then complemented with equations describing photochemistry. The ionisation fraction $x$ is $n^{+}_{\rm H}/n_{\rm H}$ and $n_{\rm H}=n^{+}_{\rm H}+n_{\rm H_{0}}$. The number of photons absorbed in a grid cell of volume d$V_{\rm cell}$ is given by the flux of photons entering the cell through the surface d$A$ multiplied by the probability of absorption d$P=\sigma n_{\rm H_{0}}{\rm d}s$, with d$s$ being the pathlength along the ray. This allows one to write
\begin{eqnarray}
{\rm d}(xn_{\rm H})={\rm d}n_{\gamma}-{\rm d}n_{\rm H_{\rm rec}},\\
{\rm d}n_{\rm H_{\rm rec}}=\alpha_{\rm B}x^{2}n^{2}_{\rm H}{\rm d}t,\\
{\rm d}n_{\gamma}=F(r){\rm d}A{\rm d}t\frac{{\rm d}P}{{\rm d}V_{\rm cell}}=\sigma Fn_{\rm H}(1-x)\frac{{\rm d}A{\rm d}s}{{\rm d}V_{\rm cell}},
\end{eqnarray}
where the last term in the last equation is a geometrical dilution factor. Once these equations have been solved, the heating due to the absorption of ionising photons and the cooling due to recombinations can be computed.\\
\indent \cite{2009MNRAS.393...21G} modelled the propagation of plane--parallel ionising radiation in SPH. Rather than drawing rays to every particle, they used an adaptive scheme, casting a small number of rays along the photon propagation direction, and recursively refining them into four subrays up to five times at locations where the separation of the rays exceeded the particle smoothing lengths.\\
\indent Similarly, but in spherical geometry, \cite{2009A&A...497..649B} used the HEALPix tessellation (\cite{2005ApJ...622..759G}) to define rays, starting with the lowest level and refining rays into four subrays. Rays are refined when their separation, given by the radius $r_{\rm ray}$ at which they are defined multiplied by the separation angle $\theta_{l}$ of the HEALPix level $l$ to which they belong, exceeds the local smoothing smoothing length $h$ multiplied by a parameter $f_{2}$ of order unity. Values of $f_{2}$ of 1.0--1.3 were found to give a reasonable compromise between speed and accuracy.\\
\begin{figure}
\centering
\includegraphics[width=0.50\textwidth]{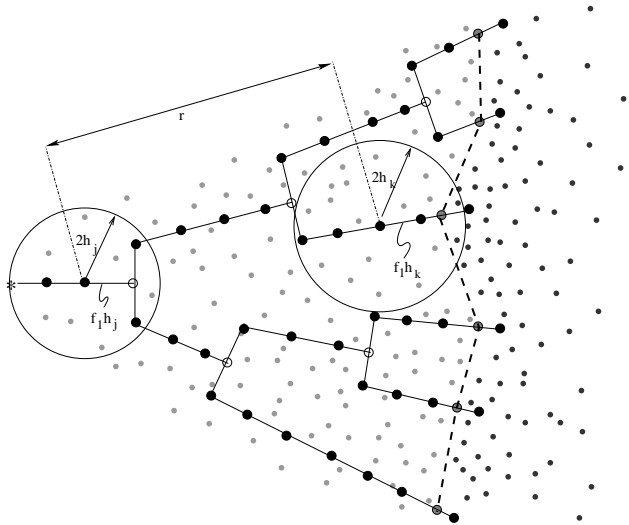}
\caption{Schematic representation of the adaptive SPH ray--tracing technique employed by \cite{2009A&A...497..649B}. The ionising source is represented by the star at the extreme left. Solid lines are rays, with black circles representing the evaluation points used to compute the discrete integral Equation \ref{eqn:rif_theta}. Grey circles are locations where they rays are split into four sub rays. The dashed line beyond which the particle density becomes abruptly larger is the ionisation front.}
\label{fig:bisbas09}
\end{figure}
\indent Once rays are defined, the discrete integral in Equation \ref{eqn:rif_theta} is computed along them by defining a series of evaluation points, each being a distance $f_{1}h$ from the previous one, with $f_{1}$ a dimensionless factor (given a value of 0.25) and $h$ being the local smoothing length. A schematic is shown in Figure \ref{fig:bisbas09}. The ionisation front is linearly smoothed over one smoothing length and the gas heated accordingly. A smiler adaptive ray--tracing scheme was presented by \cite{2002MNRAS.330L..53A} for use on Cartesian grids. This scheme differs from that of \cite{2009A&A...497..649B} in that child rays can be merged in regions where high resolution is not necessary, and that they solve a time--\emph{dependent} problem, using the results of the ray--trace to compute fluxes at cells. The algorithm is taken even further by \cite{2011MNRAS.414.3458W}, who also implement non--ionising radiation (e.g. Lyman--Werner dissociation) and radiation pressure.\\
\indent \cite{2007ApJ...671..518K} use a variant of the ray--tracing method of \cite{2002MNRAS.330L..53A}, periodically rotating the rays with respect to the Cartesian grid to avoid geometrical artefacts. To avoid spurious overcooling at the ionisation front, molecular heating and cooling processes are disabled for cells with ionisation fractions in the range $[0.01,0.99]$. They also explore the convergence of the results with changes in the update timestep for the radiation scheme, which effectively sets by how much the temperature of a cell near the ionisation front is allowed to change in one timestep. Allowing the temperature to change by a factor of 100 led to larger errors in the location of the ionisation front at early times, although the error declines as the front expands, and they caution against allowing sudden temperature jumps in photoionisation algorithms. This was also pointed out by \cite{2008ApJ...673..664W}, who explicitly  compared an algorithm employing the assumption of ionisation equilibrium embodied by Equation \ref{eqn:rif_theta} with a more sophisticated radiative transfer algorithm described in \cite{2006ApJS..162..281W}. They found that the structure of ionisation--front instabilities varied substantially between the two codes, especially at early times, and attributed the differences to the sudden heating inherent in the equilibrium method.\\
\\
\indent \cite{2010ApJ...711.1017P} implemented the ray--tracing algorithm of \cite{2006A&A...452..907R} in the {\sc flash} AMR code (\cite{2000ApJS..131..273F}). The algorithm computes column densities on nested grids using a hybrid long-- and short--characteristics method. A long characteristic is a ray drawn between a radiation source and an arbitrary cell, and may have many segments since it may pass through many intervening grid cells. A single long characteristic can transport radiation between two arbitrary points in the simulation. A short characteristic passes only across a single grid cell and only transports radiation from one cell to another (see Figure \ref{fig:chars} for an illustration). From a computational perspective, long characteristics are more amenable to  parallel computation, since each ray can be treated independently and the radiation transport equation solved along it. However, time is wasted near the source, since many rays pass though the same volume. Short characteristics cover the domain uniformly but radiation properties of cells must be updated from the source working outwards because each short characteristic must begin from the (usuallly interpolated) end solution of one closer to the source. Short characteristic methods are therefore difficult to parallelise and more diffusive.\\
\indent For a given AMR block, the \cite{2006A&A...452..907R} algorithm computes pseudo--short--characteristic rays which enter the block from the direction of the source and terminate at all the cell \emph{centres} for use local to that block, and pseudo--long--characteristic ray segments which terminate at the cell \emph{corners}. It is the latter which are shared with other processors so that what are effectively long characteristics can be stitched together across blocks. Once the rays have been defined, transfer of ionising photons is solved in a manner similar to that employed by \cite{2012A&A...538A..31T}, save that collisional ionisations are also accounted for.\\
\indent Several authors (e.g. \cite{2010MNRAS.403..714M}, \cite{2011MNRAS.414.1747A}) use the {\sc c$^{2}$--ray} algorithm of \cite{2006NewA...11..374M} to model photoionisation feedback. This method is photon--conserving and accurately solves the RT problem in the (common) case that the computational resolution elements are optically thick. For an infinitesimally thin spherical shell of radius $r$ with a radiation source of luminosity $L$ at its centre, the rate of absorption of photons per unit area is\\
\begin{eqnarray}
\Gamma_{\rm local}(r)=\frac{1}{4\pi r^{2}}\int_{\nu_{0}}^{\infty}\frac{L_{\nu}\sigma_{\nu}{\rm exp}[\tau_{\nu}(r)]}{h\nu}{\rm d}\nu.
\label{eqn:irate}
\end{eqnarray}
If, however, the shell has a finite thickness $\Delta r$, photons strike the inner surface at a rate $\dot{N}(r-\Delta r/2)$ and emerge from the outer surface at a rate $\dot{N}(r+\Delta r/2)$, and the number of atoms in the shell is $nV_{\rm shell}$. Equation \ref{eqn:irate} can then be rewritten as the photoionisation rate across a finite shell,\\
\begin{eqnarray}
\Gamma(r)=\frac{1}{4\pi r^{2}}\int_{\nu_{0}}^{\infty}\frac{L_{\nu}\sigma_{\nu}{\rm exp}[\tau_{\nu}(r)]}{h\nu}\frac{1-{\rm exp}(\Delta\tau_{\nu})}{nV_{\rm shell}}{\rm d}\nu.
\label{eqn:irate2}
\end{eqnarray}
If $\Delta\tau_{\nu}$ tends to zero, Equations \ref{eqn:irate2} reduces to Equation \ref{eqn:irate}. This issue could place severe constraints on the spatial discretisation required to model the propagation of an ionisation front. \cite{2006NewA...11..374M} also discuss issues of time discretisation. The above assumes that the optical depth does not change over the course of the computational timestep, so that either a very small timestep is required, or a time--averaged value of the optical depth should be used. They show that using at any location the time-- and space--averaged values of the neutral density, ionisation fraction and the optical depth from the source allows accurate solution using large timesteps, at least as long as the local recombination time, and volume elements with large optical depths.\\
\indent \cite{2012MNRAS.420..745C} present {\sc treecol} which solves Equation \ref{eqn:rte_2} in an SPH code using the gravity tree to speed up the calculation of optical depths. The idea behind the gravity tree is that when computing the gravitational forces acting on a given particle, groups of particles which are sufficiently far away can be amalgamated into pseudoparticles. The gravity tree groups all particles into a hierarchy, usually by recursively dividing the simulation domain into eight subdomains. When computing gravitational forces, the angle subtended at the particle by all the tree nodes is computed and compared to a parameter $\theta_{c}$, the tree--opening angle. If the node subtends an angle larger than $\theta_{c}$ at the particle in question, the node is decomposed into its children and they are tested. Once nodes subtending angles smaller than $\theta_{c}$ are found, they are treated as pseudoparticles.\\
\indent {\sc treecol} uses this formalism to save time in computing column--densities along rays. For every particle, a low--level (48-- or 192--wedge) {\sc healpix} tessellation is constructed and the contribution of tree nodes to the column density in each wedge is computed, respecting the tree--opening angle criterion. This results, for every particle, in a moderate--resolution map of the column density from its location to the edge of the simulation domain. This is ideal for computing \emph{external} heating by a radiation bath such as the ambient UV background that permeates the ISM. \cite{2015MNRAS.449.2643B} recently presented a code which hybridises the FLD model of \cite{2005MNRAS.364.1367W} with a version of {\sc treecol} to model, respectively, protostellar heating and heating from the intercloud radiation field. A method analogous to {\sc treecol} but designed specifically for AMR--based trees was recently described by \cite{2014A&A...571A..46V}.\\
\subsubsection{Moment methods}
\indent Many alternative approaches to the RT problem involve so--called moment methods. The fundamental radiative quantity is the intensity or spectral irradiance, $I_{\nu}$ which describes at a given location the rate at which energy is emitted per unit area, per steradian and per frequency interval. Integrating over frequency and integrating out the angular dependence yields the zeroth, first and second moments of the radiation field, better known as the energy density $E$, radiative flux ${\bf F}$ and the radiation pressure tensor ${\bf P}$:
\begin{eqnarray}
E=\frac{1}{c}\int_{\nu=0}^{\infty}\int I_{\nu}{\rm d}\Omega{\rm d}\nu\\
{\bf F}_{i}=\int_{\nu=0}^{\infty}\int I_{\nu}{\bf \hat{n}}.{\bf \hat{x}_{i}}{\rm d}\Omega{\rm d}\nu\\
{\bf P}_{ij}=\int_{\nu=0}^{\infty}\int  I_{\nu}({\bf \hat{n}}.{\bf \hat{x}_{i}})({\bf \hat{n}}.{\bf \hat{x}_{j}}){\rm d}\Omega{\rm d}\nu.\\
\end{eqnarray}
\indent While the meaning of $E$ and ${\bf F}$ are clear, the radiation pressure tensor needs a little explanation. Its $(i,j)$ component ${\bf P}_{ij}$ is the rate at which momentum in the ${\bf \hat{i}}$ direction is being advected by the radiation field through a surface whose normal is the ${\bf \hat{j}}$ direction.\\
\indent Once these transformations are done, the radiation field can be treated like a fluid, coupled to the matter density field by the equations of radiation hydrodynamics (RHD). In a frame following the flow of matter and under the assumption of local thermodynamic equilibrium, the RHD equations can be written (\cite{2001ApJS..135...95T}):
\begin{eqnarray}
\frac{{\rm D}\rho}{{\rm D}t}+\rho\nabla.{\bf v}=0\\
\rho\frac{{\rm D}{\bf v}}{{\rm D}t}=-\nabla p+\frac{1}{c}\chi_{F}F\\
\rho\frac{{\rm D}}{{\rm D}t}\left(\frac{E}{\rho}\right)=-\nabla.F-\nabla{\bf v}:{\bf P}+4\pi\kappa_{P}B-c\kappa_{E}E\\
\rho\frac{{\rm D}}{{\rm D}t}\left(\frac{e}{\rho}\right)=-p\nabla.{\bf v}-4\pi\kappa_{P}B+c\kappa_{E}E\\
\frac{\rho}{c^{2}}\frac{{\rm D}}{{\rm D}t}\left(\frac{{\bf F}}{\rho}\right)=-\nabla{\bf P}-\frac{1}{c}\chi_{F}{\bf F}
\end{eqnarray}
where $\chi_{F}$ is the frequency--integrated mean opacity including components due to absorption and scattering, and $\kappa_{P}$ and $\kappa_{E}$ are the Planck mean and energy mean absorption opacities, and the colon operator represents a double dot product operation.\\
\indent A popular approach to solving these equations, owing to its conceptual simplicity, is the flux--limited--diffusion (FLD) method. FLD simplifies the evolution of the radiation flux by first assuming a steady state, so that the derivative on the LHS of Equation 23 vanishes, then asserting that the radiation field is approximately locally isotropic, so that ${\bf P}=E/3$ and
\begin{eqnarray}
{\bf F}=-\frac{c}{3\chi}\nabla E
\label{eqn:fld1}
\end{eqnarray}
As the opacity becomes small, this approximation for the radiation flux tends to infinity, whereas in fact ${\bf F}$ cannot exceed $cE$. In the optically thin limit, the radiation field can in principle be strongly anisotropic, so that the assumptions behind the derivation of the moment equations break down. However, this problem can overcome by inserting an additional parameter into Equation \ref{eqn:fld1}:\\
\begin{eqnarray}
{\bf F}=-\frac{c\lambda(E)}{3\chi}\nabla E
\label{eqn:fld2}
\end{eqnarray}
where $\lambda(E)$ is the \emph{flux--limiter}, whose purpose is to prevent the energy flux becoming unphysically large. The flux limiter can be defined via the radiation pressure tensor as follows. ${\bf P}$ can be written in terms of $E$ as ${\bf P}={\bf f}E$, with ${\bf f}$ being the Eddington tensor, which simply encodes the local directionality of the radiation field. Formally, 
\begin{eqnarray}
{\bf f}=\frac{1}{2}(1-f){\bf I}+\frac{1}{2}(3f-1){\bf \hat{n}}{\bf \hat{n}}
\label{eqn:edt}
\end{eqnarray}
with ${\bf \hat{n}}=\nabla E/|\nabla E|$, ${\bf \hat{n}}{\bf \hat{n}}$ is a tensor formed by the vector outer product of ${\bf \hat{n}}$ with itself, and $f$ is the dimensionless Eddington factor. The flux limiter is then defined via $f=\lambda+\lambda^{2}R^{2}$, where $R=|\nabla E|/(\chi E)$. The function $\lambda$ must then be chosen so that the radiation field is always reconstructed smoothly.\\
\indent \cite{2005MNRAS.364.1367W} and \cite{2009MNRAS.392.1363B} implemented the effects of accretion heating from low--mass protostars in SPH using the FLD approximation. These models account for conversion of gravitational potential energy to heat in the accretion flows (i.e. it is radiated away in a physical manner as opposed to be being entirely lost, as in an isothermal model, or entirely retained as in an adiabatic model). However, the protostars have no intrinsic luminosity of their own, so that the feedback in these calculations is effectively a lower limit (\cite{2009ApJ...703..131O}).\\
\indent \cite{2007ApJ...656..959K} report on a FLD method implemented in the {\sc orion} AMR code to model radiative feedback in massive molecular cores, considering sink particles as radiation sources. The detailed protostellar models are derived from \cite{2003ApJ...585..850M} and account for dissociation and ionisation of infalling material, deuterium burning, core deuterium exhaustion, the onset of convection and hydrogen burning.  They set an opacity floor at high temperatures, since FLD schemes do not deal well with sharp opacity gradients where the radiation field can be strongly anisotropic.\\
\indent To avoid the issues that can be encountered in FLD with steep opacity gradients, \cite{2010A&A...511A..81K} present a novel hybrid method which combines FLD and ray--tracing. They split the radiation field into two components. Diffuse thermal dust emission is computed using an FLD method, whereas the direct stellar radiation field is handled by doing ray--tracing on a spherical grid either in the grey approximation, or using (typically $\approx 60$) frequency bins to capture frequency--dependent opacities.\\
\indent  \cite{2009Sci...323..754K} used the {\sc orion} code with FLD to model accretion onto high--mass protostars. They simulated a 100M$_{\odot}$, 0.1pc radius rotating core which collapsed to form a disc with a central massive object. Once the star achieved sufficient mass, Kelvin--Helmholtz contraction raised its luminosity to the point where radiation pressure became dynamically important. Radiation dominated bubbles inflated along the rotation axis and infalling material landed on the bubbles, travelled around their surfaces and was deposited in the accretion disc. The accretion rate onto the massive star was thus little altered. A second star grew in the disc resulting in a massive binary, and the radiation--inflated bubbles became Rayleigh--Taylor unstable, rapidly achieving a steady turbulent state.
\indent \cite{2012arXiv1211.7064K} and \cite{2012A&A...537A.122K} simulated essentially the same problem -- accretion onto a high--mass protostar -- using their hybrid FLD/ray--tracing approach and arrived at qualitatively different results. In the latter paper, they performed a comparison in which they operated their code using the FLD solver only. Both radiation transport schemes drove radiation--dominated cavities, but those produced by the hybrid scheme continued to grow until leaving the simulation domain, whereas those from the pure--FLD run collapsed along the rotation axis. They found that the cavity in the FLD case was unable to resist accretion onto it, which they attribute to the radiative flux in the FLD method tending to point in directions that minimise the optical depth, allowing photons to escape and depressurising the cavities. The hybrid scheme does not suffer from this problem, because the stellar radiation field is transported by direct ray tracing. They did not observe the kind of instability seen by \cite{2009Sci...323..754K}.\\
\indent A second alternative to FLD is to compute the Eddington tensor directly. These are usually known as variable Eddington tensor (VET) techniques, and differ in the ways they compute or estimate the tensor. \cite{2001NewA....6..437G} give a very clear description of the basis of these techniques. They define the VET as ${\bf f}_{ij}={\bf P}_{ij}/{\rm Tr}({\bf P}_{ij})$ with
\begin{eqnarray}
{\bf P}_{ij}=\int_{V}{\rm d}^{3}x_{1}\frac{L({\bf x}_{1}){\rm e}^{-\tau({\bf x},{\bf x}_{1})}}{|{\bf x}-{\bf x}_{1}|^{2}}\frac{{\bf x}^{i}-{\bf x}_{1}^{i}}{|{\bf x}-{\bf x}_{1}|}\frac{{\bf x}^{j}-{\bf x}_{1}^{j}}{|{\bf x}-{\bf x}_{1}|},
\end{eqnarray}
where ${\bf x}_{1}$ are the position vectors of the radiation sources and could consist of a set of $\delta$--functions modelling stellar sources, or could include every point in the domain if diffuse radiation fields are of interest. $L({\bf x}_{1})$ is the luminosity function of the sources and, ${\bf x}$ is the location at which the radiation field is to be computed. This integral is very expensive to evaluate because of the optical depth term ${\rm e}^{-\tau({\bf x},{\bf x}_{1})}$.\\
\indent \cite{2001NewA....6..437G} make the integral tractable by dropping this term. This means that the radiation flux is computed at each point in the simulation domain assuming that the radiation reaching it from all sources suffers only $r^{-2}$ geometric dilution, with no opacity effects. This is referred to as the optically thin variable Eddington tensor, or OTVET, scheme. They stress that this latter approximation is \emph{not} the same as in standard diffusion methods, because in these the Eddington tensor is computed strictly locally. They also point out that radiation need not be propagated at the speed of light, as long as its characteristic velocity is much larger than any dynamical velocities present. They suggest that even 100 km s$^{-1}$ may be adequate. Schemes making use of this approximation are known as reduced--speed--of--light (RSL) methods. OTVET methods have also been implemented in SPH codes, e.g. \cite{2009MNRAS.396.1383P} in {\sc gadget}--3, and \cite{2014MNRAS.439.2990S} implement an improved version of the \cite{2009MNRAS.396.1383P} scheme in the {\sc arepo} code.\\
\indent Some tension has recently emerged between FLD and OTVET schemes. \cite{2012ApJS..199....9D} use a short--characteristics VET method in {\sc athena} and perform explicitly comparisons with an FLD solver and a Monte Carlo solver implemented in the same code. The Monte Carlo algorithm is much too slow to be used in a dynamical simulation, so they instead use the three schemes to obtain an equilibrium solution on a single snapshot from a shearing--box simulation. They find that the VET and MC methods agree well, with the FLD solver being the odd one out, with the discrepancies largest in optically--thin regions.\\
\indent \cite{2012ApJ...760..155K} use a two--temperature FLD approximation on a 2D Cartesian grid to study the evolution of radiation pressure driven winds in a gravitationally--stratified atmosphere (intended as an approximation to ULIRGs and bright, dense young star clusters). Simulations are characterised in terms of whether the radiation pressure forces are greater or smaller than the gravitational forces. Where the radiation pressure forces are smaller, the gas undergoes vertical oscillations which eventually die out. Otherwise, an instability resembling the Rayleigh Taylor instability develops, with columns of dense gas falling into the low--density material at the base of the atmosphere where the radiation pressure forces are greatest. These columns contain most of the mass, but the low density and low optical depth gas occupies most of the volume, allowing radiation to escape in the vertical direction. The simulation reaches a steady turbulent state with nearly constant velocity dispersion and density scale height.\\
\indent \cite{2014ApJ...796..107D} also modelled radiation pressure feedback in ULIRGs. RT was implemented in 2D in the {\sc athena} code using either FLD or VET, and substantial differences exist between the results. In the low--flux case, both radiative transfer schemes (and \cite{2012ApJ...760..155K}) agree that the gas undergoes stable vertical oscillations. Their high--flux FLD case rapidly becomes Rayleigh--Taylor unstable, as does that of \cite{2012ApJ...760..155K}, and most of the gas sinks back towards the z=0 plane, where it remains in a turbulent state. However, in the VET calculation, the behaviour is very different. The RTI also develops, but most of the dense gas is nevertheless accelerated upwards out of the simulation domain. The volume--averaged Eddington factor in the VET run is generally larger than in the FLD run and exceeds unity for most of the time, while in the FLD run it is mostly just under unity. The difference is modest, but crucial. Deeper analysis shows that the two schemes agree well in the dense gas, where FLD should be a good approximation, but disagree on the magnitude and direction of radiation fluxes in low--optical depth regions where the diffusion approximation is likely to fail. In some regions, the FLD fluxes point in the opposite direction to the VET fluxes, accelerating the gas downwards instead of upwards, reinforcing the development of low--density channels and accelerating the development of the RTI.\\
\subsubsection{Monte Carlo methods}
\indent Monte Carlo methods solve the RT problem by emitting `photon packets' in randomly--chosen directions from the radiation source and following their paths through the simulation domain. In each fluid element through which the packet passes the probability of it being absorbed or scattered is computed and more random numbers are used to decide its fate. This is repeated with a large enough number of packets to sample the radiation field and iterated until some convergence criterion is satisfied. Monte Carlo codes are good at modelling processes such as scattering and re--emission, which are cumbersome to compute in ray--tracing schemes, and covering the frequency domain of the radiation transport problem is relatively simple, although it usually entails emitting more photon packets. The drawback of Monte Carlo methods is that they converge slowly and there is inevitably noise in the resulting temperature field owing to the discrete emission of energy.\\
\indent Monte Carlo methods are expensive and have traditionally been used to post--process the interaction of non--dynamically--important radiation fields with fixed pre--computed matter density distributions. However, improvements in computer power and algorithms have recently led owners of Monte Carlo codes to implement hydrodynamical algorithms in the codes (the reverse of what is usually done).\\
\indent The {\sc sphray} algorithm is presented by \cite{2008MNRAS.386.1931A}. Not a Monte Carlo code in the strict sense, {\sc sphray} uses the {\sc c$^{2}$--ray} method to solve the radiation transport equation on randomly--cast rays in an SPH simulation. All particles through which a ray passes are identified using the Axis--Aligned Bounding Box method acting on an octal tree. In solving the photoionisation problem, the OTS approximation is made (for both hydrogen and helium ionisation). The impact parameters of particles intersected by the ray is computed and the smoothing kernel integrated through accordingly to compute the particle's contribution to the optical depth. Photon packets are then propagated along rays and a fraction of their energy $(1-e^{-\tau_{\nu}})$ is subtracted as they pass through each particle.\\
\indent \cite{2008MNRAS.389..651P} present the {\sc traphic} SPH RT code. Radiation packets are emitted from sources every simulation timestep and propagated through the gas until a stopping criterion is satisfied. Each source (which can be a star particle or a gas particle) emits photons into an array of cones that covers the sky. Virtual SPH particles are placed into any cones which do not contain any real gas particles. Photon packets are distributed amongst the real and virtual neighbours of a source, and are then passed on in the radial propagation direction in a cone with the same opening angle as the original emission cone. The cones therefore subtend smaller and smaller solid angles at the source as one moves further away. Gas particles can receive and retransmit multiple photon packets, and packets coming from similar directions are merged to improve efficiency.\\
\indent Particles absorb energy from photon packets according to their opacities, removing a fraction ($1-$exp$^{-\tau}$) of the energy from the packet. Alternatively, photon packets can be reemitted, treating the gas particle as a radiation source, to model scattering. The absorption and reemission process continues until one of two stopping criteria are reached. If a state of radiative equilibrium is desired, the process is continued until all packets have been absorbed or have left the simulation domain. Otherwise, photon propagation is stopped when the packets have travelled a distance set by the speed of light and the timestep.\\
\indent \cite{2009MNRAS.397.1314N} present an algorithm for modelling radiation pressure in SPH using a Monte Carlo method. They track the trajectories of photon packets as ${\bf r}(t)={\bf r}_{0}+{\bf v}_{\rm phot}t$, where the propagation speed $|{\bf v}_{\rm phot}|$ need not be the speed of the light. They also reduce the packet energies $E_{\rm phot}$ and momenta $p_{\rm phot}=E_{\rm phot}/c$ continuously, rather than discretely, using\\
\begin{eqnarray}
\frac{1}{v_{\rm phot}}\frac{{\rm d}p_{\rm phot}}{{\rm d}t}=-p_{\rm phot}\rho\kappa.
\end{eqnarray}
Packets are destroyed when their momentum drops below $10^{-4}$ of its initial value.\\
\indent \cite{2011MNRAS.416.1500H,2012MNRAS.420..562H,2015MNRAS.448.3156H} implemented a grid--based finite volume hydrodynamical scheme into the pre--existing TORUS Monte Carlo code. Photon packets have constant total energy and are initially given a frequency chosen randomly using the source emission spectrum. The frequency determines the number of photons the packet represents. For a source luminosity $L$, integration time interval $\Delta t$ and number of packets $N$, the energy per packet is just $\epsilon=L\Delta t/N$. The packet propagates in a randomly--selected direction for a randomly--chosen pathlength $l$, at the end of which it is absorbed. A new packet is immediately emitted at that location in the same fashion, but with a frequency determined by the emission spectrum of the gas at the absorption point. This continues until the packet leaves the grid or its propagation time becomes equal to $\Delta t$.\\
\indent As packets travel through the grid, they contribute to the energy density in every cell through which they pass. Each packet contributes d$U$ which is estimated from
\begin{eqnarray}
{\rm d}U=\frac{4\pi J_{\nu}}{c}{\rm d}\nu=\frac{\epsilon}{c\Delta t V}\sum_{{\rm d}\nu}l,
\label{eqn:mc_u}
\end{eqnarray} 
with $V$ being the cell volume. In the photoionisation calculations presented by \cite{2012MNRAS.420..562H}, this can be used immediately to solve the ionisation balance equations (\cite{2006agna.book.....O}). For a given species $X$ with ionisation states $X^{i}$, $X^{i+1}$, etc., the balance equation reads
\begin{eqnarray}
\frac{X^{i+1}}{X^{i}}=\frac{1}{\alpha(X^{i})n_{\rm e}}\int_{\nu_{1}}^{\infty}\frac{4\pi J_{\nu}a_{\nu}(X^{i}){\rm d}\nu}{h\nu}=\frac{\epsilon}{\Delta t V\alpha(X^{i})n_{\rm e}}\sum\frac{la_{\nu}(X^{i})}{h\nu},
\end{eqnarray}
where $\alpha(X^{i})$ are recombination coefficients, $a_{\nu}(X^{i})$ are absorption cross sections, $n_{\rm e}$ is the electron number density, and $\nu_{1}$ is the ionisation threshold frequency for species $X^{i}$ respectively. {\sc torus} continues iterating the radiation transfer until an equilibrium temperature is achieved in each cell such that the heating and cooling rates balance. In radiation hydrodynamics calculations, the radiation transport problem is solved by iteration first, since the solutions for subsequent timesteps are likely to be relatively small perturbations on the initial state, and the radiation and hydrodynamics problems are then solved one after the other, with the radiation always being done first.\\
\indent \cite{2015MNRAS.448.3156H} discusses two methods by which radiation pressure forces may be computed in Monte Carlo schemes. The simplest is to compute the change in momentum $\Delta {\bf p}_{\rm phot}$ suffered by a packet in a cell and add that impulse to the gas in the cell\\
\begin{eqnarray}
\Delta{\bf p}_{\rm gas}=-\Delta{\bf p}_{\rm phot}=\frac{\epsilon}{c}({\bf \hat{u}}_{\rm in}-{\bf \hat{u}}_{\rm out}),
\end{eqnarray}
where ${\bf \hat{u}}_{\rm in}$ and ${\bf \hat{u}}_{\rm out}$ are the unit vectors of the packet's trajectory as it enters and leaves the cell respectively. The radiation force can then be computed at the end of the iteration from ${\bf f}_{\rm rad}=\sum\Delta{\bf p}_{\rm gas}/(\Delta t V)$.\\
\indent The above method suffers problems in optically thin gas, however, where the number of absorptions and scatterings can be small or zero. This can be overcome by computing the radiation pressure directly from the radiation flux. Equation \ref{eqn:mc_u} can be rewritten to give an expression for the radiation intensity\\
\begin{eqnarray}
I_{\nu}{\rm d}\Omega{\rm d}\nu=\frac{\epsilon l}{\Delta t V},
\end{eqnarray}
leading to a Monte Carlo estimate of the radiation flux
\begin{eqnarray}
{\bf F}_{\nu}=\int I_{\nu}{\rm d}\Omega=\frac{1}{\Delta t V {\rm d}\nu}\sum_{{\rm d}\nu}\epsilon l {\bf \hat{u}}.
\end{eqnarray}
The force may be computed by converting the above expression into one for momentum flux and multiplying it by an appropriate opacity $\rho\kappa_{\nu}$:
\begin{eqnarray}
{\bf f}_{\rm rad}=\frac{1}{c}\int {\bf F}_{\nu}\rho\kappa_{\nu}{\rm d}\nu =\frac{1}{\Delta t V {\rm d}\nu}\sum_{{\rm d}\nu}\epsilon l \rho\kappa_{\nu}{\bf \hat{u}}.
\end{eqnarray}
A much smoother estimate of the radiation pressure is recovered by this last expression, since all packets passing through a cell contribute to the estimate, and not just those that are absorbed or scattered.\\
\indent In principle, Monte Carlo methods are very easy to parellelise, since each photon packet can be treated independently. On a shared memory machine where all the processors can access the entire computational domain, parallelisation is than almost trivial. However, most problems are run on distributed--memory machines using the message--passing interface (MPI). The {\sc moccasin} code (\cite{2003MNRAS.340.1136E}) gets around this problem by giving a copy of the whole domain to each processor, but this is very memory intensive and it is more usual to decompose the domain into subdomains, as is done in {\sc torus}. When a photon packet leaves one sub--domain belonging to one processor for another belonging to a second processor, the first processor sends an MPI message containing the details of the packet to the second. In {\sc torus}, photon packets are communicated in stacks to cut down on communication overhead.\\
\subsubsection{Alternative methods}
\indent \cite{2007A&A...475...37S} devised a novel method for estimating the mean optical depth at the location of an SPH particle from the local density, temperature and gravitational potential, reasoning that the optical depth and the gravitational potential are both non--local properties determined by each particle's location within the simulation as a whole. Each particle is regarded as being inside a spherically symmetric polytropic pseudo--cloud at an unspecified location. For any location within the pseudocloud, the central density and scale length are adjusted to reproduce the \emph{actual} density and potential (neglecting any stellar contribution) at the SPH particle's location within the real gas distribution. The optical depth for any given location is computed by integrating out along a radius to the edge of the pseudo--cloud. The target particle is then allowed to move anywhere in the pseudo--cloud and a mass--weighted average optical depth over all possible positions is computed. Radiation transport is then conducted in the diffusion approximation.\\
\indent Another approach, taken for example by \cite{2009ApJ...698.1341U}, is to treat the detailed radiation transport problem as subgrid physics, and to parameterise it in some way. \cite{2009ApJ...698.1341U} used detailed one--dimensional {\sc dusty} (\cite{2000ASPC..196...77N}) models or analytic one--dimensional approximations to compute the temperature distribution near protostars owing to their accretion and intrinsic luminosities, and imprint the temperatures on their 3D SPH simulations.\\
\indent \cite{2015A&A...573A.112M} use the {\sc mcrt} Monte Carlo RT code (\cite{2004MNRAS.348.1337W}) in a novel way to essentially post--process the galactic--scale dynamical simulations of \cite{2013MNRAS.430.1790B}. Snapshots from the SPH simulation are interpolated onto a grid and the ionising radiation field from massive stellar sources is calculated using {\sc mcrt}. The solution is mapped back onto the SPH particles and a list of ionised particles generated. The masses of any of these that were later accreted by sinks in the dynamical simulation are then docked from the mass of the relevant sink, so that the influence of ionisation on the star formation rate may be inferred.\\
\subsection{Winds}
\indent Main--sequence O--star winds have received less attention than photoionisation in the context of simulations of star formation, probably owing to theoretical estimates suggesting that photoionisation is likely to be a more important feedback mechanism (e.g. \cite{2002ApJ...566..302M}). There have been many papers written analysing the evolution of wind bubbles in 1D (e.g. \cite{1975ApJ...200L.107C}, \cite{2012MNRAS.421.1283A}, \cite{2013ApJ...765...43S}), dealing in detail with the microphysics at the interface between the hot shocked wind and the cold ISM. However, relatively few authors have addressed this problem in 3D.\\
\indent Modelling the interaction of stellar winds with the ISM in SPH is difficult because the total mass of the wind is much smaller than the mass of molecular gas with which its interaction is to be studied. SPH is most stable when all the particles have the same mass. However, this is very difficult to achieve when attempting to model winds, since the wind gas may then be represented by too small a number of particles to be adequately resolved.\\
\indent Since winds inject momentum as well as matter and energy, \cite{2008MNRAS.391....2D} took the view that a lower limit to their effects could be established by injecting momentum alone. Treating stars as sources of momentum flux, they employed a Monte Carlo method in which wind sources were imagined to emit large numbers of `momentum packets' in random directions, which were then absorbed by the first gas particle which they struck. A similar technique was recently employed by \cite{2015ApJ...798...32N}, although the Monte Carlo element was avoided by distributing momentum into the wedges of a HEALPix grid which was recursively refined to ensure that the width of the wide end of the wedges was comparable to the local particle resolution at the location where the wind was interacting with the gas, in a similar fashion to the ray--casting technique employed by \cite{2009A&A...497..649B}.\\ 
\indent \cite{2012MNRAS.420.1503P} use the modular {\sc amuse} code to model embedded clusters. Mass and energy from winds and supernovae is injected into the gas, which is treated as adiabatic and is modelled using an SPH code. Mass loss rates and mechanical luminosities are computed from \cite{1992ApJ...401..596L}. Gas particles injected to model feedback have the same mass as those used to model the background gas and are injected at rest with respect to the injecting star. The gas particle masses used are $\sim10^{-3}-10^{-2}$M$_{\odot}$ and the wind mass loss rates vary between $\sim10^{-8}$ and $\sim10^{-5}$M$_{\odot}$yr$^{-1}$ per star. At the highest wind mass loss rates, the simulation timestep is such that 10s--100s of particles are injected per timestep. For the less powerful winds, there are some discretisation issues, but they do not affect the global results substantially. The internal energy carried by the wind particles is determined by the mechanical luminosity multiplied by a feedback efficiency parameter which accounts for unmodelled radiative losses. The values of the parameter are taken to 0.01--0.1 (see Section 5 for discussion of the results).\\
\indent Winds are somewhat more straightforward to model explicitly in grid--based codes and several authors have accomplished it, e.g. \cite{2008ApJ...683..683W}, \cite{2011ApJ...731...13N} and \cite{2013MNRAS.431.1337R}. Winds are modelled by explicitly injecting gas (and therefore momentum and kinetic energy) from the location of the sources with mass fluxes determined as functions of time from stellar evolution models. \cite{2013MNRAS.431.1337R} for example simulate the effects of winds on turbulent GMCs using models appropriate for the three O--stars present (with initial masses of 35, 32 and 28M$_{\odot}$), with wind terminal velocities fixed at 2 000 km s$^{-1}$ during the main sequence, dropping to 50 km s$^{-1}$ when each star enters its Wolf--Rayet phase (accompanied by dramatic increases in mass fluxes).\\
\subsection{Jets and outflows}
\index The modelling of jets and outflows in simulations of low--mass star formation has become increasing popular in the last decade, although the effort has been almost entirely confined to grid--based codes, even though jets can be adequately modelled by the injection of momentum and so are less problematic to model in SPH codes than main--sequence winds.\\ 
\indent In their fixed--grid MHD calculations, \cite{2006ApJ...640L.187L} and \cite{2007ApJ...662..395N} assume that every sink particle injects momentum instantaneously into its immediate surroundings. The injected momentum is taken to be $fM_{*}P_{*}$, with $f=0.5$ and $P=100$km s$^{-1}$, and $M_{*}$ being the stellar mass. In the earlier paper, the impulse is distributed isotropically over the 26 cells neighbouring the cell containing the sink, but in the later work, the outflows have two components. Using the direction of the local magnetic field to define the jet axis, they distribute a fraction $\eta$ of the outflow momentum into the neighbouring cells within 30$\degree$ of the axis. The remainder is distributed as before in a spherical component. \cite{2009ApJ...695.1376C} adopt a similar mechanism, in which they inject momentum into regions with 5$\degree$ opening angle ten cells across, but with no spherical component.\\
\indent \cite{2011ApJ...740..107C} describe in detail the implementation of an algorithm to model jets in the {\sc orion} AMR code, using sink particles as jet sources which inject momentum continuously as the sinks accrete. The mass injection rate of the jet is determined by the sink accretion rate in the absence of outflows $\dot{M}_{\rm acc}$ by $\dot{M}_{\rm jet}=f_{\rm w}/(1+f_{\rm w})\dot{M}_{\rm acc}$. Conservation of mass results in a modified accretion rate of $1/(1+f_{\rm w})\dot{M}_{\rm acc}$, and the jet velocity is set to a fraction $f_{\rm v}$ of the Keplerian velocity at the protostellar surface, the protostellar model being derived from \cite{2003ApJ...585..850M}. The total momentum injected by a star of mass $M_{*}$ is then $f_{\rm w}f_{\rm v}M_{*}v_{\rm k}$.\\
\indent The momentum is introduced over a range of radii $\chi_{\rm w}$ between four and eight grid cells from the source falling off as $r^{-2}$, and is modulated by a function of the polar angle $\theta$ and jet opening angle $\theta_{0}$, $\xi(\theta,\theta_{0})$ given by
\begin{eqnarray}
\xi(\theta,\theta_{0})=\left[{\rm ln}\left(\frac{2}{\theta_{0}}\right)({\rm sin})^{2}\theta+\theta_{0}^{2})\right],
\end{eqnarray}
derived from \cite{1999ApJ...526L.109M}. The connection with the simulation is made by inserting source terms into the density, momentum and energy equations.\\
\indent An analogous model is implemented in {\sc flash} by \cite{2014ApJ...790..128F}. Outflows are launched in spherical cones with opening angle $\theta_{0}$ about the sink particle rotation axes, with $\theta_{0}$ taken to be 30$\degree$. The outflow mass inserted in a timestep $\Delta t$ is scaled to the sink particle accretion rate so that $M_{\rm out}=f_{\rm m}\dot{M}\Delta t$, with $f_{\rm m}$ taken to be 0.3. The outflow mass is inserted uniformly inside the cones, but the outflow velocities are smoothed both with distance from the sink, and in an angular sense so that they drop to zero on the surfaces of the outflow cones, avoiding numerical instabilities. The chosen smoothing functions are as follows:\\
\begin{eqnarray}
\begin{array}{rlrl}
\mathcal{R}(r,r_{\rm out}) &=  {\rm sin}[\pi(r/r_{\rm out})] & {\rm for} & r \le r_{\rm out} \\
 &=  0  & {\rm for} & r > r_{\rm out} \\
 \Theta(\theta,\theta_{\rm out})&= {\rm cos}^{p}[\pi/2(\theta/\theta_{\rm out})]& {\rm for} & |\theta|\le\theta_{\rm out}\\
 &=  0  & {\rm for} & |\theta|>\theta_{\rm out}\\
\end{array}
\end{eqnarray}
The outflow characteristic velocity is set to the Keplerian value appropriate for a star with the mass of the sink and a radius of 10R$_{\odot}$. In order to produce a highly--collimated jet and a more extended `wind', the velocity profile is further modified to 
\begin{eqnarray}
\mathcal{V}(\theta,\theta_{\rm out})=\frac{1}{4}\Theta(\theta,\theta_{\rm out})+\frac{3}{4}\Theta(\theta,\theta_{\rm out}/6).
\end{eqnarray}
The authors stress that they take particular care to correct the mass and momentum fluxes in the two outflow cones to ensure that global momentum is exactly conserved. They also transfer a fraction of the accreted \emph{angular} momentum to the outflow.\\
\subsection{Supernovae}
\indent Few authors have attempted to model the effects of supernovae on well--resolved individual clouds, probably because most GMC--scale simulations do not form any O--stars, or do not progress far enough in time for massive stars that do form to reach the ends of their lives. However, there are some notable exceptions.\\
\indent \cite{2013MNRAS.431.1337R} extend their study of the impact of stellar winds on a turbulent molecular clump through the late and terminal stages of their three embedded O--stars. They follow the WR phase of each star and allow them to detonate as supernovae one after another by depositing 10$^{51}$ erg and 10M$_{\odot}$ of ejecta instantaneously.\\
\indent \cite{2012MNRAS.419..465D} set out to compare the relative efficacy of injecting supernova energy in thermal and kinetic form in SPH simulations, distributing energy amongst 32 particles nearest the explosion site. Both methods reproduced the Sedov solution well when global timesteps were employed, but poorly when individual particle timesteps were used. The two methods did agree in the case of individual particle timesteps if they were updated immediately after the energy release, and if the timesteps of neighbouring gas particles were forbidden from differing by more than a small factor (they suggest 4).\\
\indent Using an SPH code, \cite{2014arXiv1410.0011W} investigate the detonation of single explosions in clouds of mass $10^{5}$M$_{\odot}$ and radius 16pc, represented by 10$^{6}$ particles. They take the ejecta mass to be 8M$_{\odot}$ and represent the ejecta using particles of the same mass as those from which the cloud is built, resulting in there being 80 particles in the supernova remnant initially. The ejecta particles are randomly distributed in a 0.1pc sphere around the explosion source and given a radial velocity of 3 400 km s$^{-1}$, to give an explosion energy of 10$^{51}$ erg (they also perform a simulation in which the ejecta particles are not given outward initial velocities, but instead carry the 10$^{51}$ erg as thermal energy, finding similar results provided small timesteps are used for the ejecta particles and their neighbours). Thermodynamics are handled using a constant heating rate and a cooling rate constructed from the table in \cite{1995MNRAS.275..143P} and the analytical formula in \cite{2000ApJ...532..980K}. Particle energies are integrated using substeps in cases when the cooling time becomes shorter than the particle's dynamical time. They are able to accurately reproduce the Sedov--Taylor phase of the remnant evolution, as well as the transition to the radiative pressure--driven snowplough stage.\\ 
\subsection{Galactic--scale models}
\indent Simulations at the scale of a galactic spiral arm or disc can in general not model most of the feedback processes described above for want of resolution. An illustrative problem, for example, comes with the inclusion of supernovae in SPH simulations. On galactic timescales, a supernova is the instantaneous point release of a quantity of energy which must then be distributed in the gas near the explosion site. The \emph{mass resolution} of an SPH simulation places a strict lower limit on the amount of material over which the explosion energy can be distributed, which in turn sets the temperature of the gas. If the mass is too high, the temperature can be much lower than that expected in a supernova remnant and is likely to lie in the thermally--unstable regime of the cooling curve. The energy will therefore be quickly radiated away. This issue has traditionally been circumvented by temporarily disabling cooling at supernova sites. Other problems arise from the tremendous dynamic ranges that need to be modelled. As discussed by \cite{2006MNRAS.371.1125S}, poor resolution of the interface between hot diffuse material and cold dense clouds, particularly in SPH, artificially raises the density and decreases the cooling time in the hot material.\\
\indent \cite{2003MNRAS.339..289S} implement a feedback model in the {\sc gadget} SPH code designed to circumvent these problems. They implicitly assume that the gas consists of two phases -- a hot rarefied phase and a cold dense phase. Mass exchange between the phases occurs via star formation, the evaporation of cold material to form hot material, and the cooling and condensation of hot material to form cool material.\\
\indent Star formation occurs on a timescale $t_{*}$, and a fraction $\beta$ of the stellar mass is immediately recycled into the hot phase via SNe, so that
\begin{eqnarray}
\frac{{\rm d}\rho_{*}}{{\rm d}t}=(1-\beta)\frac{\rho_{\rm c}}{t_{*}}.
\end{eqnarray}
The heating rate from supernovae, expressed in terms of the specific internal energy stored in the hot phase, is taken to be
\begin{eqnarray}
\frac{{\rm d}}{{\rm d}t}(\rho_{\rm h}u_{\rm h})=\epsilon_{\rm SN}\frac{{\rm d}\rho_{*}}{{\rm d}t}=\beta u_{\rm SN}\frac{{\rm d}\rho_{*}}{{\rm d}t},
\end{eqnarray}
with $\epsilon_{\rm SN}$=10$^{48}$erg M$_{\odot}^{-1}$.\\
\indent The cold phase is assumed to evaporate into the hot phase by a process similar to thermal conduction. The mass evaporated from the cold phase is taken to be proportional to the mass released from SNe, so that
\begin{eqnarray}
\frac{{\rm d}\rho_{\rm c}}{{\rm d}t}=A\beta\frac{\rho_{c}}{t_{*}}.
\end{eqnarray}
The constant $A$ is environmentally dependent, with its functional form taken to be $A\propto\rho^{-\frac{4}{5}}$ and the normalisation treated as a parameter.\\
\indent The cold phase is assumed to grow by thermal instability:
\begin{eqnarray}
\frac{{\rm d}\rho_{\rm c}}{{\rm d}t}=-\frac{{\rm d}\rho_{\rm h}}{{\rm d}t}=\frac{1}{u_{\rm h}-u_{\rm c}}\Lambda(\rho_{\rm h},u_{\rm h}),
\end{eqnarray}
where $\Lambda$ is a cooling function, and the thermal instability is only permitted to operate in gas whose density exceeds a threshold. \cite{2003MNRAS.339..289S} showed that this model leads to self--regulated star formation, since star formation increases the evaporation rate of the cold clouds, which increases the density and cooling rate in the hot gas, thereby increasing the rate of formation of cold gas. A reasonable choice of $A$ and the star formation timescale results in a star formation law resembling the SK law.\\
\indent However, \cite{2003MNRAS.339..289S} point out that the assumed tight coupling between the hot and cold phases does not permit them to model star--formation driven galactic winds. These are a vital component of galaxy formation, since they further suppress star formation by ejecting, or at least cycling, baryons into diffuse regions where star formation does not occur, and they also assist disc formation by expelling low--angular momentum material from haloes. They therefore additionally implement a parameterised wind model. The wind mass loss rate is taken to be proportional to the star formation rate, $\dot{M}_{\rm w}=\eta\dot{M}_{*}$ and the wind carries a fixed fraction of the total supernova energy output. Gas particles are entrained in the wind in a probabilistic fashion, so that in a timestep $\Delta t$, the probability of entrainment is 
\begin{eqnarray}
p_{\rm w}=1-{\rm exp}\left[-\frac{\eta(1-\beta)x\Delta t}{t_{*}}\right],
\end{eqnarray}\\
where $x$ is the mass fraction in the cold phase. In order to prevent wind particles being trapped inside thick discs, their hydrodynamic interaction with other particles is disabled for a period of 50Myr.\\
\indent \cite{2006MNRAS.371.1125S} improve upon this model with a more sophisticated treatment of SNe. Star particles are assigned two smoothing lengths, enclosing equal masses of the hot and cold gaseous phases only. Supernova energy is divided between the hot and cold phases, weighted by the corresponding smoothing kernel. The fraction $\epsilon_{\rm h}$ assigned to the hot phase is injected as thermal energy. A second fraction $\epsilon_{\rm r}$ is assumed to be radiated away by the cold phase and lost. The remainder, $1-\epsilon_{\rm h}-\epsilon_{\rm r}$ is injected into the cold phase. Cold particles accumulate supernova energy until their thermodynamic properties are similar to their hot neighbours', at which point they are `promoted' to the hot phase. $\epsilon_{\rm h}$ and $\epsilon_{\rm r}$ are then treated as free parameters and adjusted to obtain the desired ISM properties.\\
\indent A novel approach to the inclusion of supernova feedback is presented by \cite{2013MNRAS.429.3068T}, who introduce a new variable $\epsilon_{\rm turb}$ which represents the density of non--thermal energy and whose evolution is followed via\\
\begin{eqnarray}
\rho\frac{{\rm D}\epsilon_{\rm turb}}{{\rm D}t}=\dot{E}_{\rm inj}-\frac{\rho\epsilon_{\rm turb}}{t_{\rm diss}},
\end{eqnarray}
where $\dot{E}_{\rm inj}=\dot{\rho}_{*}\eta_{\rm SN}\epsilon_{\rm SN}$ is the feedback energy injection term composed of the star formation rate, an efficiency factor and the energy per supernova, and $t_{\rm diss}$ is a dissipation timescale, which they set to 10Myr, comparable to the typical GMC lifetime (although they discuss several other possibilities). Feedback is connected to the hydrodynamics by defining $\epsilon_{\rm turb}=\rho\sigma_{\rm turb}^{2}/2$ and modifying the total pressure to include thermal and turbulent terms, so that $P_{\rm tot}=P_{\rm therm}+\rho\sigma_{\rm turb}^{2}$. A similar scheme has been explored by \cite{2013ApJ...770...25A}.\\
\indent Other authors attempt to model supernovae in a more explicit manner in similar fashion to simulations at GMC scales. In {\sc flash} \cite{2015MNRAS.449.1057G} inject 10$^{51}$ erg of supernova energy into a region containing 10$^{3}$M$_{\odot}$ of gas or eight grid cells across, whichever is larger. If thermalisation of this energy would result in temperatures above 10$^{6}$K, the energy is inserted in thermal form. However, at higher densities and masses, the overcooling problem would be encountered, essentially because the Sedov--Taylor phase of the remnant cannot be resolved. \cite{2015MNRAS.449.1057G} get around this problem by computing, given the actual density in the injection region, the expected bubble radius and momentum at the end of the Sedov--Taylor phase and injecting this momentum into the cells instead, while also heating them to 10$^{4}$K.\\
\indent \cite{2011MNRAS.417..950H} and \cite{2012MNRAS.421.3488H} point out that, on galactic scales, most of the gas is so dense that it cools efficiently -- only $\sim$10 percent of the total ISM pressure comes from the hot ISM. In denser regions, momentum dominates and radiation pressure, stellar winds and supernovae are all comparable when averaged over galactic dynamical timescales. \cite{2011ApJ...731...41O} make a similar point. \cite{2011MNRAS.417..950H} identify star--forming clumps around the densest gas particles and compute the stellar bolometric luminosity within the clump using {\sc starburst}--99 (\cite{1999ApJS..123....3L}) models and a Kroupa IMF. They assume that the momentum is distributed equally amongst all gas particles within a clump and for a given particle $j$ in a clump, the imparted momentum flux is then
\begin{eqnarray}
\dot{p}_{j}=(1+\eta_{p}\tau_{\rm IR})\frac{L_{j}}{c},
\end{eqnarray}
with $L_{j}=(M_{\rm gas,j}/M_{\rm gas,clump})L_{\rm clump}$. The first factor in the brackets represents momentum deposited in the gas by dust absorption of the optical and UV photons from the massive stars. The dust reradiates in the infrared and the second term allows for absorption of this radiation. $\tau_{\rm IR}$ is the optical depth across the clump, equivalent to a trapping factor, and $\eta_{p}\approx1$ is a parameter which can be adjusted to allow for other sources of momentum, e.g. jets, winds and SNe ($\eta_{p}>1$), or for photon leakage ($\eta_{p}<1$).  A similar model is presented in \cite{2013ApJ...770...25A} but their cosmological--scale simulations do not have the resolution to estimate the infrared optical depths, so these are estimated from subgrid models.\\
\indent As well as the momentum imparted by winds and SNe, they include the thermal energy released by the associated shocks, again tabulated from  {\sc starburst}--99 models, and including AGB winds as well as main--sequence and WR winds. This energy is deposited over the SPH smoothing kernels of the dense gas particles defining the star--forming clump centres. Photoionisation heating is implemented by finding particles within the Str\"omgren spheres centred on the clumps and heating the gas inside to 10$^{4}$K (or preventing it from cooling below this temperature).\\
\indent \cite{2011MNRAS.417..950H} and \cite{2012MNRAS.421.3488H} also allow for the fact that feedback partially or completely disrupts GMCs, so that substantial quantities of the IR and UV photons released inside them escape and are only absorbed at larger distances. \cite{2012MNRAS.421.3488H} allow the energy to spread over the larger of the local gas smoothing length or the local gravitational softening length.\\
\indent \cite{2014MNRAS.442.1545C} discuss a similar scheme for including radiation pressure from ionising photons where the gas is assumed to be locally optically thin. The radiation pressure is taken to be one third the radiation energy density, $P_{\rm rad}=4\pi I/(3c)$ and isotropic. The intensity $I$ is computed using {\sc starburst}--99 (\cite{1999ApJS..123....3L}) as a function of the stellar mass, spread over a reference area $A$, so that $I=\Gamma m_{*}/A$ and $P_{\rm rad}=\Gamma m_{*}/(R^{2}c)$, $R$ being set to half a grid cell size for cells containing stellar mass, and one grid cell size for neighbours of such cells. If the gas density in the cell exceeds 300cm$^{-3}$, the radiation pressure is boosted by a factor of ($1+\tau$), where $\tau=n_{\rm cell}/300 {\rm cm}^{-3}$ to account approximately for the trapping of infrared radiation in optically--thick gas. Using {\sc cloudy} (\cite{1998PASP..110..761F}) models, they also account of the change in the local heating and cooling rates resulting from irradiation by stellar populations of different ages with different SEDs.\\
\indent \cite{2014MNRAS.444.2837R} model radiation fields at galactic scales using an escape probability formalism and splitting the radiation field into UV and IR components. The energy absorbed from the UV field is taken to be $E_{\rm UV}=E_{\rm rad}[1-{\rm exp}(-\kappa_{\rm UV}\rho_{\rm dust}\Delta x)]$, where $E_{\rm rad}$=10$^{52}$ erg M$_{\odot}^{-1}$ is the total specific energy released by a 10M$_{\odot}$ star. It is then assumed that the absorbed UV radiation is reemitted in the infrared, so that $E_{\rm IR}=E_{\rm UV}[1-{\rm exp}(-\kappa_{\rm IR}\rho_{\rm dust}\Delta x)]$. The dust opacity is treated as a free parameter. This energy is then added to the supernova feedback energy, and effectively deposited as momentum.\\
\section{What we have learned from including feedback in simulations}
\begin{figure}
\centering
\includegraphics[width=0.75\textwidth]{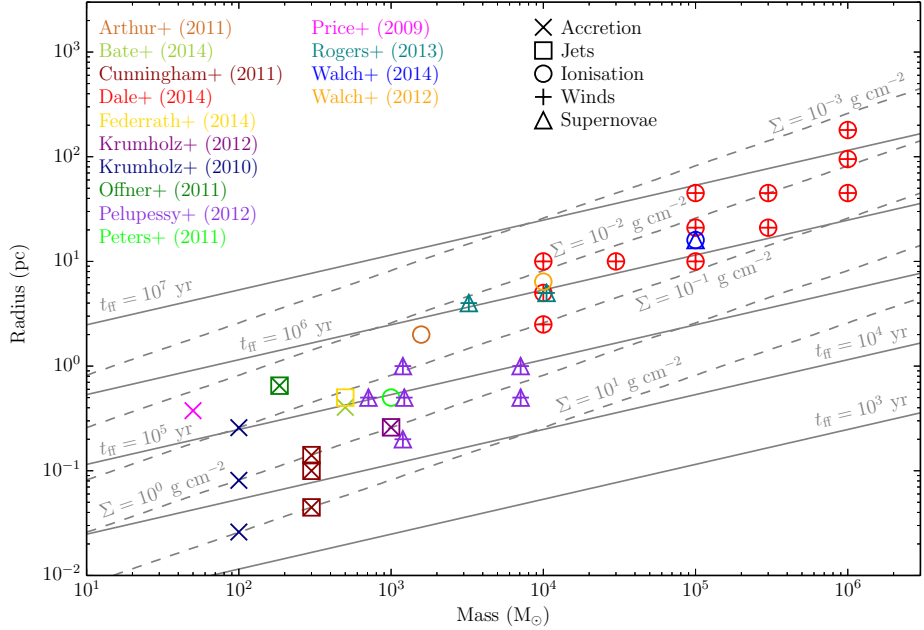}
\caption{Mass--radius parameter space with most of the simulations discussed in this Section overplotted. Colours refer to different papers, given in the left and middle columns of the key. Symbols indicate which feedback mechanisms are modelled, given in the right column of the key. Lines of constant freefall time (solid) and constant surface density (dashed) are also plotted.}
\label{fig:all_sims}
\end{figure}

\indent The previous section concentrated on technical descriptions of algorithms and is intended to be mainly of use to researchers who are considering writing their own feedback prescription and wish to get an overview of how it has been done before. This section is aimed at a different readership and will concentrate on the science results of simulations run using the algorithms and codes described. There will inevitably be a small amount of repetition and overlap between these two sections, so some forbearance on the part of reader is requested.\\
\indent The first five subsections deal with simulations at GMC scales or below. Figure \ref{fig:all_sims} gives an overview of the mass and size scales covered by a selection of these simulations. The last subsection deals with simulations at galactic scales and above.\\
\subsection{Fragmentation, the IMF, and star formation rates and efficiencies}
\indent One of the most urgent questions that simulations involving feedback hope to answer is, what is the effect of stellar feedback on the star formation process itself? At the smallest scales, as modelled by \cite{2009Sci...323..754K}, \cite{2012A&A...537A.122K} (discussed in Section 4.1.2) and, in the case of primordial star formation, by \cite{2011Sci...331.1040C} and \cite{2011MNRAS.414.3633S}, feedback affects the rate at which individual stars build up mass by interacting with accretion flows and circumstellar discs, altering their propensity to fragment. At somewhat larger scales, concentrations of dense gas will be disrupted, and heating of the gas will raise the Jeans mass and suppresses fragmentation. However, shocks driven by expanding bubbles and outflows can also locally increase the gas density and cooling rate. Triggered star formation is a very popular topic in observational astronomy, and triggering of star formation by stellar feedback is even more intriguing because it should be directly observable in star--forming regions, and gives rise to the attractive idea that star formation  may be self--propagating. There is a wealth of observational literature on this topic (see \cite{2015MNRAS.450.1199D} for a recent survey) and it has recently also started to receive increased attention from modellers.\\
\indent There are two popular models of triggering -- radiation--driven implosion (or cloud--crushing -- \cite{1982ApJ...260..183S,1989ApJ...346..735B}) and the collect--and--collapse process (\cite{1978ApJ...220.1051E}) -- both of which are now amenable to simulation.\\
\subsubsection{Radiation--driven implosion}
\indent The RDI or cloud--crushing regards feedback as an external agent which perturbs a stable or quasi--stable equilibrium of some kind. It can thus be very rapid and need not involve large masses of material. Many authors have considered the effects of irradiating objects of only a few to a few tens of solar masses. Such objects are commonly observed around the borders of HII regions, so their evolution is of obvious interest. They are usually modelled as Bonner--Ebert spheres. Unless the radiation is strong enough to ionise the entire BES, it heats a curved layer, thickest at the point closest to the radiation source. The photo--heated material flows away, driving a shock back into the cloud, a phenomenon known as the rocket effect. It is this which may drive the BES to gravitational collapse.\\
\indent \cite{2009MNRAS.393...21G}, \cite{2009A&A...497..649B}, \cite{2011ApJ...736..142B} and later \cite{2015ApJ...798...32N} all employed SPH simulations and present similar results. \cite{2011ApJ...736..142B} performed a parameter study in which they varied the mass of the BES and the ionising flux. Low fluxes lead to slow but efficient star formation. Higher fluxes produce faster evolution but less efficient production of stars, with star formation concentrated in a pillar--like structure behind the ionisation front. This structure is formed by a shock--focussing phenomenon originating in the curved outer surface of the BEF. For large fluxes or low BES masses, the cloud is destroyed by photoevaporation before any stars form.\\
\indent \cite{2010MNRAS.403..714M} illuminated groups of triaxial clumps and found that small groups of sufficiently massive and dense clumps produced long--lived elongated structures. Three nearly collinear clumps or three clumps close together in a triangular configuration were particularly successful in forming pillars by the smearing of the clumps away from the ionising source and the filling up of shadowed regions behind the clumps with low--density material.\\
\indent Other authors have instead turned to irradiating turbulent clouds or boxes. \cite{2009ApJ...694L..26G} and \cite{2010ApJ...723..971G} created a turbulent box which they illuminated with plane--parallel radiation from the negative x--direction. The inhomogenous density field allowed the radiation to penetrate to different depths at different locations, producing an irregularly--shaped mass of hot gas which then expanded in the face of the ram--pressure of the remaining turbulent cold gas. The subsequent evolution was found to depend strongly on the Mach number of the initial turbulence. Low Mach numbers presented little resistance to the HII region, which expanded like a piston, producing a rather flat ionisation front. Higher Mach numbers allowed progressively longer and more prominent pillar structures to project into the ionised gas. In the tips of several of these objects, collapsing cores and discs were found, although the simulations could not be run far enough to follow their evolution.\\
\indent In a series of papers, \cite{2012A&A...538A..31T}, \cite{2012A&A...546A..33T} and \cite{2013A&A...560A..19T} thoroughly examine the irradiation of perturbed ionisation fronts and turbulent boxes with a view to understanding pillar formation. They begin by imposing on an ionisation front a single sine--shaped perturbation of the same length, but different widths perpendicular to the radiation field to study the influence of the curvature induced in the photoevaporative shock. Narrower perturbations result in longer pillar structures, with shock curvature concentrating material along the pillar axis, while a concave pit is excavated around the base of the pillar. They next reintroduce a flat ionisation front but with a spherical overdensity just behind it. Shock curvature around the obstacle causes material to meet behind it, forming a pillar structure which was observed to be longer the higher the density contrast. Unlike in the case of the perturbed fronts, the pillars produced by this process develop pronounced heads shaped roughly like umbrellas.\\
\indent In the second paper, a turbulent velocity field with mean Mach number 1, 2 or 4 is irradiated. In common with \cite{2009ApJ...694L..26G}, they observe that the higher Mach number turbulence is better able to resist the compression by the hot ionised gas, and the formation of pillar structures. The pillars have much more complex shapes than those formed in the previous paper. In the high Mach number simulation, they also observe globules of dense cold gas isolated inside the ionised gas, which are thrown there by the ram pressure of the turbulence. They also observe a characteristic double--peaked structure in the gas column--density PDF, with one peak corresponding to the turbulent velocity field, and the second to gas compressed by feedback. In the third paper, they show that just such PDFs are observed in the neighbourhood of the Pillars of Creation in M16.\\
\begin{figure*}
     \centering
   \subfloat[Monchromatic]{\includegraphics[width=0.30\textwidth]{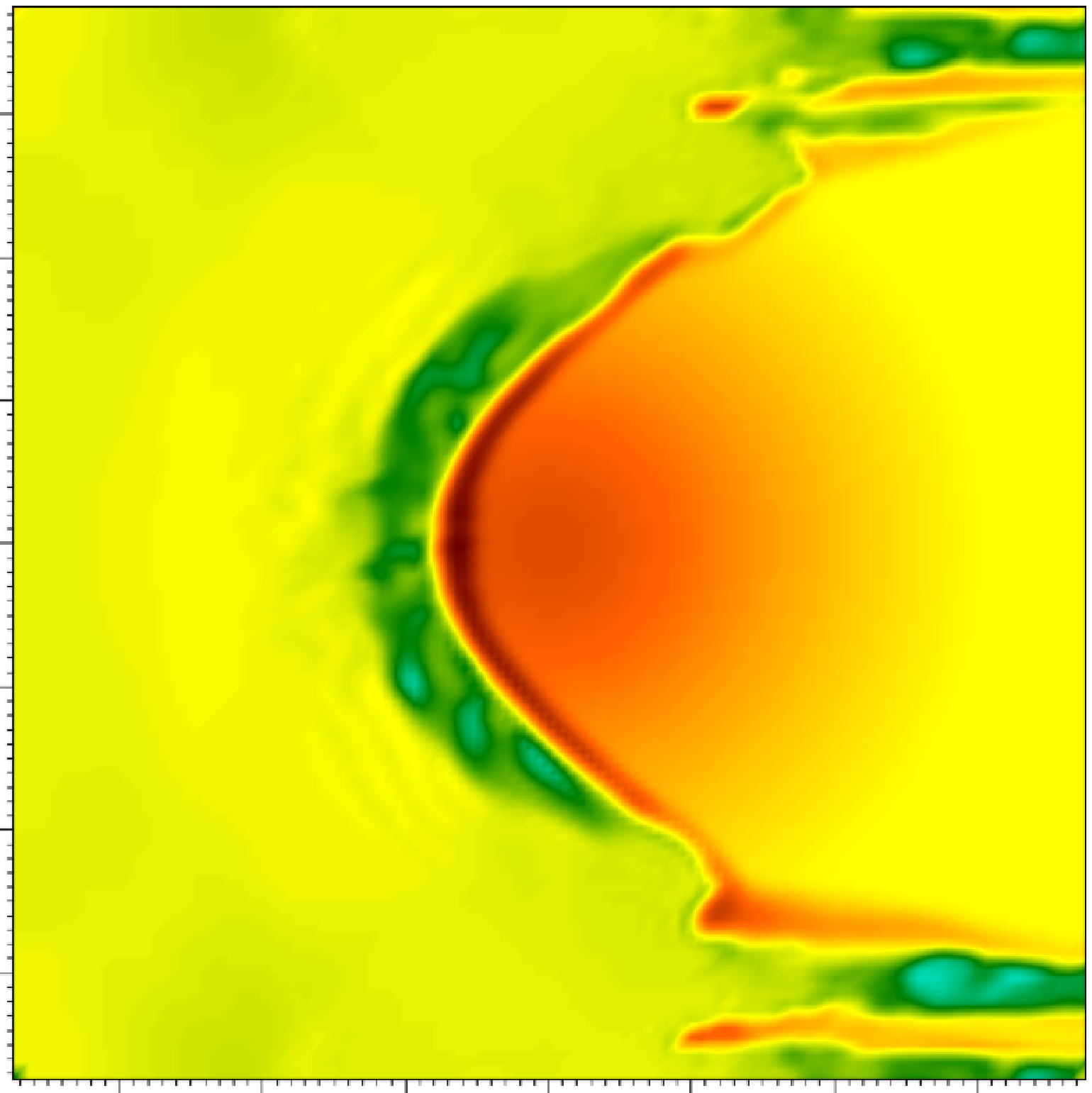}}
   \hspace{0.1in}
      \subfloat[Polychromatic]{\includegraphics[width=0.30\textwidth]{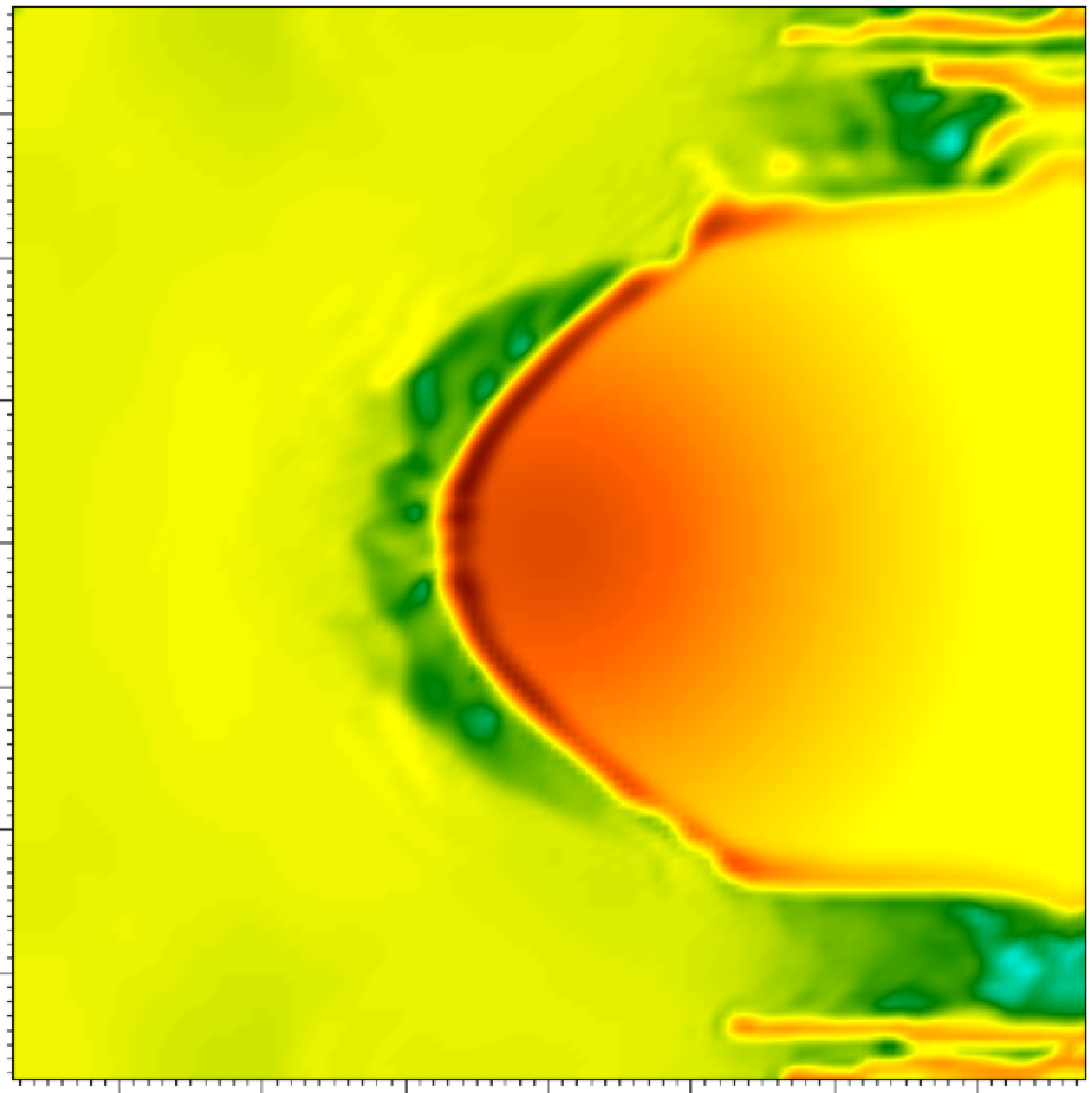}}
         \hspace{0.1in}
      \subfloat[Polychromatic diffuse]{\includegraphics[width=0.31\textwidth]{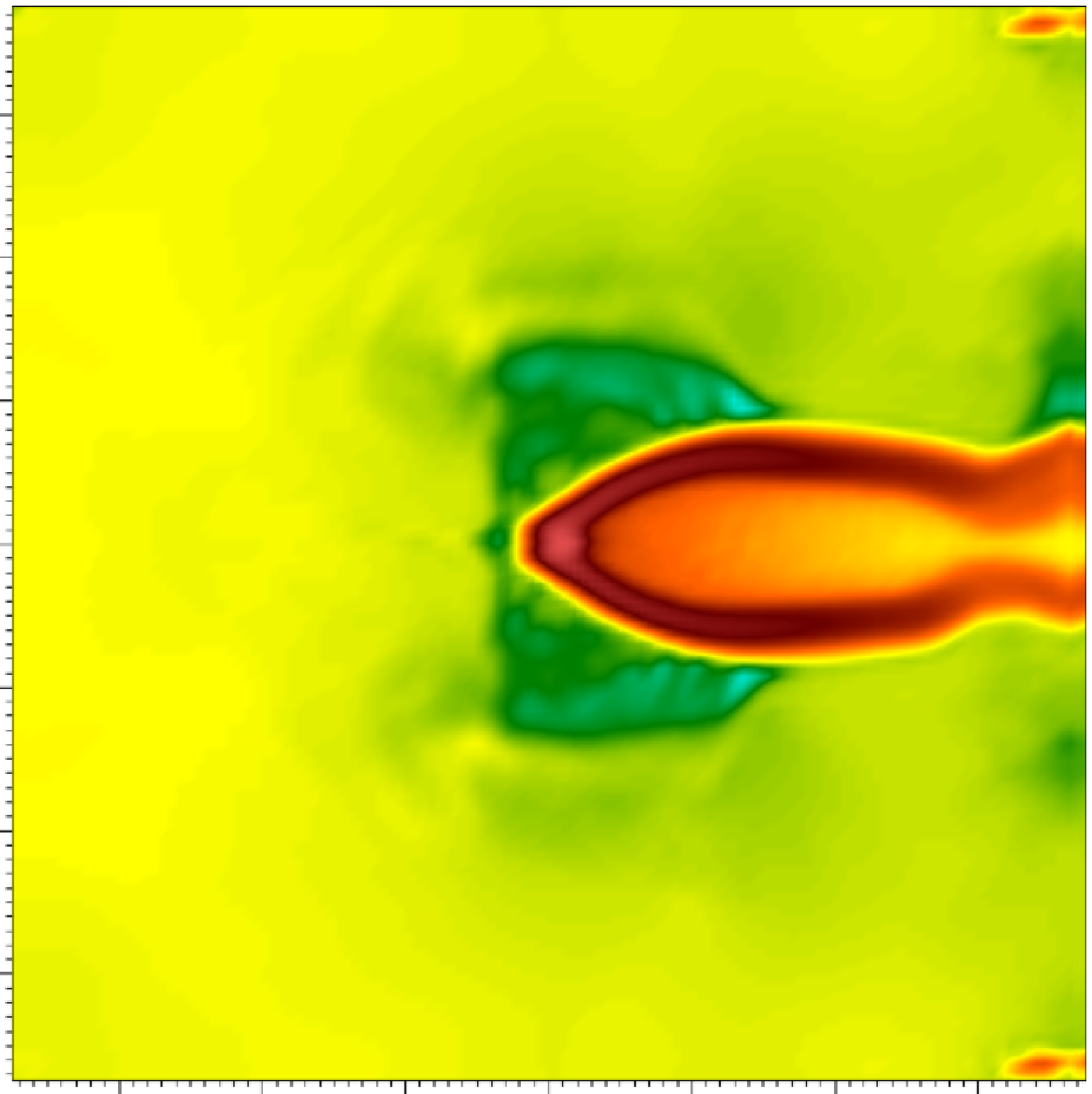}}
\caption{Comparisons of the end result of the medium--flux radiation--driven implosion simulations from \cite{2012MNRAS.426..203H}. The left panel shows the result of a monochromatic direct radiation field with the diffuse field treated using the OTS approximation. The middle panel shows the result of using a polychromatic radiation radiation field and again employing the OTS approximation. The right panel depicts the effect of the polychromatic radiation field with the diffuse field calculated self--consistently.}
\label{fig:haworth12}
\end{figure*}
\indent \cite{2012MNRAS.420..562H}, \cite{2012MNRAS.426..203H} and \cite{2013MNRAS.431.3470H} examine the subject of triggered star formation in bright--rimmed clouds (BRCs) using the {\sc torus} hybrid AMR/Monte Carlo RT code. Their detailed treatment of the RT problem allows them to include several physical mechanisms that have been left out of previous studies, such as the effect of the diffuse ionising radiation field, and to produce synthetic observational images on--the--fly, rather than through post--processing. They find that the diffuse field alters the character of the RDI, particular in cases where the radiation field is of moderate strength, where the diffuse field compresses the target clump very effectively from the sides, resulting in a much denser and more bullet--shaped configuration, as shown in Figure \ref{fig:haworth12}. They relate these results to the BRC classification scheme proposed by \cite{1991ApJS...77...59S}, where gently--curved rims are classified as type A, tightly--curved as Type B and cometary as Type C. High and low radiation fields both produce Type A BRCs, but the strong lateral compression in medium--flux cases reliably leads to Type B or C morphology, as shown in Figure \ref{fig:haworth12}. Using standard observational techniques on their synthetic datacubes, they find that the dynamical states of BRCs can be reasonably well recovered, although the synthetic observations systematically underestimate electron densities due to line of sight contamination. This would lead to systematic underestimation of the pressure in the ionised flows, and hence to underestimation of the degree of shock compression and of the effectiveness of the RDI in real objects. The interpretation of molecular line profiles originating in the cold gas is more complex. They are able to reproduce line profiles similar to those observed, but find that some interpretations of those profiles are likely to be erroneous, in particular the two--component envelope--expansion--with--core--collapse model invoked to explain the lack of blue asymmetry (which would imply infall), lack of any asymmetry, or even red asymmetry (implying expansion or outflow) often seen in optically--thick self--absorbed lines in BRCs. \cite{2013MNRAS.431.3470H} rule out the EECC model because the material moving towards the observer is ionised and not molecular.\\
\indent Moving to larger scales, \cite{2007MNRAS.377..535D} considered the influence of an O--star on a nearly $10^{4}$M$_{\odot}$ turbulent molecular cloud. They used a globally unbound SPH model GMC whose star formation rate and efficiency were expected to be low. By comparison with a control simulation, the Lagrangian nature of SPH allowed them to show that some of the stellar objects formed in the feedback simulation \emph{did not form} in the control simulation, but the enhancement in star formation rate and efficiency was modest, at 30--40$\%$ over $\approx0.3$ cloud freefall times. The also found that some objects which did form in the control simulation were destroyed, aborted or suffered reduced growth due to the ionisation of some of the potentially star--forming material.\\
\indent \cite{2012MNRAS.422.1352D} performed a similar calculation, except that they irradiated a \emph{bound} GMC. they found that the impact of feedback was even more underwhelming. Although some of the low--density material on the outskirts of the cloud was destroyed, and some intermediate--density gas was driven towards the cloud centre by the rocket effect, the dense core of the GMC where most of the star formation was taking place remained largely unaffected. The number and total mass of stars were little changed and the stellar mass functions proved to be statistically indistinguishable.\\
\subsubsection{The collect--and--collapse process}
\indent In contrast to the RDI model, the collect--and--collapse process is driven internally by stars that have formed inside a given cloud and is a large--scale process taking place on relatively long timescales, since it takes time for a sufficient mass of gas to be swept up and to become unstable. The mass of gas required is also generally large enough to form many stars.\\
\indent \cite{2007MNRAS.375.1291D} used calculations of an HII region expanding in a uniform medium to test a theoretical model of the collect--and--collapse process derived by \cite{1994A&A...290..421W} from a perturbation--theory analysis of the gravitational stability of a shocked gas shell. They obtained reasonably good agreement with the model in terms of the time-- and length--scales at which the shell became unstable, and of the mass of the fragments produced, and also showed that the results were immune to noise of a factors of a few in the initial density field. They found fragment masses in the range 10--100 M$_{\odot}$ but were not able to follow the simulations long enough for many fragments to collapse.\\
\indent A similar approach was taken by \cite{2013MNRAS.435..917W} who instead controlled the quantity of structure in their initial density field by constructing fractal clouds with fractal dimensions in the range 2.2 (highly--substructured) to 2.8 (rather smooth). Clouds with small fractal dimensions (corresponding to large--scale structure) resulted in semi--coherent shell structures punctured by large holes through which ionised gas was able to vent (the authors refer to these calculations as \emph{shell--dominated}). Large fractal dimensions, which generate small--scale substructure, instead resulted in large numbers of pillarlike--objects pointing towards the ionising source, created by dense clumps of material shielding or shadowing gas behind them from the ionising photons (these calculations are hence referred to as \emph{pillar--dominated}). The fractal dimension had a concomitant effect on the fragmentation induced, with low fractal dimensions leading to a small number of large fragments and large fractal dimensions producing many small objects. The subsequent evolution of the clumps masses is governed by a competition between destruction of the clumps by photoevaporation, and their acceleration away from the ionising source by the rocket effect. Very strong differences in the clump mass functions result, with the mass function slope being -0.18 for a D=2.0 cloud and -0.91 for a D=2.8 cloud. Regarding the stars that form, those in the low--D clouds tend to acquire high radial velocities from the acceleration of the large coherent clumps in these runs by the rocket effect, so they tend to be found ahead of the ionisation front. Conversely in the high--D clouds, the stars are usually found at the tips of pillars and are left behind in the HII region.\\
\indent \cite{2011ApJ...731...13N} examined a similar process, but acting at still larger scales. They modelled the energy and momentum injection by winds of two star clusters 500 pc apart in initially uniform or initially turbulent boxes containing 8 000K \emph{atomic} gas. Their intention was to study the formation of molecular material, rather than assuming its preexistence, with the transition from atomic to molecular gas handled by piecewise heating/cooling functions. The expansion of the $\sim10^{8}$K wind bubbles drove shocked shells into the warm gas which were simultaneously susceptible to the non--linear thin shell, thermal, and Kelvin--Helmholtz instabilities, generating very complex structure. The former two instabilities combined to form cold overdense clumps in the shells which accounted for 85$\%$ of the total mass by the end of the simulations after $\approx7$Myr. However, the total numbers of clumps was large, so their individual masses were equivalent to moderately massive stars. The calculations were rather similar except that additional inhomogenieties in the background density field created by turbulence induced the two shells to fragment at different times, by approximately 2Myr. The non--uniform emergence of the NTSI also produced larger shear and amplified the KH instability, leading to longer filamentary structures in this calculation. Filamentary structures were also generated by shear when the two shells collided, resulting a turbulent layer in which the ISM phases mixed and interacted.\\
\subsubsection{Collapsing turbulent clouds}
\indent Much recent work has concentrated on implementing feedback of one kind or another as additional physics in simulations of collapsing turbulent clouds. These simulations do not fit neatly into the categories of collect--and--collapse or RDI, although both processes may be taking place within them at different times and places. However, most of these simulations find that the global effect of feedback on these objects is to \emph{reduce} the star formation rates and efficiencies.\\
\indent \cite{2009MNRAS.392.1363B} use the FLD scheme of \cite{2005MNRAS.364.1367W} to model the influence of accretion feedback in a 50M$_{\odot}$ turbulent cloud. Heating strongly suppresses the formation of new objects after about one freefall time, mostly by preventing disc fragmentation. This reduces the numbers of stars formed by a factor of $\approx4$ compared to a purely barotropic calculation. The decrease in disc fragmentation also results in fewer dynamical interactions, sharply decreasing the numbers of brown dwarfs formed. More importantly, the accretion feedback decouples the mean stellar mass from the cloud initial Jeans mass. These calculations were extended by \cite{2014MNRAS.442..285B} to a 500M$_{\odot}$ cloud to obtain improved statistics. The IMFs produced are statistically indistinguishable from the Chabrier IMF, and this result is robust against variations of factors of 300 in the gas metallicity.\\
\indent This problem was approached in the {\sc orion} code by \cite{2009ApJ...703..131O} using the FLD implementation described by \cite{2007ApJ...656..959K}. Protostellar heating again comes to dominate by about one freefall time in the radiation transfer calculation. The regions heated by the protostars are small, of order 0.05 pc in radius, and fragmentation in most of the cloud proceeds unaffected. The main effect of the feedback is on accretion onto the protostars and on their discs. Warmer discs are able to transfer material onto their central stars at higher rates, so can absorb larger quantities of infalling gas without becoming unstable and fragmenting. They point out that radiation emitted from the protostellar surface originating from, e.g., deuterium burning or Kelvin--Helmholtz contraction, is an essential component of feedback in simulations of this kind, and that simulations such as those by \cite{2009MNRAS.392.1363B} which neglect it are likely to underestimate the effects of feedback in fragmentation.\\
\indent \cite{2010ApJ...713.1120K} use the {\sc orion} code to investigate how clouds of the same mass, virial ratio and internal structure but with different surface densities respond to radiative feedback from accreting protostars. They find that low surface density (0.1 g cm$^{-2}$) clouds analogous to Taurus fragment into a large number of stars of roughly equal masses, whereas clouds with surface densities 1 to 10 g cm$^{-2}$ fragment very little and most of the mass ends up in either a single massive binary or a single massive star. The root cause is that the higher density clouds support higher accretion rates and therefore higher accretion luminosities, and are also more optically thick so absorb the energy released more efficiently. This has the net effect of raising the Jeans mass over substantial fractions of the cloud volume, suppressing further fragmentation.\\
\indent \cite{2011ApJ...740...74K} modelled feedback from protostars in a massive (10$^{3}$M$_{\odot}$), dense (1g cm$^{-2}$) core using the prescription from \cite{2009ApJ...703..131O}, which includes both the energy released by Kelvin--Helmholtz contraction and from deuterium burning. By direct comparison with an isothermal calculation, they found that feedback left the stellar mass largely unchanged, but led to a smaller number of stars forming. In fact, the warming of the gas eventually shut down fragmentation entirely, while allowing accretion onto already--existing stars to continue, resulting in the peak of the mass function moving continuously to higher masses. They attribute this to the high density of their clump, which allows regions heated by different protostars to overlap, so that virtually all the gas in the clump becomes warm, and none of it is able to fragment.\\
\indent In a subsequent calculation, \cite{2012ApJ...754...71K} tried to remedy this problem in two ways. Firstly, they included outflows in addition to thermal feedback. However, this had relatively little impact, reducing star formation rates by around 20 percent, entraining very little material and having no major influence on the cloud structure. However, they also compared different initial conditions. The earlier calculations were initialised with smooth spherical gas distributions seeded with turbulence. Additionally, they constructed initial conditions from fully developed turbulence created in a periodic box with no gravity. This produces a density and velocity field which are \emph{initially} self--consistent, leading to more distributed star formation and a slower overall star formation rate, since the cloud does not immediately begin to collapse. They found that this lower star formation rate alleviated the overheating issue, leading to a consistent mass function at all times. From this study, they conclude that two apparently only weakly connected characteristics -- the star formation rate and the shape of the mass function -- become intimately coupled by the inclusion of feedback, and that the latter cannot be correct if the former is too high.\\
\indent \cite{2014ApJ...790..128F} model the influence of outflow feedback on $500$M$_{\odot}$ turbulent magnetised clumps with virial parameters $\alpha=0.4$. They evolve the clumps for almost two freefall times, with a control run used for comparison. The feedback simulation forms twice as many sink particles and its star formation efficiency reaches 75$\%$, whereas that in the control simulation reaches 100$\%$, and the difference in the SFEs is larger at earlier epochs, being over a factor of two at one freefall time. On average, the star formation rate per freefall time was 0.57 and 0.30 in the control and feedback runs. About 40$\%$ of the stellar mass in the feedback run was accreted, ejected and re--accreted at least once. The average stellar mass is reduced by a factor of $\approx$3 by the combined effect of outflows causing more stars to form, and reducing the accretion rates onto individual objects.\\
\indent Most simulations of accretion--driven feedback assume that accretion is smooth or continuous. Building on several observational studies suggesting otherwise (e.g. \cite{1996ARA&A..34..207H}), \cite{2011ApJ...730...32S} and \cite{2012MNRAS.427.1182S} examine the influence of \emph{episodic} accretion on the fragmentation of circumstellar discs. The effects of the energy released by accretion are computed using the radiation transport method of \cite{2007A&A...475...37S}, but the accretion \emph{rate} is determined using a more sophisticated method than in other studies. They divide the accretion discs into two zones -- an outer zone, which they can resolve and in which angular momentum transport occurs primarily through gravitationally--driven spiral wave formation, and inner disc which they do not resolve but instead parameterise, where angular momentum transport is driven by the magneto--rotational instability (MRI). The MRI can only operate, however, if the inner disc becomes hot enough to generate the ionisation required to couple it to the protostellar magnetic field. The authors show that the ability of the disc to fragment and form secondary stars or brown dwarfs is determined by two factors. The first is the `base rate' of accretion through the inner disc when it is \emph{not} experiencing an outburst. This sets the temperature to which the disc (very quickly) relaxes when an accretion outburst is complete. The second factor is the duration of the intervals between outbursts, which determines whether the disc has time to become gravitationally unstable between stabilising accretion outbursts. If the intervals are sufficiently long and the disc is able to cool to low enough inter--burst temperatures, they find that accretion feedback is much less effective in suppressing disc fragmentation than inferred by, for example, \cite{2009MNRAS.392.1363B}.
\indent \cite{2010ApJ...715.1302V} use the {\sc art} code (\cite{1997ApJS..111...73K}) with energetic feedback from O--stars implemented by the deposition of energy in single grid cells where star particles are located, with energy deposition rates adjusted to obtain HII regions of reasonable sizes, temperatures and internal velocity dispersions. They simulate the formation of a flattened configuration of molecular gas from two colliding streams. Feedback \emph{increases} the mass of dense material while \emph{decreasing} the star formation efficiency, by factors up to $\approx10$. Feedback is unable to destroy the clouds themselves, nor the atomic streams from which they are forming, but it is able to destroy the smaller--scale dense clumps where star formation is actually occurring. The HII regions do produce more dense clumps, but these generally disperse and fail to form stars of their own.\\
\indent The effects of photionizing feedback on embedded clusters formed in artificially--constructed turbulent clouds was examined by \cite{2012MNRAS.427.2852D} and \cite{2013MNRAS.431.1062D}. They found that star formation rates and efficiencies were reduced by factors of up to two by the disruption of the dense filaments of gas feeding the clusters. Feedback was largely unable to unbind the clusters themselves though and had little effect on the stellar mass functions. They were able to demonstrate, by comparison with control simulations, the triggering of stars in the sense of the formation of stars that would not otherwise have been born. However, they found that the triggered objects were spatially and dynamically mixed with the spontaneously--formed objects and were therefore very difficult to identify.\\
\indent Many simulations of turbulent clouds still rely on simple equations of state or optically thin heating and cooling functions to compute the background gas temperature, when in fact this is likely to be set by the external bath of radiation and cosmic rays in which the clouds sit, and by the complex chemistry driven thereby. Several papers make use of the {\sc treecol} (\cite{2012MNRAS.420..745C}) algorithm to compute the optical depth from the outside of the cloud to any point in its interior, and therefore the heating rate from the external radiation field. \cite{2012MNRAS.421....9G} used the combination of {\sc treecol} and the chemical network of \cite{2007ApJ...659.1317G} and \cite{2012MNRAS.421..116G} to evaluate the importance of molecular cooling on star formation. They showed that in fact cooling from C$^{+}$ and dust were sufficient to allow star formation to proceed, once sufficiently dense gas is formed. They stress that this does not mean that molecular cooling is irrelevant, but that it is not \emph{necessary}.\\
\indent \cite{2012MNRAS.424.2599C} use similar numerics to answer the question of how long an observable molecular cloud takes to form. They model colliding atomic flows at either 6.8 or 13.6 km s$^{-1}$. They find that molecular hydrogen appears early on, respectively 10 and 3Myr before the onset of star formation, so that the clouds are in this respect `molecular' long before they begin manufacturing stars. However, detectable amounts of CO form much later, about 1--2 Myr before star formation commences, supporting the idea that there could exist a population of undetectable molecular clouds. \cite{2014MNRAS.441.1628S} extended this work to galactic--disc scale simulations using {\sc arepo} and showed that 42$\%$ of the molecular mass in their model spiral galaxy was in CO--dark form.\\
\indent \cite{2014MNRAS.444.2396C} examine the question of whether there is a column--density threshold for star formation, as suggested by, e.g. \cite{2004ApJ...609..667S} and \cite{2010ApJ...724..687L} (who find a value of 116 M$_{\odot}$ pc$^{-2}$. They find that the correlation between column-- and volume--density in their model clouds is very poor, so that the latter cannot be safely inferred from the former, and the star formation rate in volumetrically--dense gas is much higher than the star formation rate in gas at high column densities. They do infer that there is a minimum mean column below which molecular clouds are sterile, but is roughly one order of magnitude lower than the threshold discussed by \cite{2010ApJ...724..687L}.\\
\subsubsection{Star formation from reinserted gas}
Under certain circumstances, the matter injected into star clusters by the combined winds and supernovae of their massive stars can itself become the raw material for a subsequent round of star formation. This is a particularly intriguing idea, given that many globular clusters are observed to have multiple main sequences.\\
\indent Two--dimensional simulations of super star cluster winds were performed by \cite{2008ApJ...683..683W} using the {\sc zeus} code. Matter and energy are inserted at the centre of a spherical grid. Above a threshold value for the mass injection rate and mechanical luminosity, the cluster winds transition from a smooth outflowing state to one in which the matter inside a critical stagnation radius is subject to thermal instability. Clumps of gas inside the stagnation radius cool catastrophically and collapse under the thermal pressure of the surrounding wind. Most of the collapsing clumps are trapped inside the cluster volume and are obvious candidates for forming a second generation of stars.\\
\subsection{Gas expulsion and cloud destruction}
\begin{figure}
\centering
\subfloat[]{\includegraphics[width=0.31\textwidth]{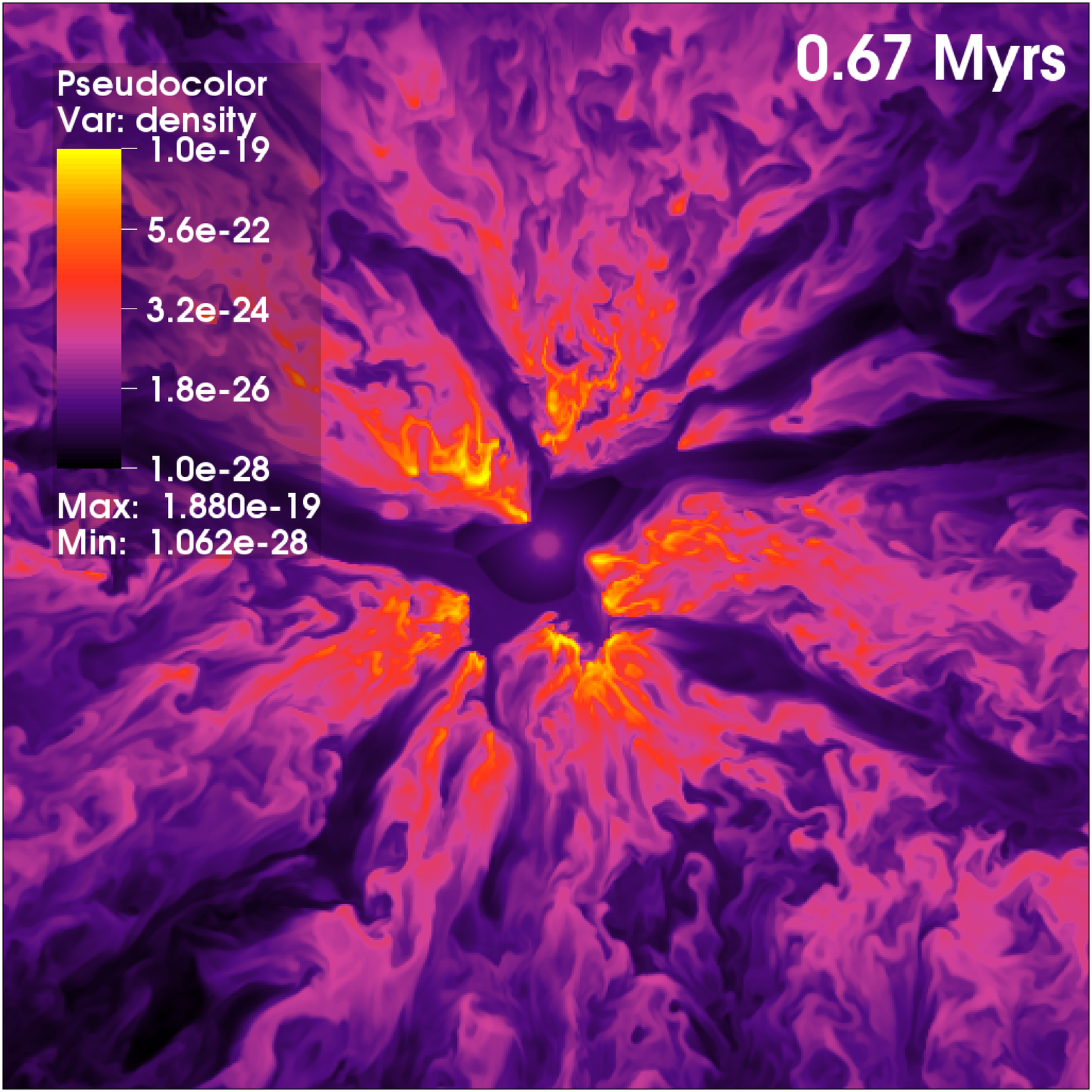}}
   \hspace{0.1in}
\subfloat[]{\includegraphics[width=0.31\textwidth]{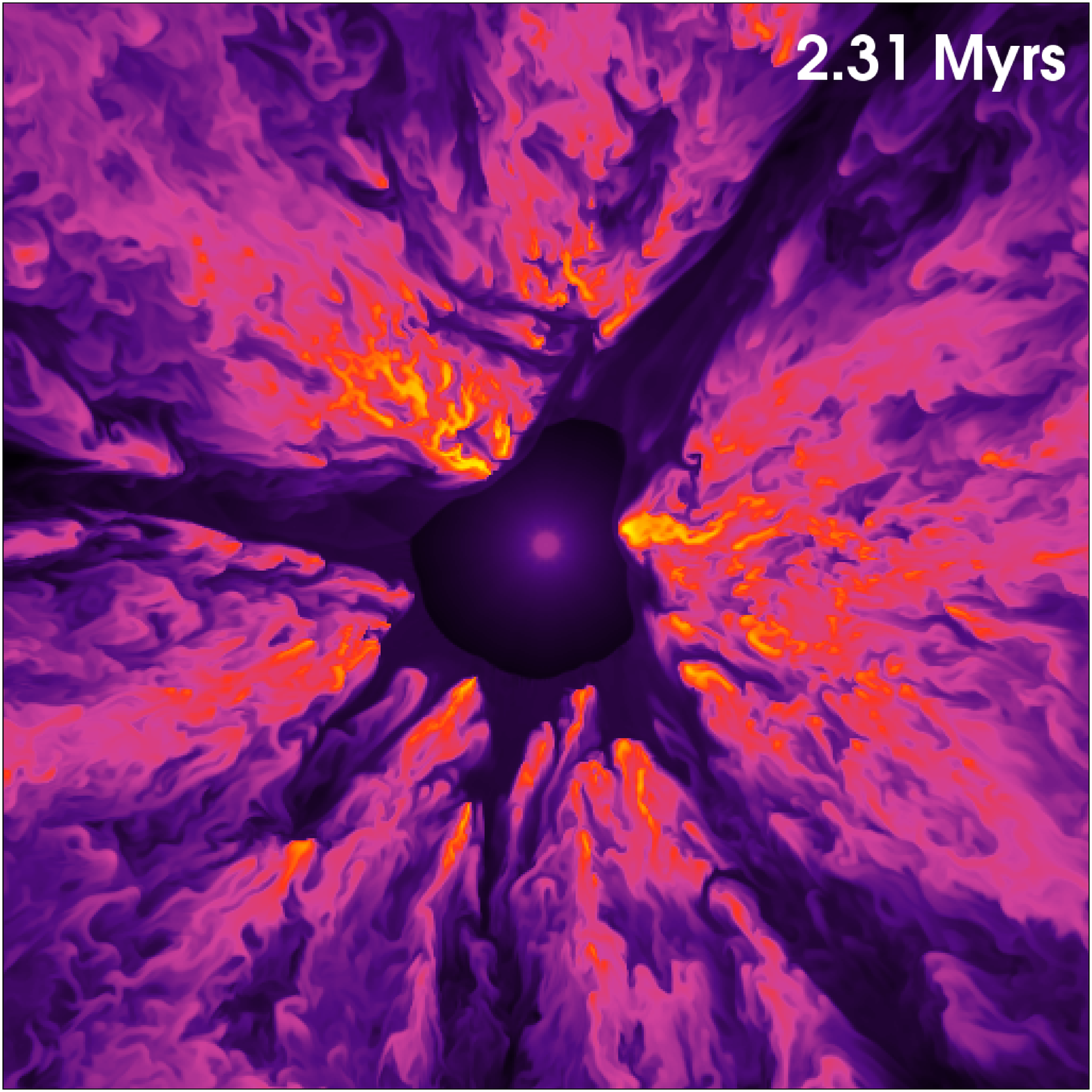}}
   \hspace{0.1in}
\subfloat[]{\includegraphics[width=0.31\textwidth]{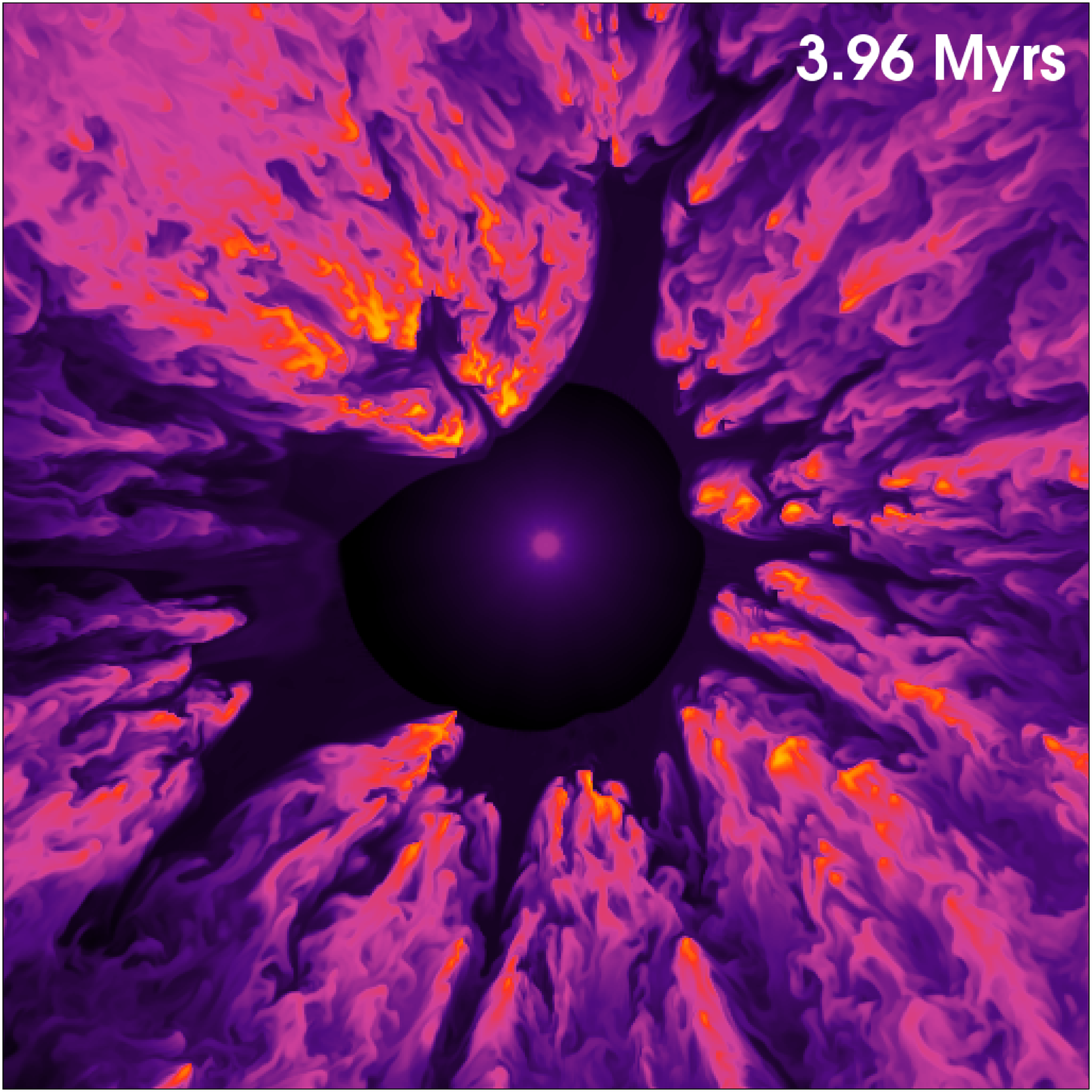}}
\caption{Time evolution of the stellar wind simulation of \cite{2013MNRAS.431.1337R}. The images show density slices though the midplane of the simulations 0.67 Myr (left panel), 2.31 Myr (centre panel) and 3.96 Myr (right panel) after the ignition of the wind--driving stars located at the centres of the images.}
\label{fig:rogers13}
\end{figure}

\indent A long--standing problem in star formation is explaining why it is such a slow and/or inefficient process. The Galaxy's molecular clouds cannot be forming stars on their freefall timescales, because this would result in a Galactic star formation rate about two orders of magnitude higher than is observed, and would have left the Milky Way devoid of gas to form stars out of several Gyr ago. Stellar feedback has long been called upon to solve this problem, by slowing the collapse of GMCs, or destroying them before they are able to convert more than a few percent of their gas to stars.\\
\indent Analytical work by, e.g., \cite{2002ApJ...566..302M} indicates that expanding HII regions are likely to be the main source of energy on GMC scales, at least until the detonation of the first SNe. \cite{2005MNRAS.358..291D} simulated the impact of the HII region driven by a single very massive star into a non--turbulent cloud. Gravitational collapse gave the cloud a filamentary structure, with the filaments meeting at a common hub at the centre of mass, where the massive star was to be found. The dense filaments strongly retarded the growth of the HII region, partly due to the deposition of neutral gas into the HII region, causing it to collapse and regrow, or flicker. Also, as in the simulations by \cite{2010ApJ...723..971G}, the ram pressure of the flows resisted the expansion of the ionised gas in many directions. Much of the ionised gas escaped from the region near the radiation source through moderately collimated outflows. Although sufficient kinetic energy was nominally deposited by ionisation to unbind the cloud, much of it escaped through the outflows and collapse continued largely unimpeded.\\
\indent In a series of papers, \cite{2011MNRAS.414..321D}, \cite{2012MNRAS.422.1352D} and \cite{2013MNRAS.430..234D} investigated the ability of photoionisation to disrupt instead turbulent GMCs with a range of initial radii, masses and turbulent velocity dispersions. The clouds were allowed to form a small number of O--stars, or subclusters large enough to host O--stars. The effects of feedback varied strongly with cloud properties, in particular with the escape velocity. GMCs in the Milky Way all have very similar column densities, i.e. $M_{\rm GMC}\propto R_{\rm GMC}^{2}$. Since the escape velocity $v_{\rm ESC}\propto(M_{\rm GMC}/R_{\rm GMC})^{1/2}$, it follows that $v_{\rm ESC}\propto M_{\rm GMC}^{1/4}$. This is weak scaling, but the escape velocities of Milky Way clouds lie in the range 1--10 km s$^{-1}$, which is significant because the speed of sound inside an HII region at around solar metallicity is fixed at $\approx10$ km s$^{-1}$. \cite{2013MNRAS.430..234D} showed using a very simple model that this leads to a very steep relation between the mass of a cloud and the fraction of material that can be unbound by HII regions in the period before supernova detonation.\\
\indent Other authors have investigated the effects of feedback on different geometries. Ionisation feedback form massive stars forming inside a rotating 10$^{3}$M$_{\odot}$ clump is studied by \cite{2010ApJ...711.1017P}. The clump contracts to a disc--like structure with massive objects forming at the centre. As the simulation progresses, the formation of smaller objects further out in the disc starves the massive objects of gas. This in turn allows HII regions to grow, preferentially out of the disc plane, although what accretion continues causes them to fluctuate and flicker. They propose that this behaviour may explain the well--known ultra compact HII region problem, and find that all the HII region morphologies catalogued by \cite{1989ApJS...69..831W} appear naturally in their simulations.\\
\indent \cite{2013MNRAS.435.1701C} simulate the formation of molecular clouds and clusters in colliding flows of warm neutral gas, and follow the effects of feedback from the cluster stars. The clouds formed by their colliding streams are flattened, turbulent, and have masses of order 10$^{4}$M$_{\odot}$. Ionising feedback \emph{is} effective at bringing star formation to a halt at star formation efficiencies of around ten percent, in contrast to the simulations of \cite{2012MNRAS.424..377D}, who found that even the lowest--mass clouds were able to continue forming stars, albeit slowly, under the influence of ionisation. \cite{2013MNRAS.435.1701C} suggest that the reason for the discrepancy may be that their clouds, being flattened, are less gravitationally bound than those of \cite{2012MNRAS.424..377D}, and thus easier for feedback to disperse. The fractal clouds modelled by \cite{2012MNRAS.427..625W} are also readily destroyed by photoionisation on timescales of 1--2 Myr, despite a very inefficient uptake efficiency of kinetic energy of well under 1 percent. Since these clouds are also spherical, this explanation cannot work here. It is possible that the fact that the O--stars in \cite{2012MNRAS.424..377D}'s calculations first have to destroy the dense filaments and accretion flows in which they born impedes them in disrupting the clouds. It is also possible, as suggested by \cite{2012A&A...546A..33T}, that the turbulence initially present in the cold gas in \cite{2012MNRAS.424..377D} (but not in those of \cite{2012MNRAS.427..625W}) plays a role.\\
\indent While winds are generally regarded as being subordinate to HII regions, they still inject significant quantities of energy into clouds and, in very dense gas, are more efficient at gas dispersal. SPH simulations by \cite{2008MNRAS.391....2D} and \cite{2013MNRAS.436.3430D} modelled momentum input from O--star winds on turbulent model clouds. As with ionisation, they found that the impact of the winds depended strongly on the escape velocity of the clouds, despite there being no obvious limit to the rate at which wind bubbles can expand. Winds were able to slow the star formation rate somewhat by disrupting the accretion flows feeding stellar clusters, and in general spatially separating the stars from the gas.\\
\indent As well as the damage they themselves do to GMCs, winds and ionisation are important because of the way in which they set the environments in which the eventual supernovae of the massive stars detonate. \cite{2012MNRAS.420.1503P} simulate winds and supernova more self--consistently in an SPH simulation by injecting hot gas into embedded clusters modelled as Plummer spheres consisting of 10$^{3}$ stars mixed with 10$^{5}$ SPH gas particles whose mass is set to give various SFEs in the range 0.05--0.5, and with total system masses in the range 700--8000M$_{\odot}$. Winds and SNe are introduced adiabatically with feedback efficiencies (i.e. fraction of the emitted stellar energy retained by the gas) of either 0.01 or 0.1. For the higher star formation efficiencies, both feedback models are able to efficiently expel gas from the clusters, the difference being that with efficient coupling, this is achieved by the winds whereas with weak coupling, the supernovae are required to complete the task. Where the SFE is 0.05, the clouds are more resistant to the winds owing to the higher gas masses, but are unable to survive the supernovae. In these runs, the expulsion of the gas also promptly disrupts the clusters.\\
\indent This problem was tackled in an Eulerian context by \cite{2013MNRAS.431.1337R}, who modelled the winds (and eventual supernovae) from a trio of massive stars in preformed turbulent clouds. The winds rapidly cleared out the central region near the stars, but hot wind gas streamed out of the cloud through low--density channels, entraining little neutral material, as shown in Figure \ref{fig:rogers13}. Much of the coldest densest material was able to survive the winds for several Myr, producing pillar--like structures pointing back towards the O--stars. The supernovae did eventually prove sufficient to destroy the cloud. Similar models using the {\sc nirvana} code of multiple stellar winds and SNe evolving in a smooth background are discussed by \cite{2013A&A...550A..49K}. The energy input from the supernovae is rather unimportant, with that accumulated in the wind bubbles beforehand being much more significant. The evolution of the supernovae does depend on the state of the wind bubble, with larger wind bubbles giving longer cooling times.\\
\indent \cite{2014arXiv1410.0011W} take the simulations of \cite{2012MNRAS.427..625W} further by detonating supernovae inside their fractal clouds, both those which have suffered ionising feedback and those which have not. The clouds' fractal structures distort the expanding shell and allow leakage of hot gas though low--density channels, advecting large quantities of energy away. Particularly in the realistic radiative cooling cases, most of the explosion energy is lost and only a few percent is imparted to the surrounding cloud. In models where the cloud has experienced photoionisation prior to the supernova, the energy uptake by the cloud is more efficient, but only by a factor of $\sim2$, because the lower average density inside the remnant encounters delays the onset of radiative cooling somewhat.\\
\indent {\sc ramses} simulations by \cite{2015A&A...576A..95I} investigate the influence of the SN detonation site (which other workers find to be of crucial importance in galactic--scale simulations -- see Section 5.6.2) on a turbulent 10$^{4}$M$_{\odot}$ cloud. SNe are modelled as purely thermal energy injected in a region four cells across, and detonated deep inside the cloud, on the cloud border, or outside the cloud. The effect of exploding a supernova outside the cloud was minimal, with only a very small fraction of the explosion momentum being transmitted to the cold dense material. Conversely, the internal explosion deposits roughly half of its momentum in the cloud and unbinds about half the cloud's dense gas.  
\subsection{Driving turbulence}
\begin{figure}
\includegraphics[width=0.95\textwidth]{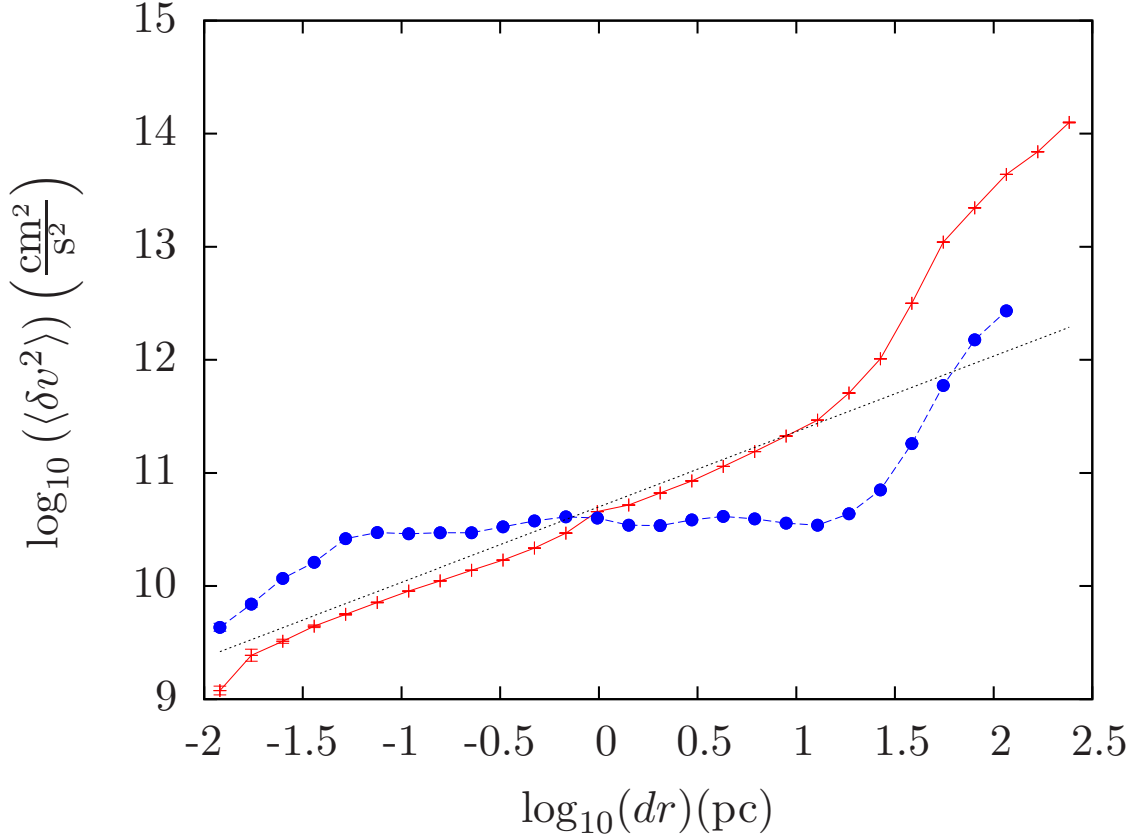}
\caption{Comparison of the velocity structure function from simulations of turbulent clouds with (red lines and symbols) and without (blue lines and symbols) photoionising feedback, 2.2 Myr after the epoch when feedback is enabled in the photoionised simulation. The dashed black line is the structure function expected from Kolmogorov turbulence. The deviations at large values of d$r$ are due to global expansion of the unconfined clouds.}
\label{fig:boneberg15}
\end{figure}

\indent GMCs exhibit supersonic turbulent velocity fields with characteristic dispersions typically in the range 1--10 km s$^{-1}$ (e.g. \cite{2009ApJ...699.1092H}). The turbulence provides an additional means of support against gravitational collapse and has been incorporated into the very successful gravo--turbulent model of star formation (e.g. \cite{2004RvMP...76..125M}). However, the cause of these velocity fields is much debated, since supersonic turbulence, even in a magnetised cloud, will die away due to energy dissipation in shocks in about one crossing time (\cite{1998Ap&SS.261..195M}). Many authors have  suggested that, whatever the original source of the turbulence, it is continuously replenished by feedback.\\
\indent The combined action of multiple jets has been championed in particular by \cite{2007ApJ...659.1394M}, who showed analytically that momentum injection from jets could maintain turbulence in $\sim10^{3}$M$_{\odot}$ clumps. Other authors (e.g. \cite{2004RvMP...76..125M}) note that the rate of turbulent dissipation is comparable to the expected energy injection rate from jets. \cite{2007ApJ...662..395N} confirmed the expectation from \cite{2007ApJ...659.1394M} that turbulence driven by collimated outflows decays more slowly than that resulting from multiple isotropic motions. This is a consequence of the collimated jets penetrating to greater depths into the surrounding cloud, driving turbulence on large length scales, for which the decay times are longer. They observed that even a modest rate of star formation was able to reproduce a turbulent--like velocity power spectrum with a power--law form $E_{k}\propto k^{-2.5}$ over about one and a half decades in wavenumber. The turbulent velocity field reached a steady state with gravitational collapse.\\
\indent \cite{2009ApJ...695.1376C} adopted a similar model with 192 randomly--oriented narrow--angle jets inside a periodic box, which they also observed to reach a steady state after a few crossing times. They compared the resulting field explicitly to an artificially--driven isotropic field with a power spectrum of $E_{k}\propto k^{-2.5}$, finding that the jets produced a spectrum with a slightly steeper exponent of about -2.75. Again, the power--law only extended over approximately one decade in $k$. In these simulations and those of \cite{2007ApJ...659.1394M}, multiple jets rapidly achieve an equilibrium and produce a velocity field resembling turbulence, with a power--law power spectrum over moderate ranges of wave numbers. However, in both cases, periodic boundaries are used, and it is not clear what affect this has on the results. Periodic boundaries make it difficult for injected energy to escape the grid, and makes a steady state a more likely outcome.\\
\indent \cite{2012ApJ...747...22H} explicitly examine the driving of turbulence in their simulations of radiative and outflow feedback in low--mass star--forming regions. The action of outflows reverses the decay in the velocity dispersion of the gas but the turbulence driven by the outflows is of a different form to the that with which their clouds were seeded. In fully developed isotropic hydrodynamic turbulence, the ratio of the energy density in solenoidal versus compressive modes is 2, but shear at the edges of their relatively long outflow cavities preferentially drives solenoidal modes, increasing this ratio to values of up to 10. Similar results are reported by \cite{2014ApJ...784...61O}.\\
\indent However, not all simulators have arrived at the conclusions that jets are efficient at driving turbulence. \cite{2007ApJ...668.1028B} modelled in great detail in {\sc flash} the evolution of a single jet and they concluded that the jet was not able to propagate the supersonic motions required to drive turbulence to large enough distances, nor was it able to entrain sufficient material, for this to be a viable driving mechanism. Jets are in any case not likely to be able to maintain turbulence on GMC scales because their combined filling factors are too low. Several groups have instead looked at the possibility of driving by expanding HII regions, again inspired by analytic calculations (e.g \cite{2002ApJ...566..302M,2006ApJ...653..361K}) suggesting that expanding photoionised bubbles should be able to supply energy at a high enough rate to compensate for turbulent decay.\\
\indent \cite{2006ApJ...647..397M} simulate the evolution of single HII regions in pre--existing turbulent clouds and find that the velocity dispersion in the neutral gas is not only prevented from decaying but raised to values higher than in the original velocity field. Similar results were obtained by \cite{2011MNRAS.414.1747A}, who also showed that this result is little affected by the presence of magnetic fields of realistic strength\\
\indent \cite{2012MNRAS.427..625W} in contrast model HII region expansion in fractal clouds with no initial velocities. They subtract the kinetic energy due to the radial expansion of the HIIR and find that the remaining random component, which they equate with turbulence, is driven more strongly by photoionisation than by gravitational collapse, although the \emph{efficiency} of energy uptake is very low at $\approx 0.05\%$.\\
\indent In their simulations of an irradiated turbulent box, \cite{2009ApJ...694L..26G} measure compressive, solenoidal, and total power spectra and compare the simulations including feedback to a control run in which the initial turbulence, which has power--spectrum close to Kolmogorov, is allowed to die away. They find significant driving, particularly of compressional modes, in the cold gas, but more so at smaller spatial scales, resulting in a flatter power spectrum.\\
\indent \cite{2012ApJ...760..155K} and \cite{2014ApJ...796..107D} both find that radiation pressure illuminating an isothermal atmosphere from underneath is able to drive turbulence provided that the radiative flux is large enough to overwhelm gravity in the atmosphere. This leads to rapidly development of the Rayleigh--Taylor instability and the outbreak of turbulence, as discussed in Section 4.1.2.\\
\indent Starting from the photoionisation calculations of \cite{2012MNRAS.422.1352D}, \cite{2015MNRAS.447.1341B} used structure functions and power spectra to assess whether the multiple bubbles expanding from different locations were able to replenish otherwise decaying turbulence. The simulations were initially seeded with Burgers turbulence, but the velocity field had transitioned to the Kolmogorov slope by the onset of star formation. They found that, particularly in the lower--mass and smaller clouds, the structure function in the control simulations subsequently became much flatter, with large quantities of power being lost on intermediate scales. By contrast, in the photoionised runs, structure functions closely resembling that expected of developed Kolmogorov turbulence were maintained, or restored, as shown in Figure \ref{fig:boneberg15}. Feedback also regenerated the characteristic ratio of power in compressive to solenoidal modes of 0.5.\\
\indent All the work cited in this section so far has concentrated on the study of turbulence in the cold, neutral star--forming gas. \cite{2014MNRAS.445.1797M} demonstrated the development of turbulence in the \emph{ionised} gas inside their HII regions. A statistically--turbulent velocity field takes approximately 1.5 crossing times of the HII region to establish itself, but the power spectrum slopes are substantially shallower than would be expected for compressible or incompressible turbulence. They suggest that this is due to the driving of turbulence on all scales, contrary to the classical model where it is driven only on the largest scale.\\
\indent Many authors modelling feedback on larger scales report producing velocity fields with power spectra resembling turbulence. However, in many cases, feedback is not the only driver of turbulence and it competes with shear in galactic discs, self--gravity and various fluid instabilities, particularly the thermal instability. Some simulations find that the influence of feedback on turbulence diagnostics such as the gas density PDF is minimal (e.g. \cite{2007ApJ...660..276W}). A prominent model of galactic dynamics simulated by, for example, \cite{2011ApJ...731...41O}, \cite{2011ApJ...743...25K} and \cite{2012ApJ...754....2S}, has as one of its main components a vertical equilibrium in the galactic disc between supernova--driven turbulence and turbulent dissipation, which they find is achieved on timescales shorter than a galactic rotation period.\\
\subsection{Effects of different feedback mechanisms combined}
\begin{figure}
\centering
\includegraphics[width=0.65\textwidth]{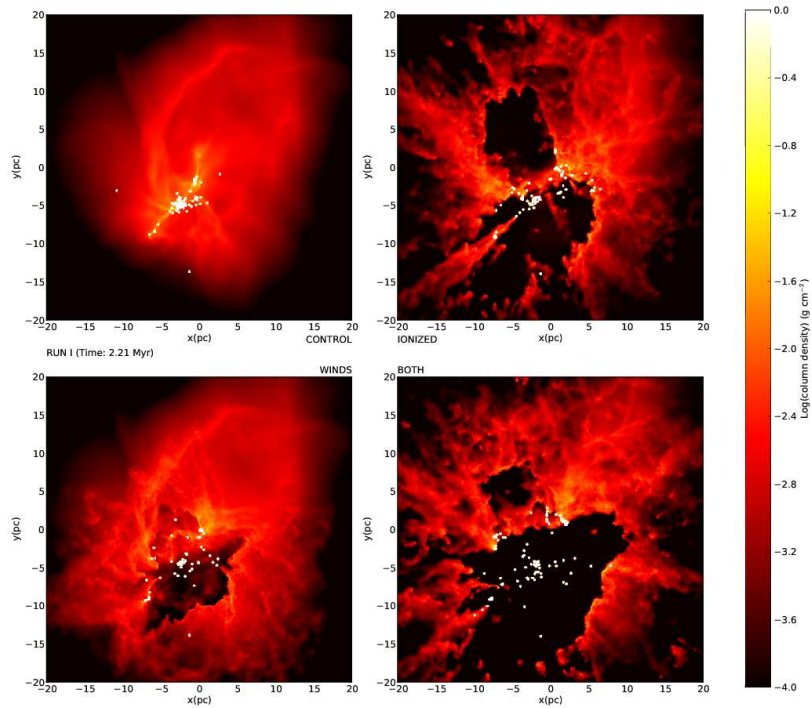}
\caption{Comparison of the control (top left), ionisation--only (top right), winds--only (bottom left) and dual--feedback (bottom right) Run I calculation from \cite{2014MNRAS.442..694D}. Colours represent gas column density and white dots are sink particles used to model stars. Screenshots are from an epoch 2.2 Myr after the time at which feedback is enabled. While the winds do have some effect on the cloud, that of the HII regions is much more severe. This is even more true of other simulations in \cite{2014MNRAS.442..694D}.}
\label{fig:dale14}
\end{figure}

\indent Stellar feedback mechanisms obviously do not act alone but in concert and, even if one is likely to dominate, they are nevertheless likely to influence each other to some extent. It is often not obvious how such interactions will proceed and more and more authors are now investigating the combined effects of several mechanisms.\\
\indent Accretion of material by protostars produces two very different but almost always concurrent feedback mechanisms, namely accretion heating and jets, which affect the star's surroundings in very different ways. \cite{2011ApJ...740..107C} simulated the evolution of a 300M$_{\odot}$ turbulent core including the effects of stellar jets, and using FLD to follow the radiative cooling of the shocked gas and radiative feedback from accretion and from the protostars themselves. They observe that the outflows reduce the efficacy of radiative feedback by providing low--density channels through which photons can escape.\\
\index In common with \cite{2012ApJ...754...71K}, \cite{2012ApJ...747...22H} simulate the interaction of radiative protostellar feedback and outflows, using the same model as \cite{2011ApJ...740..107C}. They simulate a 0.65 pc turbulent box containing 185M$_{\odot}$ of material. Simulations including radiation \emph{and} outflows form substantially more stars than those with radiative feedback alone because the outflows reduce the accretion rates which contribute much of the protostellar luminosity. The dual--feedback simulation in fact results in very similar numbers of stars and total stellar mass to a simulation with outflows and a barotropic equation of state. This result is difficult to reconcile with that of \cite{2012ApJ...754...71K}, who found that the addition of winds was \emph{not} able to strongly reduce the protostellar accretion rates or luminosities in their calculation. \cite{2012ApJ...754...71K} attribute the ineffectiveness of outflows in their simulation to the higher densities of their clouds (1 g cm$^{-2}$ as opposed to \cite{2012ApJ...747...22H}'s 0.1 g cm $^{-2}$.\\
\indent Another complicating factor in the evolution of combined HII regions and wind bubbles is the possibility of the rapid motion of the driving star through its natal cloud. \cite{2006ApJS..165..283A} and \cite{2015A&A...573A..10M} study the evolution of wind bubbles inside HII regions driven by runaway stars. The wind bubbles become asymmetrical almost immediately, while the HII regions take much longer to do so, owing to their much lower sound speed. \cite{2015A&A...573A..10M}'s wind bubbles fill about twenty percent of the HII region volume, appropriate for the relatively low--luminosity star modelled, and have relatively little influence on the champagne--flow--like internal dynamics of the HII region. \cite{2006ApJS..165..283A} use higher wind momentum fluxes and the winds dominate the flows in their calculations. \cite{2015A&A...573A..10M} find, as suggested by \cite{2014MNRAS.442.2701R}, that much of the wind energy is radiated away via microturbulent mixing driven by Kelvin--Helmholtz instabilities at the contact discontinuity between the wind and the HII region. However, they stress that the contact discontinuity is likely modified by numerical diffusion, and that they include neither physical diffusion nor magnetic fields, so urge caution in interpreting this result.\\
\indent From the point of view of individual massive stars, it is clear that photoionisation and the different kinds of stellar wind expelled during different stellar evolutionary stages will interact in complex ways. \cite{2003ApJ...594..888F} and \cite{2006ApJ...638..262F} investigate the coupled evolution of the HII regions and wind bubbles of (respectively) a 60 and a 35 M$_{\odot}$ star in a uniform background. In the case of the lower--mass object, the effect of the wind is largely to compress the HII region into a thick shell lining the inside of the feedback--blown cavity. By contrast, the stronger wind of the more massive star sweeps up a shell inside the HII region which becomes dense enough to fragment, casting shadows on the ionisation front. This leads to a spikelike structure in the ionised gas, although this does not last very long as the HII region is quickly swept up into a thin shell by the wind.\\
\indent \cite{2011ApJ...737..100T} use 1-- and 2D RHD models to study the interaction of the fast WR wind with the previously--ejected and slower YSG or RSG wind around 40 and 60 M$_{\odot}$ stars, comparing two different stellar evolution models. The results depend strongly on the nature of the mass loss in the slower wind phase, on which the two stellar evolution models do not agree. A short RSG phase leads to a thin dense wind shell which becomes strongly unstable and breaks up into clumps when hit by the WR wind, whereas a long RSG/YSG phase leads to a much thicker and lower--density wind which is stable to interaction with the WR wind. Detailed 3D {\sc ramses} simulations investigate the mutual interaction between the HII region, wind bubble and eventual supernova explosion of a 15M$_{\odot}$ star in smooth ambient media at a range of densities. They observe that the expansion of the wind bubble is particularly sensitive to the evolution of the ionising photon flux and wind luminosity of the star as it ages. In the denser media, they find that the wind bubble is prevented from expanding by the pressure in the HII region until the star moves onto the horizontal branch, and that in general winds from the 15M$_{\odot}$ object considered do not inject significant energy into the ISM. They also find, in common with \cite{2014arXiv1410.0011W}, that the fact that the supernova expands into a rarefied medium increases its effectiveness by delaying the transition to the radiative phase.\\
\indent The above already complicated scenarios are made richer still by the likelihood that a massive star will have a close binary companion, often another massive star. \cite{2014Natur.512..282M} present a model of the circumstellar material around the red supergiant star Betelgeuse. Since massive stars tend to be found in one anothers' company, the cool wind emanating from a red supergiant is likely to be photoionised by its companions. In the case of a star moving supersonically through the ISM, the wind will be terminated by a bow--shock, leading to a layered structure where a parabola--shaped outer shock encases an ionised wind separated from the supergiant's neutral wind by a spherical neutral shell confined by the ionised wind. The net effect is to confine a large quantity of material (up to about one third of the total wind mass) in a dense shell near the star. This shell becomes important when the star eventually explodes, since the blast wave will promptly collide with the shock, potentially radiating a large fraction of its energy.\\
\indent From the point of view of the effects feedback on the surrounding ISM, \cite{2014MNRAS.442..694D} and \cite{2015ApJ...798...32N} studied, at very different scales, the combined effects of photoionisation and momentum--driven winds.  \cite{2014MNRAS.442..694D} essentially repeated the simulations of \cite{2012MNRAS.424..377D} and \cite{2013MNRAS.430..234D} with the inclusion of stellar winds, having already established in \cite{2013MNRAS.436.3430D} that winds acting on their own were not a very effective means of dispersing any of their model clouds. The results of their simulations were in fact little different from the photoionisation--only simulations, as shown in Figure \ref{fig:dale14}. At very early stages, the winds helped to clear away the dense filamentary gas in which the O--stars in these calculations are born, but the structure of the cold gas rapidly comes to be dominated by the effects of the expanding HII regions. The structures of the HII regions themselves were observed to be different, since the winds expel much of the ionised gas through low--density channels, excavating large holes inside the HII regions and leaving the ionised gas as a thin skin lining the inside of the shocked bubbles of cold gas in a fashion reminiscent of the simulations of \cite{2006ApJ...638..262F}.\\
\indent \cite{2015ApJ...798...32N} instead modelled the effect of including winds in the radiation--driven implosion models of \cite{2011ApJ...736..142B}. They also observed that, while there are some visible impacts on the morphology of the ionised gas, the cold gas is almost entirely insulated from the winds by the HII region.\\
\indent \cite{2014ApJ...788...14P}'s {\sc flash} simulations include radiative heating from ionising and non--ionising photons, and from jets. They start from the same initial state as \cite{2010ApJ...711.1017P} and the collective outflow driven by the stars, being directed along their rotation axes, act perpendicularly to the disc, rapidly exiting the simulation box. The leading fronts of the outflow cones are Rayleigh--Taylor unstable, exhibiting elongated fingers, and the mass and momentum contained in the outflow are substantially larger than those driven by the HII regions. They find in fact that only jets are able to reproduce observed levels of mass and momentum entrainment in outflows in their simulations.\\
\subsection{Combining feedback and magnetic fields}
\begin{figure}
\centering
\includegraphics[width=0.75\textwidth]{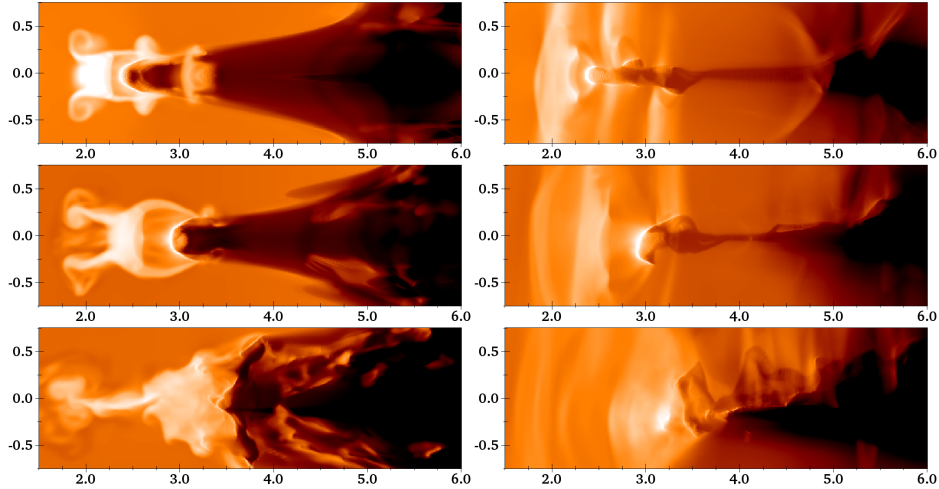}
\caption{Projections along the $z$--axis (left) and $y$--axis (right) of RMHD pillar simulations from \cite{2011MNRAS.412.2079M} at 100 (top) 250 (middle) and 400 (bottom) kyr. The magnetic field is aligned with the $z$--axis. The images are rendered in energy flux in recombination line radiation, arbitrarily normalised.}
\label{fig:mackey11}
\end{figure}

\indent The role of magnetic fields in ISM evolution remains unclear, partly due to the difficulty of measuring their strengths and geometries. Nevertheless, the typical flux densities present in molecular clouds are known to order--of--magnitude accuracy, sufficient for their effects to be meaningfully simulated. This is not the place to discuss their general effects, but it is clear that jets and bubbles will interact with magnetic fields, and some authors have begun to explore this issue.\\
\indent \cite{2007ApJ...671..518K} considered the effect of a uniform magnetic field on HII region expansion. They model the case where the initial gas density and magnetic field strength are such that the thermal sound speed in the promptly--ionised gas is much larger than the Alfv\'{e}n speed $v_{\rm A}$ in the undisturbed neutral gas, and $v_{\rm A}$ is much greater than the thermal sound speed in the neutral gas. At early times, the effect of the magnetic field is minimal, but it becomes significant when the magnetic pressure inside the swept--up shell becomes comparable to the thermal pressure inside the HII region, which occurs when the shell expansion velocity drops to near $v_{\rm A}$. They define a critical radius $r_{m}$ at which this occurs, give by $r_{m}=(c_{\rm II}/v_{\rm A})^{4/3}R_{s}$. From this point on, the HII region and the shell are deformed from the usual spherical shape. Gas motions along the field lines are unaffected, so that the expansion in these directions continues in a similar way to that in a normal HII region. Perpendicular to the field, expansion is resisted and the shell is initially supported by magnetic pressure. However, as the expansion proceeds, the field lines deform to become perpendicular to the shell surface, so the shell loses magnetic support and becomes thinner. Very similar results were obtained by \cite{2011MNRAS.414.3458W} in testing their RMHD implementation in {\sc enzo}, and by \cite{2011MNRAS.412.2079M}.\\
\indent \cite{2012ApJ...745..158G} build on the work of \cite{2007ApJ...671..518K} by considering also blister HII regions. The orientation of the magnetic field, not surprisingly, strongly affects the results, in particular the degree to which the magnetic field lines are distorted and the distribution of the gas kinetic energy in the swept--up shock. In particular, orienting the field parallel to the edge of the cloud results in the strongest compression of field lines, and in almost all the shell kinetic energy being stored around the lip of the depression in the neutral gas. In all runs, they observe that the gas kinetic energy is smaller than in the case with no magnetic fields, but that the \emph{total} energy is much larger, owing to energy stored in the magnetic fields. They propose that this could be an effective driver of turbulence.\\
\indent \cite{2011ApJ...729...72P} present an extension to the simulations of \cite{2010ApJ...711.1017P} in which they combine HII region expansion and magnetic fields in simulations of a 10$^{3}$M$_{\odot}$ rotating cloud. The cloud is initially threaded with a uniform magnetic field parallel to its rotation axis, with a mass--to--flux ratio fourteen times larger than that required to support it, consistent with observed field strengths. As expected, while the field cannot prevent gravitational collapse, it slows it, reducing the global star formation rate in the calculations by a factor of order unity. However, the most massive star, which forms at the centre of the disc, acquires a larger mass because magnetic braking drains the region of angular momentum, enhancing infall into the centre. As in their earlier calculations, the central regions of the cloud collapse into a flattened rotating disc--like structure, with the rotation and collapse dragging the magnetic field into a toroidal configuration. Once a sufficiently massive star has formed to drive an HII region, the influence of the magnetic field is largely to slow the HII region's expansion.\\
\indent \cite{2011MNRAS.414.1747A} in contrast examined the effects of magnetic fields on HII regions expanding in already--turbulent clouds. They simulate a 4pc magnetised turbulent box in which the mean atomic number density is 1000 cm$^{-3}$ and the ratio of thermal to magnetic pressure is 0.032. Into the centre of the box, they place either an O9 or a B0.5 star, whose ionising photon fluxes differ by 2.5 orders of magnitude. In the case of the brighter star, their box size is smaller than the critical radius defined by \cite{2007ApJ...671..518K} at which the HII region expansion should start to feel the influence of the magnetic field. However, in the case of the fainter star, whether or not this is true depends on whether the RMS or mean magnetic field strength is used to compute the representative Alfv\'{e}n speed in the neutral gas. They find that, in the former case and as expected, the presence of the magnetic field has little effect on the gas morphology produced by the HII region expansion, aside from some smoothing out of small--scale structures. In the B--star simulation, they find that the magnetic field does become critical during the simulation, but that it still fails to have a marked effect on the gas structure. They attribute this result to the disordered state of the field, which has no preferred direction except on small scales, and cannot therefore deform the HII region in a systematic or global fashion. In fact, because the pressure in the HII region exceeds the magnetic pressure for most of the duration of the simulations, the HII region expansion deforms the magnetic field into a more ordered configuration, with field lines approximately parallel to the ionisation fronts.\\
\indent \cite{2013MNRAS.436..859M} consider the influence of magnetic fields on the HII regions driven by runaway O--stars (using $\zeta$--Oph as an example). As in other studies, HII region expansion and the accumulation of dense material perpendicular to the field lines are strongly retarded. The opening angle of the cone trailed by the star is increased by the presence of the magnetic field.\\
\indent The interaction of supernovae and magnetic fields in turbulent 10$^{4}$M$_{\odot}$ clouds is investigated by \cite{2015A&A...576A..95I}, who find that the magnetic field has only a rather weak effect, modestly increasing the efficiency of momentum coupling between the supernova and the cold gas in the case where the explosion occurs deep inside the cloud.\\
\indent Some of the clearest signs of the action of feedback in real astrophysical systems are the production of unusual structure, such as pillars. If these are to be correctly interpreted, the role of magnetic fields in their formation and evolution must be understood. \cite{2009MNRAS.398..157H} conduct RMHD simulations of the formation and evolution of pillars using the C$^{2}$--RAY algorithm. Spherical globules with Gaussian density profiles are embedded in an initially uniform magnetic field aligned at a given angle to the x--axis, along which lies a point source of ionising photons. With magnetic pressures of order 100 times the initial gas pressure, the effects on the evolution of the system are profound and depend strongly on the orientation of the field relative to the direction of the source. When the field is perpendicular, the globule evolves to a flattened structure, whereas a 45 degree angle results in a very asymmetric globule which the authors describe as comma--shaped. Thin--shell instabilities caused by rocket acceleration cause the heads of the pillars to fragment into smaller objects. Similar results are obtained by \cite{2011MNRAS.412.2079M} (an example is shown in \ref{fig:mackey11}), who also observe that radiation--driven implosion and rocket acceleration both tend to align the magnetic field with the radiation field. The morphologies in the simulations can be used indirectly to constrain the field strengths in real objects, such as the famous M16 pillars. High field strengths produce sheet and ribbonlike structures which should be easily observable, but are not seen in M16. Emission in that region is instead consistent with that seen in the simulations where the accretion flow from the pillar head is roughly spherically symmetric.\\
\indent Regardless of their direct interaction with a given feedback mechanism, magnetic fields will always influence to some extent the environment in which feedback operates. This issue is explored by \cite{2009MNRAS.398...33P}, who extend the work of \cite{2009MNRAS.392.1363B} by repeating their calculations with magnetic fields of various strengths included. They find that magnetic fields and thermal accretion feedback (which do not interact directly) are additive, in the sense that they act at different scales. The magnetic fields, if they are strong enough, affect the large--scale collapse or contraction of the cloud, exerting an influence on gas of all densities. Accretion feedback, by contrast, only operates on the smallest scales by suppressing fragmentation in the densest material. The average stellar mass formed in their calculations is lower than in pure hydrodynamic calculations.\\
\indent When considering feedback from jets, however, the interaction with the magnetic field is more direct. \cite{2010ApJ...709...27W} model the effect of jets on magnetised clouds, where the jet emission is aligned with the local field direction. Magnetic fields on their own do not much influence the denser material from which their sink particles are accreting, but the inclusion of outflows partially disrupts the dense gas and slows accretion substantially. Combining the two process increases this effect, since the magnetic fields couple gas over much large distances than local pressure forces are able to, reducing total mass accretion rates and those onto individual stars by around an order of magnitude.\\
\indent \cite{2014MNRAS.439.3420M} perform simulations of the formation of a star cluster including radiative and outflow feedback from protostars, and magnetic fields. Contrary to \cite{2009MNRAS.398...33P}, they find that their mean stellar mass increases. They interpret this as being due to the mechanism which actually terminates accretion in many cases in the \cite{2009MNRAS.398...33P} simulations, which is the ejection of sink particles from the densest gas by N--body interactions with other sinks, in a globally contracting cloud. This shuts off accretion onto the sink and fixes its final mass. \cite{2014MNRAS.439.3420M} do not simulate a globally--infalling system, and most of their sinks form from fragmentation. As well as slowing the star formation rate by factors of around two, the magnetic fields also suppress fragmentation, decreasing the numbers of sinks and raising the average fragment mass.\\
\subsection{Large--scale simulations}
\indent In larger--scale simulations, the star formation process itself usually cannot be directly resolved, certainly in terms of individual stars. It is often modelled by assuming that all the gas above some threshold  density generates stars with some efficiency on some timescale, which is often taken to be the freefall time at the threshold density, i.e.
\begin{eqnarray}
\dot{\rho}_{*}=\eta\frac{\rho(>\rho_{\rm crit})}{t_{\rm ff}(\rho_{\rm crit})}
\end{eqnarray}
With this prescription, the star formation rate automatically reproduces something close to the Schmidt--Kennicutt (SK) law. Some simulations do not impose such a law but try to recover it. The SK law is the result of regulation of star formation. What is the regulator is unclear, but clearly little can be said about the role of feedback if the SK law is imposed from the outset. However, large--scale simulations have other goals, such as reproducing the multiphase ISM and driving turbulence, and reproducing the structure of galaxies, especially spirals.\\
\subsubsection{Vertical slices or columns through a disc and shearing boxes}
\indent Here, simulators model a vertical slice or column through a galactic disc, often including tidal and Coriolis forces as external perturbations. These simulations are generally most concerned with the dynamical equilibrium between the vertical gravitational field and the thermal and dynamic pressure on the gas.\\
\indent The effects of expanding HII regions on galactic disc scales are explored by \cite{2009ApJ...693.1316K}. They produce a turbulent, multiphase ISM, although the velocity dispersion of the turbulence (2--4 km s $^{-1}$) is lower than would be expected on these scales (7--10 km s$^{-1}$). They estimate the Toomre Q parameters of their model discs and find that they equilibrate into marginally stable states. The star formation rate surface densities also settle into equilbria which are mostly well described by power laws resembling the SK law (although they also find this for their purely hydrostatic test case, so caution against over--interpretation of this result).\\
\indent \cite{2011ApJ...731...41O} and \cite{2011ApJ...743...25K} discuss a model of galactic evolution in terms of dynamical equilibria, where heating balances cooling, pressure balances gravity and turbulent driving balances dissipation. They inject momentum from star--forming events and take SNe to be the main sources of feedback. Radiative heating, which is assumed to come from massive stars proportionally to the star formation rate surface density, and from a constant galactic radiation field, is introduced via a volumetric heating rate. An equilibrium between the vertical gravitational acceleration and feedback--driven turbulence is rapidly achieved, heating and cooling and star formation rates settle into equilibrium, and thermal and turbulent pressures settle into nearly linear relations with the star formation rate surface density. This result, combined with the vertical dynamical equilibrium straightforwardly implies that the star formation rate surface density is proportional to the weight of the ISM.\\
\indent Similar results are obtained in models of ULIRGs by \cite{2012ApJ...754....2S}, who include momentum input from supernovae. A dynamical equilibrium establishes itself on a timescale short compared to the orbital time within the disc in which the input momentum flux balances the vertical gravitational field. The SNe occur only in the densest gas in the midplane but are able to accelerate gas to altitudes of more than 100pc, driving turbulence and resulting in an ISM structure dominated by large shells.\\
\indent \cite{2013ApJ...775..109K} extend the work of \cite{2012ApJ...754....2S} and \cite{2011ApJ...743...25K} to three--dimensional shearing--box calculations. They observe the formation of cold cloudlets by thermal instability which gravitate toward and merge with one another in the disc midplane. The models again globally reach a dynamical equilibrium, which includes formation and destruction of gravitationally bound clouds.\\
\indent Numerous authors have come to the conclusion that the locations and times at which feedback is inserted into simulations are at least as important, if not more so, as the detailed physics implemented. \cite{2005MNRAS.356..737S} simulate a three--dimensional periodic non--driven turbulent box 1.28kpc on a side. Star formation is taken to occur in collapsing and rapidly--cooling regions whose density exceeds a threshold. Matter and energy released by winds and SNe are injected, smeared out over a characteristic timescale of either the local dynamical time, or 10Myr (whichever is smaller). The manner in which the supernovae are detonated strongly affects the resulting star formation rates. With the employed time delays, hot gas fills a progressively larger and larger fraction of the simulation box, driving up the density and star formation rate in the cool gas. However, if supernovae are detonated immediately in star forming gas, this does not occur and the star formation rates were observed to be up to two orders of magnitude lower. More recently, several other groups have explored the issue of the importance of supernova detonation sites. \cite{2015MNRAS.449.1057G} obtain radically different results from detonating SNe randomly in their turbulent ISM, or at peaks in the cold gas density (arguably more likely explosion sites). Random driving produces an ISM almost entirely filled with very hot (10$^{6}$--10$^{7}$K) gas with cold material being confined to clumps with small filling factors. Peak driving instead produces almost no very hot gas, although there are substantial volumes filled with gas at 10$^{4}$K. The cold material has a much larger filling factor and a filamentary topology. \cite{2014arXiv1412.2749W} use the supernova prescription of \cite{2015MNRAS.449.1057G} and examine two additional means of specifying SN sites, namely a mixture of random and peak driving, and clustered random driving where the SNe are correlated with other but not with the dense gas. They find that clustering the SNe has only a minor effect. Peak and mixed driving both leave the galactic discs strongly concentrated at their mid planes, again with a filamentary distribution of cold gas. Random driving results in an outflow perpendicular to the disc.\\
\indent \cite{2014A&A...570A..81H} examine the influence of the supernova detonation scheme on star formation in their {\sc ramses} models. Peak--driving reduces the star formation rates by factors of 20--30 owing to the rapid destruction of dense star--forming material. However, they also find that the details of the implementation of the supernova explosions themselves have a strong influence. Introducing the supernova energy in purely thermal form results in a star formation rate twice as high as injecting just five percent in kinetic form.\\
\subsubsection{Whole--galaxy models of spirals}
\indent Although they allow one to achieve better resolution with given computational resources, vertical--slice or shearing box simulations require substantial physical components of spiral galaxies, such as spiral arms, to be parameterised. Truly satisfying simulations should of course reproduce these features self--consistently. Besides this, self--regulation of star formation in the disc plane is a different (though obviously related) question to the vertical equilibria between gravity and pressure discussed in the previous subsection.\\
\indent In fact, the formation of realistic spiral galaxies, either individually or in the context of cosmological simulations is a long--running problem in computational astrophysics. The issue is catastrophic angular momentum loss, in which overcooling of collapsing baryons leads to the formation of too many low--angular--momentum spheroidal galaxies, and too few spirals. Early attempts to solve this problem include \cite{1991ApJ...377..365K}, \cite{1992ApJ...391..502K} and \cite{1994MNRAS.267..401N}, which still produced discs that were too small and centrally concentrated. \cite{2003ApJ...591..499A} suggested that the resolution of this issue lay in the regulation of star formation by feedback. The most important form of feedback at galactic scales has traditionally been considered to be supernovae.\\
\indent Notwithstanding these issues, several authors reached the conclusion that many properties of spiral galaxies could be reproduced \emph{without} the need to invoke feedback and that various instabilities, rotational shear and gravitational torques were of comparable or greater importance. \cite{2001ApJ...547..172W} find that gravitational and thermal instabilities produce a two--phase turbulent ISM in the absence of feedback. After implementing stellar wind and SN feedback from Lagrangian star particles, they observe that the overall morphology of their galactic discs and the gas density PDF are little changed. Both models produce velocity power--spectra indicative of turbulence, although the model without feedback has a steeper slope of -3, compared to -2 in the feedback model, which they attribute to the increased importance of shocks in the feedback calculation. Turbulence in both simulations is also driven by shear and self--gravity. \cite{2003ApJ...590L...1K} found that their simulations produce a good approximation to the SK law with or without feedback. In common with \cite{2001ApJ...547..172W}, they find that the effect on the density PDF on galactic scales is small, although more low--density gas is produced by feedback. Similar results were reported by \cite{2007ApJ...660..276W}, despite their rather high supernova rate of 1.5$\times10^{-3}$yr$^{-1}$kpc$^{-2}$. \cite{2013MNRAS.436.1836R}, who include SN and photoionising feedback in their {\sc ramses} models of a Milky Way--type galaxy, concluded that feedback was not very efficient in destroying the clouds because of drift between the clouds and the star particles, ensuring that most supernovae do not explode inside clouds. Star formation was regulated more by large--scale turbulent support than directly by feedback.\\
\indent Other groups have concluded that supernova feedback acting alone can be important, but only if either the supernova rate is increased to larger values than observed, as in the aforementioned studies, of if the individual supernova energies are increased. \cite{2008ApJ...684..978S} perform two--dimensional simulations, some with an external potential to induce the formation of spiral arms, including the effects of SNe modelled as injection of momentum to avoid the overcooling problem. Star formation is restricted to gas above a density threshold and SNe are introduced probabilistically each timestep. Models with and without spiral potentials both required very strong levels of feedback to influence the evolution of these structures, resulting in positive feedback from the collisions of adjacent bubbles and a rough equilibrium between cloud formation and destruction. The \emph{overall} effect is to decrease star formation rates, even if positive feedback occurs locally. In AMR simulations of disc galaxy formation, \cite{2011MNRAS.410.1391A} included Type Ia and II supernova feedback, treating the star formation threshold density, efficiency and the supernova energy as free parameters. They found that the disc surface density profiles and relative contributions of gas and stars to the surface density depended strongly on the parameterisation of the star formation rate, whereas the self--regulation of star formation was controlled by the supernovae energies. They were only able to achieve self--regulation for explosion energies of $5\times10^{51}$ergs, rather higher than canonical values.\\
\indent Many simulations are, however, significantly influenced by supernova feedback, both in terms of the rates of star formation and the structure of the resulting galaxies. These two issues are increasingly found to be connected -- the distribution of star formation in time controls the final structure of simulated galaxies. \cite{2004ApJ...606...32R} employ the \cite{2003MNRAS.339..289S} multiphase ISM model to model a set of galaxy haloes. One of the ten most massive haloes in their simulation hosts a well--resolved disc galaxy with no strong bulge component, exhibiting instead an exponential surface brightness profile. Similar results are reported by \cite{2005MNRAS.363.1299O} and \cite{2008MNRAS.389.1137S} who see that feedback suppresses early collapse, leaving large quantities of hot gas to cool at late times, from which an extended disc of young stars eventually form. More powerful feedback favoured the disc over the spheroid as expected, although the disc--to--spheroid mass ratio never much exceeded unity.  \cite{2012MNRAS.423.1726S} present the Aquila cosmological code comparison project, in which different implementations of feedback are tested against each other in different Eulerian and Lagrangian codes, as well as the {\sc arepo} hybrid code. All the codes implement SN feedback, but using different schemes, although most inject SN energy in thermal form. Galaxy morphology is largely independent of which hydro scheme is employed and depends instead on the distribution of star formation in time. Forming realistic discs requires that feedback delay star formation so that it continues to late times. However, all the codes formed stellar discs which are too concentrated, implying a failure to prevent the accretion of low--angular--momentum material. This in turn is attributed to the inability of the feedback mechanisms employed to prevent a strong initial burst of star formation. \cite{2013MNRAS.434.3142A} use {\sc gadget}--3 simulations to study the formation of spiral galaxies in cosmological simulations. Supernova feedback is implemented along with radiation pressure, as in \cite{2011MNRAS.417..950H}. Feedback, in particular pre--supernova feedback, helps in production more realistic galaxies by delaying feedback to later times.\\
\indent Galactic winds have a role to play in the angular momentum problem, since they preferentially expel low--angular momentum material from haloes. Additionally, real galactic discs have much smaller baryon quantities than simulations without feedback predict. IGM enrichment indicates that much of the IGM must have been inside halos at some point to be enriched, then expelled again. Galactic super winds with mass loss rates much larger than the SFR are generally required.\\
\indent \cite{2006ApJ...641..878T} perform three--dimensional simulations using the {\sc enzo} code, with similar star formation criteria and a similar technique to smear out the effects of Type II SNe feedback in time as \cite{2005MNRAS.356..737S}. The major effect of feedback is to drive a kpc--scale galactic fountain, cycling gas out of the galactic plane, and causing the disc in general to puff up. Feedback also increases the star formation timescales so that it is at least to some extent self--regulating, but the star formation rates are still substantially higher than observed in real galaxies. \cite{2008A&A...477...79D} simulate the formation of stellar feedback--driven galactic winds in {\sc ramses}. They allow star formation to proceed in the centre of an isolated rotating halo, forming a disc structure. Haloes with different masses, spin parameters, virial velocities and star formation timescales are considered, under the influence of SN feedback. In the low--mass ($10^{10}$M$_{\odot}$) haloes, the combined SNe blow bipolar wind cavities, with compact discs forming in haloes with low spin parameters being the most efficient wind drivers. In the high--mass $10^{11}$M$_{\odot}$ haloes, the feedback--driven outflow is stalled by continuing accretion, leading to an accretion shock around the disc, and no galactic winds. They conclude that the infalling gas from the halo is a critical determinant in the formation of the wind. However, even in the low--mass haloes, the mass ejection efficiency is about an order of magnitude lower than observations suggest it should be, and they are unable to solve the overcooling problem in their simulations. \cite{2012MNRAS.424.1275B} use the {\sc gasoline} SPH code to study the formation of galaxies over a wide range of masses. Early radiative feedback and supernovae are both implemented. Feedback--driven outflows remove low--angular momentum material, resulting in more realistic extended discs and removing dark matter cusps.\\
\indent Supernovae are generally regarded as the most important feedback mechanism at galactic scales, but several authors have concluded that they cannot on their own reproduce observed galactic properties and that in fact other forms of feedback, such as HII regions and radiation pressure do have important contributions to make. \cite{2015MNRAS.450.1937C} make the point that the energy, momentum and mass fluxes into the ISM from star--forming regions and supernovae are not known, and nor are the magnitudes of the prompt losses of energy by radiation, and of momentum by cancellation. In any case, these losses occur at scales too small to be resolved.\\
\indent Many groups have tackled this problem by varying the `strength' of feedback in their simulations and exploring the consequences for observable properties such as molecular cloud properties or the star formation rate.  \cite{2011MNRAS.417.1318D}, for example, simulate whole galactic discs in SPH with a background potential to model spiral arms. They include energy from supernovae split 2:1 between kinetic and thermal components, and with various assumed star formation efficiencies in dense gas. The higher star formation efficiencies (and therefore levels of feedback) produce populations of GMCs which are predominantly unbound and irregularly shaped, although they note that collisions between clouds also contribute substantially to their internal velocity dispersions. \cite{2013MNRAS.432..653D} quantify this result further and find that it is mostly the smaller ($\sim10^{4}$M$_{\odot}$) clouds that are strongly affected by feedback, whereas the larger $\sim10^{6}$M$_{\odot}$ are more affected by shear. \cite{2015MNRAS.447.3390D} resimulate portions of \cite{2013MNRAS.432..653D}'s calculations at higher resolution and explore different prescriptions for the timing of the injection of supernova feedback, including delayed and stochastic models and smooth and instantaneous energy injection, but they find that in practice their choices have little impact on the outcome of their simulations.\\
\indent \cite{2014MNRAS.443.2092U} explicitly test the effect of varying the strength of combined--feedback models. Their strong feedback model is derived from \cite{2013MNRAS.434.3142A} and includes SNe kinetic and thermal feedback in a multiphase decoupled ISM as described by \cite{2003MNRAS.339..289S}, as well as radiation pressure, while their weak feedback model includes only SNe. The star formation efficiency of the weak--feedback model is always too high, and the resulting galaxy is spheroidal, whereas the strong feedback scheme gives disc--like stellar distributions and credible baryon conversion efficiencies at z=2--3, but too few stars at z=1. Higher star formation rates in the weak feedback case also result in older, redder stellar populations.\\
\indent \cite{2014MNRAS.442.1545C} improve on the basic thermal feedback from winds and SNe in the {\sc art} code (\cite{1997ApJS..111...73K}) by implementing radiation pressure feedback. They find substantial differences, with radiation pressure driving additional turbulence in galactic discs and destroying the most compact and high--column density clumps, which are similar to normal GMCs and substantially less massive than the giant clumps of the kind observed in moderate redshift galaxies (e.g. \cite{2011ApJ...733..101G}). The stellar distribution is correspondingly more extended and less concentrated, and star formation rates drop by factors of a few. The formation of smaller satellite galaxies is almost entirely terminated.\\
\indent The potential of feedback to reduce the over--formation of stars in cosmological SPH simulations is explored by \cite{2013MNRAS.428..129S}, who found that supernova feedback on its own is not sufficient and turn to pre--supernova radiative feedback. They do not include kinetic feedback as, for example, in \cite{2011MNRAS.417..950H}, because molecular clouds in their simulations are not resolved, being represented by only a few particles. They instead introduce thermal feedback over a 4 Myr period into these few particles. They do not disable cooling, so the gas rapidly cools to $10^{4}$K. The large--scale dynamical effects are negligible, but star formation is locally suppressed. They adjust the efficiency of energy uptake to reproduce the required stellar-- to halo--mass ratio, and find that the resulting star formation rate is highly sensitive to this parameter.\\
\indent A suite of AMR simulations of an isolated disc galaxy is executed by \cite{2013ApJ...770...25A} in which the feedback prescription is varied. Momentum is injected either by momentum kicks or as a non--thermal form of pressure and different values of the IR optical depth for momentum absorption are tested, cooling delays of 10 or 40 Myr after energy injection are implemented, and a passively--advected decaying energy variable akin to that used by \cite{2013MNRAS.429.3068T} is also explored. They conclude that pre--supernova feedback is a crucial ingredient in their models, capable of reducing star formation rates and efficiencies by an order of magnitude on its own, and averting the overcooling problem by destroying dense gas before SNe can detonate inside it. This very rich suite of feedback models was used by \cite{2015ApJ...804...18A} to show that feedback is only effective in regulating star formation rates if local star formation efficiencies are high enough that there are strong spatial and temporal correlations between the injection of momentum and energy, but that in such cases, the results are sensitive to the details of the feedback prescription.\\
\indent \cite{2014MNRAS.444.1389M} test two models of feedback in their simulations of the formation of galaxies, and in particular of the very massive (10$^{7}$--10$^{9}$M$_{\odot}$) and extended ($\sim1$kpc) clouds observed in star--forming galaxies at redshifts $z\sim1-3$ (e.g. \cite{2011ApJ...733..101G}). The first model encompasses only thermal feedback from stellar winds and SNe, modelled as a constant heating rate for 40Myr after a star formation event. The second model includes radiation pressure from the ionising radiation fields of the sources, suppressing the formation of lower--mass clumps, and slowing star formation by factors of around two. The addition of radiation feedback produces a much stronger decrease in the star formation rate in their simulations than SN feedback on its own, but they still find that they overproduce stars by factors of a few. They also find that stellar feedback reduces the bulge:disc ratio and produces discs with flat rotation curves, but which are too thick and only poorly rotationally--supported.\\
\indent \cite{2015MNRAS.446..521S} implement star formation stochastically via a scheme proposed in \cite{2008MNRAS.383.1210S} in which the rate is set by the local pressure, and which automatically reproduces the SK relation. Star formation events inject energy and momentum from radiation, winds and SNe. Energy is injected via a temperature increment to avoid spurious overcooling, with an efficiency factor which sets how much energy is available for feedback. They adjust parameters such as the density threshold for star formation and the efficiency of energy uptake, calibrating the results using the $z=0.1$ galactic stellar mass function as a yardstick. Since feedback regulates the supply of gas to the ISM, the strength of feedback influences a wide variety of properties of the formed galaxy population, including central black hole masses, spatial concentration of stars, star formation rates and metallicity.\\
\indent Other authors, in particular \cite{2011MNRAS.417..950H}, \cite{2012MNRAS.421.3488H}, \cite{2013MNRAS.430.1901H} and \cite{2014MNRAS.445..581H} attempt to model feedback in a physically--motivated manner with as few adjustable parameters as possible. These papers model isolated and merging galaxies of several different types -- high--redshift objects, and local Milky Way--type and dwarf galaxies. The structures of the merging galaxies are largely controlled by gravitational torques, and the violently unstable high--redshift galaxy is less affected by feedback and more by gravitational instability. However, the ISM in their model galaxies generally reaches a steady state after a few dynamical times in which the star formation rate and feedback keep the galactic discs on the edge of Toomre instability, as observed by \cite{2009ApJ...693.1316K}. Good approximations to the SK law emerge naturally from this state. For the larger galaxies, radiation pressure is the dominant feedback mechanism, particularly in driving large--scale outflows. However, the density of their dwarf model is low enough, and the corresponding cooling time long enough, that thermal feedback from SNe and winds become important. \cite{2014MNRAS.445..581H} conclude that feedback is both necessary and sufficient to regulate star formation, but only when all their feedback mechanisms act in concert. The results of these simulations are an interesting contrast to those of \cite{2011ApJ...731...41O} and \cite{2011ApJ...743...25K}, who discuss \emph{vertical} stability maintained by an equilibrium between gravity and dynamical pressure acting perpendicular to the disc plane.\\
\indent \cite{2014MNRAS.444.2837R} study disc galaxy formation in {\sc ramses} and examine the importance of pre--supernova radiative feedback as well as supernovae. They vary the dust opacity so that the gas absorbs larger and larger quantities of momentum form the stellar radiation field. They find that radiation feedback is able to avert the overproduction of stars by expelling gas from their model halo. However, they also find that this result cannot be achieved without severely disrupting the morphology and dynamics of the disc.\\
\subsubsection{Dwarf galaxies}
\indent  Spiral galaxies have received a lot of attention from modellers, not least because we live inside one. However, dwarf galaxies are the building blocks of larger systems, are the most dark--matter dominated objects known, and are enigmatic in the sense that cosmological simulations produce many more of them than are actually observed. They are thus a rich field of study, and their smaller masses makes the effect of feedback upon them more dramatic.\\
\indent Many authors have addressed the self--regulation of star formation in dwarf galaxies which can, under the influence of feedback, operate in the the so--called breathing mode (\cite{2007ApJ...667..170S}), where gas is expelled into the halo but eventually reaccretes to form a new generation of stars, which restart the cycle. This is described in detail by \cite{2014MNRAS.438.1208K}, who perform SPH simulations including stellar winds and SNe, with radiative cooling disabled for the duration of feedback. Star formation in the feedback simulation becomes intermittent, and large fractions of the galaxy volume are cleared of gas. The expanding bubbles driven by the earliest--forming stars collide with gas still accreting onto the galaxy and a second round of star formation is induced. Feedback from the second generation of stars increases the bubble size further, temporarily terminating star formation. However, the bubbles are not able to expel gas entirely from the galaxy's halo and it eventually reaccretes for yet another round of star formation.\\
\indent \cite{2013ApJ...776....1K} model dwarfs using the {\sc enzo} code, injecting matter and energy from SNe with a characteristic time delay of 6Myr, and including transport of ionising radiation using the methods of \cite{2011MNRAS.414.3458W} and \cite{2011ApJ...743...25K}, with star particles given ionising fluxes proportional to their masses during the 6Myr interval before the SNe. Both heating and momentum from the radiation field are included. They observe that star formation becomes self--regulating after a transient period of order 10Myr, and that the addition of radiative feedback slows the star formation rate by roughly one fifth compared to models involving only SNe.\\
\indent Some dwarf galaxy simulations reveal interesting parallels with simulations conducted at much smaller molecular cloud scales. \cite{2013ApJ...767...59P} use the {\sc traphic} code to investigate the formation of the first galaxies under the influence of radiative feedback from the first stars. The primary coolant in the metal--free gas is rovibrationally excited molecular hydrogen, which is destroyed by Lyman--Werner band photons, so that stars deprive forming galaxies of their ability to cool. Secondly, hydrogen--ionising photons heat the gas, driving it out of haloes with virial temperatures lower than $10^{4}$K, and slowing accretion onto lower--mass objects. However, photoionisation also produces free electrons, catalysing molecular hydrogen formation and increasing the ability of the gas to cool. They find, as in smaller--scale simulations, that the effects of radiation are constrained by inhomogenous gas structures, in this case filamentary accretion flows feeding baryons into their simulated forming dwarf galaxy. The similarity between this and the process observed by \cite{2005MNRAS.358..291D} is striking. The reduction in accretion is sufficient to terminate star formation, eventually enabling the gas to cool and to undergo another burst of star formation, by which time the halo is too massive to lose any further baryons due to feedback, and the second starburst lasts longer. The virial temperature of the halo becomes high enough to enable atomic hydrogen cooling and the galaxy forms stars continuously, although more slowly than in the case with no radiative feedback, anf it is the photoionisation heating and not the Lyman--Werner photodissociation that makes the difference. As the simulation proceeds, the gas baryon density in the halo rises and the expansion of the HII region is retarded by the high recombination rates (as in \cite{2005MNRAS.358..291D} and \cite{2010ApJ...711.1017P}), at which point the simulation effectively ceases to respond to feedback. Radiative feedback also effects, via gas expulsion and redistribution, the \emph{dark matter} distribution, decreasing the central density. The structure of the baryonic discs is largely unchanged, although photoheating suppresses star formation in neighbouring haloes as well.\\
\indent Other authors model situations in which star formation is overall slowed or outright terminated by feedback. \cite{2012MNRAS.427..311W}, who consider radiation pressure from population III and II star formation, implemented using the scheme of \cite{2011MNRAS.414.3458W}. They focus on the evolution of the most massive halo in their simulation, finding that radiation pressure reduces the overcooling problem and reduces star formation by a factor of $\approx5$. Thermal and momentum transfer from the stellar radiation field create a $\sim100$pc expanding turbulent region centred on the main concentration of stars.\\
\indent In many of these models, \emph{external} feedback is at least as important if not more so than internal processes. \cite{2008ApJ...679..925W} examine the external irradiation of minihaloes by massive primordial stars. Their results show interesting parallels with the much smaller--scale, present--day calculations of, for example, \cite{2011ApJ...736..142B}. They show that radiative feedback accelerates star formation in relative dense and evolved haloes, but terminates it in more diffuse structures.\\
\indent \cite{2011MNRAS.412..935P} used their optically--thin variable Eddington tensor radiation transport scheme to examine the feasibility of cosmic reionisation by star--forming galaxies, and the consequences on the global star formation rate. Photo--heating of the intergalactic medium significantly slows the accumulation of baryons in dark matter haloes, particularly for the lower--mass haloes, and slows star formation by a factor approaching two. Although they are mainly interested in the effects of supernovae, \cite{2010MNRAS.402.1599S} find that background UV radiation is of crucial importance in their simulations. The combined mechanisms are sufficient to empty dwarf haloes of gas and terminate star formation by z=6. \cite{2013MNRAS.432.1989S} reached similar conclusions using {\sc enzo}. Their halo's quantity of dense star--forming material is largely regulated by SN feedback, but the total gas fraction is largely set by photoheating, and once it is cleared of gas, it cannot reaccrete.\\
\indent Reionisation and photoheating is also examined by \cite{2015arXiv150101980P}, who point out that reionisation is self--regulating. Photoheating boils gas out of low--mass haloes and shuts down accretion, lowering the ionising emissivities of galaxies. However, a positive feedback is also exerted, since the heating smooths out density inhomogenieties and reduces the recombination rate in the IGM. SN feedback also affects reionisation, since it clears out gas from low--mass haloes, shutting down star formation, although it also potentially locally \emph{triggers} star formation and opens low--density escape routes for photons, which may enhance reionisation. Star formation is suppressed by almost an order of magnitude by feedback between $z=14$ and $z=6$, with SNe being much more significant than photoheating, although the two mechanisms complement each other, suppressing star formation more when acting together than alone. SNe suppress star formation evenly across most of the range of galactic masses, whereas photoheating suppresses star formation only in the lower mass haloes. In both cases, suppression occurs by destruction of dense gas and by reducing accretion.\\
\indent The internal structures of dwarf galaxies are also sensitive to feedback. The core--cusp problem in dwarf galaxies is discussed by \cite{2013MNRAS.429.3068T}. While pure N--body models predict that dark--matter haloes have central cusps, observations suggest that they instead have constant density cores. Of many possible solutions, the least outlandish is the modification of the baryon distribution and hence the dark--matter distribution by feedback. They examine this question with the {\sc ramses} AMR code. The effect of feedback is to produce a thick, turbulent gas disc supporting outflows and emptying the central regions of baryons. The star formation rate is lowered, on average, by an order of magnitude and is much more staccato in nature. Gas cools and collapse, starting a burst of star formation which expels gas from the disc, shutting star formation down. However, gas falls back onto the disc and triggers another round of star formation, and the cycle repeats. The effect of gas removal form the central regions is indeed to transform the dark--matter distribution there from a cusp to a core after $\approx0.5$Gyr.\\
\section{Summary and outlook}
The last decade has seen the development of a tremendous variety of sophisticated algorithms to model the various kinds of stellar feedback, and a corresponding wealth of simulations employing these algorithms to answer a wide variety of questions over a huge range of scales. We have learned an enormous amount from these simulations.\\
\indent Feedback regulates or helps to regulate the rate at which gas is converted to stars at every stage of the star formation process. The background cosmic ionising radiation field controls the accumulation of baryons in primordial haloes, and supernova and radiation pressure feedback are capable of emptying haloes of gas and terminating star formation inside them (e.g. \cite{2010MNRAS.402.1599S}). Most calculations agree that the cycling of baryonic matter between the hot and cold phases of the ISM and the formation of GMCs is influenced -- if not controlled by -- feedback. Other mechanisms, such as gravitational torques, are of course involved and the relative contributions of the various processes is still a matter of debate. It seems to be increasingly clear that, whether they are dominant energetically or not, SNe are not the only \emph{important} form of feedback and other mechanisms, particularly radiation pressure, cannot be ignored. This is particularly true in simulations of dwarf galaxies, whose lower escape velocities make them vulnerable to radiative feedback (e.g. \cite{2010MNRAS.402.1599S}, \cite{2015arXiv150101980P}).\\
\indent However, the majority of galactic--scale simulations are still not able to resolve these processes. Some authors have parameterised the strength and form of feedback and varied the parameters until acceptable fits to some observable metric(s) are obtained (e.g. \cite{2015MNRAS.446..521S}). A more satisfying approach, taken for example by \cite{2014MNRAS.445..581H}, is to try to devise physically--motivated subgrid models, but even these must rely on some physics, such as the leakage of ionising photons, which cannot be resolved in the simulations themselves.\\
\indent Simulations at GMC scales have the advantage that they have much better length and mass resolution and have shown that all forms of feedback play some role in regulating the rate and efficiency of star formation in these objects, from radiation pressure on accretion flows at sub--AU scales to winds, HII regions and supernovae up to $\sim100$pc scales. However, most simulations still overproduce stars and none are yet capable of terminating star formation and reaching `completion'.\\
\indent These models are capable of realistically reproducing the geometrical structure of clouds and therefore also quantities which depend on this, again such as photon leakage, can be much more accurately measured. So far, almost no effort has been made to connect simulations performed at these smaller scales to galactic--column, galactic--disc or cosmological calculations. The GMC--scale models are presently an untapped resource which could supply more accurate parameterisations of many quantities of use in the larger--scale simulations. However, simulations by \cite{2014MNRAS.442..694D} suggest that the permeability of clouds to photons, momentum, energy and polluted ejecta, is a cloud--dependent property, making its parameterisation more difficult.\\
\indent In addition, none of the GMC--scale simulations yet include \emph{all} feedback modes. It is often said that HII regions are likely to be the most important feedback mechanism on GMC scales, at least until the detonation of the first supernova. While this is probably true, it does not mean that other types of feedback can be ignored. Some molecular clouds, such as Ophiuchus, are too small to form any OB stars and are of necessity dominated by accretion feedback. Such clouds are the most common by number and, in galaxies such as M33, which has a very steep GMC mass function, they also dominate the molecular mass. \cite{2009ApJ...703..131O} makes the point that even in clouds that are massive enough to manufacture O--stars, accretion is still the dominant mode \emph{before} these massive stars are born (and may continue to be even afterwards in places which are shielded from ionising photons). Accretion feedback therefore does help determine the environments in which the massive stars form, particularly cloud properties such as the star formation efficiency at that epoch. This can also be said with regard to magnetic fields.\\
\indent Two problems which are therefore of crucial importance are determining how all the different feedback types interact with one another in clouds with various different properties (density, escape velocity, geometry, etc.), and how magnetic fields contribute to this picture. It is clear from work already done that different feedback mechanisms are not necessarily additive (e.g. \cite{2014MNRAS.439.3420M}), and that the likely most important mechanism -- HII regions -- can be strongly affected by magnetic fields (e.g \cite{2012ApJ...745..158G}). In addition, the work of  \cite{2014A&A...570A..81H} shows that the effects of SN feedback on the largest scales is likely to be strongly dependent on the details of the environment in which the massive stars explode, which sets the relative quantities of thermal and kinetic energy deposited. Teasing out all these interactions requires a great deal of painstaking work, particularly if we wish to explain, as opposed to simply reproduce, the evolution of molecular clouds.\\
\indent They are some important disagreements between models that need to be resolved. Those that have emerged between flux--limited diffusion and variable Eddington tensor radiation transport methods are of particular concern. However, the question of why some galactic disc simulations require feedback to produce realistic galaxy properties whilst some can achieve them without feedback (e.g \cite{2007ApJ...660..276W}) is also curious and needs to be explored, as is the differing opinions of, for example, \cite{2013MNRAS.435.1701C} and \cite{2014MNRAS.442..694D} on how efficiently ionisation is able to disperse intermediate--mass GMCs. While much progress has been made, and the pace is accelerating, many of the details of the effects of feedback on star formation, the ISM and galactic structure are still murky.\\
\section{Acknowledgements}
The author is grateful for the support of the DFG cluster of excellence `Origin and Structure of the Universe'. In writing this review, he has made extensive use of the {\sc nasa/sao ads} literature search engine and of the {\sc papers} software package (\url{http://www.papersapp.com}), without either of which the process would have been much longer and more tedious. The author is also grateful to David Hubber for useful discussions of numerical methods, and to the referee, Ant Whitworth, for a characteristically careful reading of the manuscript.\\
\bibliography{narevbib}

\end{document}